\documentclass[a4paper,11pt]{article}
\pdfoutput=1 

\usepackage{jcappub}
\usepackage[capitalise, noabbrev]{cleveref} 
\usepackage{physics} 
\usepackage{graphicx}
\usepackage{subcaption} 
\usepackage{enumitem} 

\newcommand{\half}{\frac{1}{2}}
\newcommand{\recip}[1]{\frac{1}{#1}}
\renewcommand{\vec}[1]{\boldsymbol{#1}}
\newcommand{\field}{\vec{\phi}}
\newcommand{\vv}{\vec{v}}
\newcommand{\atvev}[1]{\left. #1 \right\vert_{\vv}}
\newcommand{\mitwo}{m_i^2}
\newcommand{\mitwop}{\mitwo(\field)}
\newcommand{\mitwov}{\mitwo(\vv)}
\newcommand{\renorm}{\mu_R^2}

\newcommand{\vol}{\mathcal{V}}

\newcommand{\phif}{\field_f}
\newcommand{\phit}{\field_t}
\newcommand{\phigs}{\field_{\text{gs}}}
\newcommand{\phicur}{\field_{\text{cur}}}
\newcommand{\Vext}{\vol_t^{\text{ext}}}
\newcommand{\Next}{N^{\text{ext}}}
\newcommand{\Vf}{V(\phif(T), T)}
\newcommand{\Vt}{V(\phit(T), T)}
\newcommand{\Vfz}{V(\phif(0), 0)}

\newcommand{\Vgsz}{V(\phigs(0), 0)}
\newcommand{\rhof}{\rho(\phif, T)}
\newcommand{\rhot}{\rho(\phit, T)}

\newcommand{\rhoh}{\rho_H}
\newcommand{\rhor}{\rho_R}
\newcommand{\rhov}{\rho_V}
\newcommand{\rhoz}{\rho_0}
\newcommand{\VTL}{V_0}
\newcommand{\VCW}{V_{CW}}
\newcommand{\VFT}{V_{T}}
\newcommand{\Vd}{V(\field, T)}
\newcommand{\Vphys}{\vol_{\text{phys}}}
\newcommand{\VTLd}{\VTL(\field)}
\newcommand{\VCWd}{\VCW(\field, T)}
\newcommand{\VFTd}{\VFT(\field, T)}
\newcommand{\FED}{\mathcal{F}}
\newcommand{\FEDd}{\FED(\field, T)}

\newcommand{\gev}{\ensuremath{\,\text{GeV}}}
\newcommand{\msbar}{\ensuremath{\overline{\text{MS}}}}

\newcommand{\TS}{\texttt{\allowbreak Trans\-ition\-Sol\-ver}}
\newcommand{\PT}{\texttt{\allowbreak Phase\-Tra\-cer}}
\newcommand{\CT}{\texttt{\allowbreak Cosmo\-Trans\-itions}}

\newcommand{\TWE}{\texttt{\allowbreak Thin\-Wall\-Err\-or}}
\newcommand{\TWEs}{\texttt{\allowbreak Thin\-Wall\-Err\-ors}}

\newcommand{\vextnFactor}[1]{c_{#1}} 
\newcommand{\cc}[1]{CC#1} 
\newcommand{\ccFactor}[1]{\alpha_{\text{CC}#1}} 
\newcommand{\Tmin}{T_{S_{\text{min}}}}
\newcommand{\Tmax}{T_{\Gamma}}
\newcommand{\Teq}{T_{\text{eq}}}
\newcommand{\Treh}{T_{\text{reh}}}

\newcommand{\scn}{\delta_{sc,n}}
\newcommand{\scp}{\delta_{sc,p}}

\newcommand{\aonv}{\mathcal{A}}

\newcommand{\hypergeom}{{}_2 F_1}
\newcommand{\hypergeomArgs}[1]{\hypergeom\!\left(\frac14,\frac12;\frac54;#1\right)}

\DeclareMathOperator{\sign}{sign}

\newcommand{\bench}[2]{M{#1}-BP{#2}}

\title{Supercool subtleties of cosmological phase transitions}

\author[a]{Peter Athron,}
\author[b]{Csaba Bal\'azs,}
\author[b,1]{and Lachlan Morris\note{Corresponding author.}}
\affiliation[a]{Department of Physics and Institute of Theoretical Physics, Nanjing Normal University, Nanjing, Jiangsu 210023, China}
\affiliation[b]{School of Physics and Astronomy, Monash University, Melbourne, Victoria 3800, Australia}

\emailAdd{peter.athron@coepp.org.au}
\emailAdd{csaba.balazs@monash.edu}
\emailAdd{Lachlan.Morris@monash.edu}

\abstract{We investigate rarely explored details of supercooled cosmological first-order phase transitions at the electroweak scale, which may lead to strong gravitational wave signals or explain the cosmic baryon asymmetry. The nucleation temperature is often used in phase transition analyses, and is defined through the nucleation condition: on average one bubble has nucleated per Hubble volume. We argue that the nucleation temperature is neither a fundamental nor essential quantity in phase transition analysis. We illustrate scenarios where a transition can complete without satisfying the nucleation condition, and conversely where the nucleation condition is satisfied but the transition does not complete. We also find that simple nucleation heuristics, which are defined to approximate the nucleation temperature, break down for strong supercooling. Thus, studies that rely on the nucleation temperature --- approximated or otherwise --- may misclassify the completion of a transition. Further, we find that the nucleation temperature decouples from the progress of the transition for strong supercooling. We advocate use of the percolation temperature as a reference temperature for gravitational wave production, because the percolation temperature is directly connected to transition progress and the collision of bubbles. Finally, we provide model-independent bounds on the bubble wall velocity that allow one to predict whether a transition completes based only on knowledge of the bounce action curve. We apply our methods to find empirical bounds on the bubble wall velocity for which the physical volume of the false vacuum decreases during the transition. We verify the accuracy of our predictions using benchmarks from a high temperature expansion of the Standard Model and from the real scalar singlet model.}

\begin{document}
\maketitle
\flushbottom

\section{Introduction}

The Universe underwent several cosmological transitions in rapid succession shortly after the Big Bang, according to standard predictions in particle physics and cosmology\cite{Dolan:1973qd, Weinberg:1974hy, Kirzhnits:1974as, Kirzhnits:1976ts, Dine:1992wr, Aoki:2006br, Aoki:2006we}. Cosmological transitions are particularly interesting if they are first-order phase transitions, proceeding through the nucleation and coalescence of bubbles \cite{Coleman:1977py, Callan:1977pt, Linde:1980tt, Linde:1981zj}. For instance, a first-order electroweak phase transition could provide a candidate mechanism for baryogenesis \cite{Trodden:1998ym, Cline:2006ts, Morrissey:2012db, White:2016nbo}, a probe for dark matter \cite{Bertone:2019irm}, generation of topological defects \cite{Vilenkin:1981bx, Vilenkin:1984ib, Kawasaki:2011vv, Figueroa:2012kw}, and could leave detectable traces in the form of a stochastic gravitational wave background\cite{Witten:1984rs, Hogan:1986qda, Kosowsky:1991ua, Kosowsky:1992rz}. In the Standard Model, the electroweak and QCD transitions are both crossovers rather than first-order phase transitions \cite{Kajantie:1996mn, Rummukainen:1998as, Csikor:1998eu, Aoki:2006we}. However, extensions to the Standard Model can predict a first-order electroweak \cite{Beniwal:2017eik, Kurup:2017dzf, Chiang:2017nmu, Bernon:2017jgv, Alves:2018jsw, Li:2019tfd, Athron:2019teq, Wang:2019pet, Demidov:2021lyo} or QCD phase transition \cite{Schettler:2010dp, Iso:2017uuu, Rezapour:2020mvi}, or they can introduce new phase transitions that may be first order \cite{Daniel:1980xn, Dev:2016feu, Balazs:2016tbi, Croon:2018kqn, Huang:2020bbe, Okada:2020vvb}. Gravitational wave observations or non-observations can be used to constrain extensions of the Standard Model that predict first-order phase transitions \cite{Caprini:2015zlo, Caprini:2019egz}.

Typically, strong supercooling aids the detectability of the generated gravitational waves and can be constrained by near-future detectors, such as LISA \cite{LISA:2017pwj}, for transitions at the electroweak scale \cite{Grojean:2006bp, Huber:2008hg}. Supercooling is when the Universe remains in a metastable phase, with strong supercooling suggesting this metastability persists for an extended duration. In this work, we focus on a subset of strongly supercooled transitions, where the bubble nucleation rate decreases below a certain temperature. The eventual decrease in bubble nucleation may prevent a transition from completing. For the electroweak phase transition, the metastability may persist across a temperature range of $\mathcal{O}(10 \gev)$, while conformal models can allow for even greater supercooling \cite{Jinno:2016knw, Marzola:2017jzl, Ellis:2019oqb, Ellis:2020nnr, Kierkla:2022odc, Levi:2022bzt}. To quantify the extent of supercooling, we introduce a supercooling parameter in \cref{sec:separating-events}. Strongly supercooled transitions require careful analysis to accurately predict the gravitational wave signal, and have received growing attention \cite{Huber:2015znp, Leitao:2015fmj, Jinno:2016knw, Megevand:2016lpr, Kobakhidze:2017mru, Marzola:2017jzl, Cai:2017tmh, Ellis:2018mja, Ellis:2019oqb, Wang:2020jrd, Ellis:2020nnr, Freese:2022qrl, Lewicki:2022pdb, Badger:2022nwo, Kierkla:2022odc, Levi:2022bzt} since the first observation of gravitational waves by LIGO \cite{LIGOScientific:2016aoc}. Key quantities such as the temperature, energy density, average bubble radius and separation, and speed of sound in the plasma, can all vary considerably throughout the duration of a strongly supercooled transition. Unfortunately, hydrodynamic simulations from which gravitational wave fits are extracted cannot yet probe strongly supercooled transitions \cite{Cutting:2019zws}. However, a recent numerical study \cite{Lewicki:2022pdb} provides preliminary predictions for gravitational waves generated by very strongly supercooled transitions in the thin-wall limit.

The prediction of gravitational waves from a first-order phase transition is a rapidly evolving field of study, with many new developments in refining the accuracy each year \cite{Weir:2016tov, Jinno:2016vai, Hindmarsh:2016lnk, Megevand:2016lpr, Kobakhidze:2017mru, Bodeker:2017cim, Hindmarsh:2017gnf, Cai:2017tmh, Jinno:2017fby, Megevand:2017vtb, Konstandin:2017sat, Cutting:2018tjt, Niksa:2018ofa, Ellis:2018mja, RoperPol:2019wvy, Ellis:2019oqb, Jinno:2019jhi, Cutting:2019zws, Jinno:2019bxw, Hindmarsh:2019phv, Alanne:2019bsm, Lewicki:2019gmv, Ellis:2020awk, Wang:2020jrd, Giese:2020rtr, BarrosoMancha:2020fay, Cutting:2020nla, Lewicki:2020jiv, Guo:2020grp, Hoche:2020ysm, Ellis:2020nnr, Megevand:2020klf, Croon:2020cgk, Friedlander:2020tnq, Jinno:2020eqg, Azatov:2020ufh, Balaji:2020yrx, Giese:2020znk, Cai:2020djd, Guo:2021qcq, Ekstedt:2021kyx, Gould:2021ccf, Megevand:2021juo, Megevand:2021llq, Ai:2021kak, Gouttenoire:2021kjv, Hirvonen:2021zej, Dahl:2021wyk, Dorsch:2021nje, Wang:2021dwl, DeCurtis:2022hlx, Cai:2022bcf, Cutting:2022zgd, Auclair:2022jod, Ajmi:2022nmq, Wang:2022lyd, Tenkanen:2022tly, Lewicki:2022pdb, Hijazi:2022uzc, Lewicki:2022nba}. Two of the requirements of these predictions, however, have remained constant for decades: the determination of whether a transition occurs and completes, and the use of a reference temperature at which the gravitational wave signal should be evaluated. These two requirements are the focus of this paper.

We first demonstrate scenarios in which intuitive, simplified conditions for whether a transition begins or completes break down and lead to incorrect classification of the transition and phase history. For instance, it is often stated that a transition cannot complete unless there is at least one bubble per Hubble volume: an event we call unit nucleation. We find counterexamples to this claim and identify conditions for realising this contrary scenario. In \cref{app:bubbleRadius} we show that bubbles can grow larger than the Hubble volume provided there is sufficient supercooling. We suspect the proposed rule that completion implies unit nucleation is based on the requirement that overlapping bubbles must be causally connected. The underlying intuition breaks down upon the realisation that a Hubble volume does not correspond to a causal volume, such as the particle horizon \cite{Davis:2003ad, Ellis:2015wdi}. Further, the condition of having one bubble per Hubble volume is often approximated using heuristic fits which perform poorly for strong supercooling. This compounds the issue of misclassifying the completion of a phase transition, because successful supercooled transitions may not satisfy the nucleation heuristics.

We next address the use of the nucleation temperature --- the temperature at which there is one bubble nucleated per Hubble volume --- as a reference temperature for gravitational wave predictions. If a transition can complete without a nucleation temperature, the latter cannot be an appropriate choice of reference temperature. The nucleation temperature remains a useful proxy for a gravitational wave reference temperature if it lies close to the percolation temperature, which is often the case in phase transitions without strong supercooling. We demonstrate that the nucleation temperature decouples from the transition progress when there is strong supercooling. Consequently, we advocate the avoidance of the nucleation temperature, particularly for supercooled transitions. Nonetheless, we recognise the convenience of using the nucleation temperature along with nucleation heuristics. We aim to resolve the disparity in computational complexity between determining the nucleation temperature and determining the percolation temperature, with code currently in development and used in this work. In the interim, we provide simple conditions for whether a transition percolates and completes for studies that only use the nucleation temperature. We also empirically determine conditions for whether the physical volume of the false vacuum decreases during a transition. Our analysis is based on the electroweak phase transition, but it should be simple to extend our analysis other phase transitions, such as phase transitions from the breakdown of a $U(1)_{B-L}$ symmetry (see e.g.\ Refs.\ \cite{Jinno:2016knw,Chao:2017ilw}) or some other extension of the Standard Model gauge group.

This paper is organised as follows. In \cref{sec:transitionAnalysis} we discuss the analysis of a cosmological phase transition and how to measure its progress, making note of assumptions along the way. In \cref{sec:methodology} we separate and compare the events of unit nucleation (having one bubble per Hubble volume) and percolation (having an infinite cluster of connected bubbles). The comparison allows us to consider scenarios in which unit nucleation occurs but percolation does not, and in which percolation occurs but unit nucleation does not. This is naturally extended to comparing unit nucleation and completion, as well as other completion criteria that will be discussed in \cref{sec:completion}. We provide model-independent bounds on the bubble wall velocity for which these scenarios are realised. These bounds are presented in \crefrange{sec:scenario1}{sec:completion}. In \cref{sec:models} we numerically confirm the analytic results from \cref{sec:methodology} using benchmark points from a toy model and the real scalar singlet model. We have ensured that \cref{sec:models} is self-contained so that the reader may skip the technical details of the analytic treatment in \cref{sec:methodology}. The discussion in \cref{sec:models} reveals many of the same insights provided in \cref{sec:methodology}. In \cref{sec:nucleationApplicability} we discuss the difference between the nucleation and percolation temperatures, both qualitatively and quantitatively, referring to results from a scan over a supercooling parameter. Additionally, we assess the accuracy of various approximations for the nucleation temperature. We present our conclusions in \cref{sec:theend}.

\section{Prerequisites of transition analysis} \label{sec:transitionAnalysis}

A first-order cosmological phase transition typically proceeds via the nucleation of bubbles of a phase with lower free energy within the existing higher free energy phase. Nucleation is suppressed by the potential energy barrier separating the phases. The transition may not complete if this barrier persists at zero temperature. Consequently, it is necessary to analyse the transition to determine its progress over time. A natural measure of this progress is the false vacuum fraction, $P_f$, which is the fraction of the Universe remaining in the old phase (or false vacuum). In this section we introduce the relevant concepts and equations for the determination of transition progress through the false vacuum fraction. We highlight assumptions that we make and avoid; many of which are frequently implicit in cosmological phase transition literature.

A transition first becomes possible at the critical temperature, $T_c$, where the phases have equal free energy density and $P_f = 1$. As the Universe cools, the free energy of the phases separate, increasing the nucleation rate of bubbles of true vacuum. The nucleation rate, or number of bubbles nucleated per unit volume per unit time, is given by \cite{Linde:1981zj}
\begin{equation}
	\Gamma(T) \simeq T^4 \! \left(\frac{S(T)}{2\pi} \right)^{\!\!\frac{3}{2}} \! \exp(-S(T)) \label{eq:nucleationRate}
\end{equation}
at finite temperature. The nucleation rate has a strong dependence on the bounce action $S(T) = S_3(T)/T$, where $S_3$ is the three-dimensional Euclidean action of a critical bubble. We will refer to $S$ simply as the action from hereon. A critical bubble has minimal radius while preventing collapse, with internal pressure matching the surface tension.%
\footnote{Typically, one assumes the thin-wall limit and solves the vanishing total energy condition in the bubble $E(R_c) = 0$ for the critical radius $R_c$. This is the radius at which the liberated vacuum energy matches the surface tension of the bubble.}
Critical bubbles dominate the distribution of nucleated bubbles. Another transition mechanism is zero-temperature quantum tunnelling \cite{Coleman:1977py, Callan:1977pt}, which we do not consider in this work. Instead we focus on transitions that occur due to thermal fluctuations over the potential barrier or through finite-temperature quantum tunnelling. It is expected that zero-temperature quantum tunnelling will not affect our results for the models we consider \cite{Megevand:2016lpr, Kobakhidze:2017mru, Cai:2017tmh}.

A first-order phase transition progresses through the nucleation and subsequent growth of bubbles. The growth is driven by the pressure difference between the phases, and is reduced by frictional forces from the plasma\footnote{At the relevant temperatures for transitions at the electroweak scale and above, the Universe is filled with a thermal bath of plasma made up of fermions and gauge bosons.} surrounding the bubble walls. The physical radius of a bubble that nucleated at time $t'$ and has grown until time $t$ is \cite{Linde:1981zj, Megevand:2007sv}
\begin{equation}
	R(t', t) = \frac{a(t)}{a(t')} R_0(t') + \int_{t'}^{t} dt'' v_w(t', t'') \frac{a(t)}{a(t'')} , \label{eq:bubbleRadius}
\end{equation}
where $a$ is the scale factor, accounting for expansion of the Friedmann-Lema\^{i}tre-Robertson-Walker (FLRW) space, and $R_0$ is the initial radius of the bubble.%
\footnote{The initial bubble radius can be estimated by assuming it is larger but almost equal to the critical bubble radius (see e.g.\ Refs.\ \cite{Darme:2017wvu, Ellis:2019oqb}). In the thin-wall limit, the critical bubble radius is given by $R_c = 2S_1(t)/\Delta p(t)$, where $S_1$ is the one-dimensional action, or surface tension, and $\Delta p = \Vf - \Vt$ is the pressure difference between the phases. The initial bubble radius $R_0 \approx R_c$ should be many orders of magnitude smaller than the bubble radius when collisions occur, particularly in strongly supercooled transitions where bubbles have significant time to grow.}
We assume all bubbles nucleated at the same time grow at the same rate. In general, the wall velocity of a bubble is expected to start at zero and increases until saturating at a terminal velocity. This terminal velocity is again assumed to be homogeneous throughout the Universe, and changes with time due to evolution of the plasma and the free energy. That is, the friction and the driving pressure change over time, thus giving a time-dependent terminal velocity.

It is often assumed that bubbles accelerate very quickly to their terminal velocity, that this terminal velocity remains fixed throughout the transition, and that their initial radius is negligible compared to their radius at any time of interest (e.g.\ when they collide with another bubble). The time-dependent wall velocity is then replaced with the approximately constant terminal velocity, for which we reuse the label $v_w$. This allows for a simple approximation for the bubble radius:
\begin{equation}
	R(t', t) = v_w \int_{t'}^{t} dt'' \frac{a(t)}{a(t'')} . \label{eq:bubbleRadiusSimple}
\end{equation}
The above can be further simplified if radiation domination is assumed. However, the assumption of radiation domination can be broken during a supercooled transition. In this work, we consider a constant bubble wall velocity throughout the transition, and we do not model the acceleration of bubbles. We therefore use \eqref{eq:bubbleRadiusSimple} for the bubble radius.

\subsection{False vacuum fraction}

With bubble nucleation and growth specified, we can determine the transition progress through the false vacuum fraction, $P_f$. The volume occupied by bubbles is difficult to determine without performing a stochastic field-fluid simulation (see e.g.\ Ref.\ \cite{Hindmarsh:2015qta}) and measuring the fraction of simulation volume encompassed by bubbles. Instead, we can proceed analytically to model the false vacuum fraction as \cite{Guth:1979bh, Guth:1981uk, Megevand:2020klf, Ajmi:2022nmq}%
\footnote{This model for phase transitions has in fact been known in the field of materials science since the 1930s, where it is referred to as the JMAK equation \cite{johnson1939reaction, Avrami1, Avrami2, Avrami3, kolmogorov1937statistical}. Importantly, there is a wealth of literature detailing the assumptions and caveats hidden within the derivation of Refs.\ \cite{Guth:1979bh, Guth:1981uk}. We review this literature in an upcoming paper \cite{Athron:2023}. For now, we refer the reader to Refs.\ \cite{fanfoni1998johnson, BURBELKO2005429} for a discussion of the correctness and assumptions of the JMAK equation.}
\begin{equation}
	P_f(t) = \exp(-\Vext(t)) , \label{eq:falseVacuumFraction}
\end{equation}
where $\Vext$ is the (fractional) extended volume of true vacuum bubbles, given by
\begin{equation}
	\Vext(t) = \frac{4\pi}{3} \int_{t_0}^t \! dt' \, \Gamma(t') \frac{a^3(t')}{a^3(t)} R^3(t', t) . \label{eq:extendedVolume}
\end{equation}
Here, $t_0$ is the time when the transition first becomes possible --- taken to be when the Universe cools to the critical temperature. As space expands, the number density of previously nucleated bubbles is reduced by the cubed ratio of scale factors. Note that $\Vext$ double-counts overlapping regions of bubbles and includes phantom bubbles which nucleate within pre-existing bubbles. The extended volume is not equivalent to the true vacuum fraction $P_t = 1 - P_f$ except in the early stages of a transition where bubbles nucleate and grow in isolation. To calculate the true vacuum fraction directly, the nucleation rate would need to be multiplied by $P_f$ and the volume of individual bubbles would need to account for deformation due to impingement. It is considerably more convenient to instead exponentiate the extended volume. Various alternative derivations for the false vacuum fraction exist in the literature (see e.g.\ Ref.\ \cite{ALEKSEECHKIN20113159} for a review), which also lead to \eqref{eq:falseVacuumFraction} and \eqref{eq:extendedVolume}. For example, a recent derivation in cosmological literature \cite{Megevand:2020klf} avoids the explicit nucleation of phantom bubbles but again leads to these results. One may interpret $\Vext$ as the fraction of the volume of all real and phantom bubbles over the total volume in which the phase transition occurs (i.e.\ the entire Universe).%
\footnote{Because the nucleation rate \eqref{eq:nucleationRate} is the number of bubbles nucleated per volume per time, the volume in $\Gamma$ can cancel with the total transition volume in $\Vext$. We use this in arriving at \eqref{eq:extendedVolume}.}
We emphasise that one cannot simply replace $\Vext$ in \eqref{eq:falseVacuumFraction} with the true vacuum fraction, or multiply $\Gamma$ by $P_f$ in \eqref{eq:extendedVolume}.

The false vacuum fraction is then
\begin{equation}
	P_f(t) = \exp(-\frac{4\pi}{3} v_w^3 \int_{t_0}^t \! dt' \, \Gamma(t') \! \left(\int_{t'}^t \! dt'' \, \frac{a(t')}{a(t'')} \right)^{\!\!3}) , \label{eq:falseVacuumFraction-fullTime}
\end{equation}
where we have used the assumptions in \eqref{eq:bubbleRadiusSimple} for bubble growth. These assumptions are readily justified in this work. We consider strongly supercooled transitions, in which case the terminal wall velocity $v_w \rightarrow 1$ is expected for the entire time window where nucleation is significant.%
\footnote{Some models may still permit $v_w < 1$ despite strong supercooling. Our discussion in \cref{sec:methodology} considers arbitrary bubble wall velocities. We note that for bubbles that expand as deflagrations or hybrids, reheating can affect the evolution of the transition. This effect is discussed in \cref{app:reheating}.}
The pressure difference between phases is given time to increase, permitting rapid acceleration of the bubble walls. With a low nucleation rate, bubbles can grow considerably before colliding, allowing ultra-relativistic velocities to be attained. In this case the initial radius of a bubble becomes a negligible contribution to its evolution.

The nucleation rate in \eqref{eq:nucleationRate} is expressed as a function of temperature. It will be convenient to also express \eqref{eq:falseVacuumFraction-fullTime} in terms of temperature. We can use the time-temperature relation:
\begin{equation}
	\dv{T}{t} = - T H(T) , \label{eq:adiabaticJacobian}
\end{equation}
where $H$ is the Hubble parameter defined below. This relation is an approximation based on the MIT bag equation of state \cite{Chodos:1974je}, but it holds well in the models we consider. Further details are available in \cref{app:timeTemperatureRelation}. Assuming flat space, with vanishing Gaussian curvature in the FLRW metric, the Hubble parameter can be determined from Friedmann's first equation as
\begin{equation}
	H(T) = \frac{\dot{a}(T)}{a(T)} = \sqrt{\frac{8\pi G}{3} \rhoh(T)} , \label{eq:HubbleParameter}
\end{equation}
where $G = 6.7088 \! \times \! 10^{-39} \gev^{-2}$ is Newton's gravitational constant \cite{Wu:2019pbm}.

With the relation \eqref{eq:adiabaticJacobian} between temperature and time, we can now express \eqref{eq:falseVacuumFraction-fullTime} in terms of temperature
\begin{equation}
	P_f(T) = \exp(-\frac{4\pi}{3} v_w^3 \int_T^{T_c} \! dT' \, \frac{\Gamma(T')}{T' H(T')} \! \left(\int_T^{T'} \! dT'' \recip{T'' H(T'')} \frac{a(T')}{a(T'')} \right)^{\!\!3} \, ) , \label{eq:falseVacuumFraction-withScaleFactorRatio}
\end{equation}
where we take the initial temperature of the transition to be the critical temperature, $T_c$.
Adiabatic expansion ensures
\begin{equation}
	\frac{a(T_2)}{a(T_1)} = \frac{T_1}{T_2} , \label{eq:scaleFactorRatio}
\end{equation}
as shown in \cref{app:scaleFactorRatio}, so we can simplify \eqref{eq:falseVacuumFraction-withScaleFactorRatio} as
\begin{equation}
	P_f(T) = \exp(-\frac{4\pi}{3} v_w^3 \int_T^{T_c} \! dT' \, \frac{\Gamma(T')}{T'^4 H(T')} \! \left(\int_T^{T'} \! dT'' \recip{H(T'')} \right)^{\!\!3}) . \label{eq:falseVacuumFraction-temperatureSimplified}
\end{equation}

It now remains to determine the energy density $\rhoh$. For the purposes of cosmological phase transitions, we can consider the Lagrangian density
\begin{equation}
	\mathcal{L}(\field, T) = \frac{1}{2} \partial_{\mu}\field \partial^{\mu}\field - \Vd
\end{equation}
with scalar potential $\Vd$.%
\footnote{We use the symbol $V$ for the potential and $\vol$ for the volume.}
All fields that do not gain a vacuum expectation during the transition can be ignored, leaving a set of scalar fields $\field$ on which the dynamics of the phase transition depend. The minima of this potential correspond to vacua, or phases, while the potential itself is the free energy density; $\FEDd = V(\field, T)$. The energy density for an arbitrary field configuration is 
\begin{equation}
	\rho(\field, T) = T \pdv{p}{T} - p(\field, T) , \label{eq:energyDensityPressure}
\end{equation}
by the first law of thermodynamics, with pressure $p = -\FED$. 
In terms of the potential, the energy density is
\begin{equation}
	\rho(\field, T) = V(\field, T) - T \pdv{V}{T} .
\end{equation}

During the phase transition we take the energy density $\rhoh$ in Friedmann's equation \eqref{eq:HubbleParameter} to be the difference in energy density between the false vacuum and the zero-temperature ground state of the potential
\begin{equation}
	\rhoh(T) = \rho(\phif(T), T) - \rho(\phigs(0), 0) , \label{eq:energyDensityGeneral}
\end{equation}
which expands to,
\begin{equation}
	\rhoh(T) = V(\phif(T), T) - T \pdv{V(\phif(T), T)}{T} - \Vgsz . \label{eq:energyDensityReduced}
\end{equation}
Here, $\phif$ and $\phigs$ are the field configurations corresponding to the false vacuum and the ground state, respectively. In general this ground state may not be the vacuum that describes our current Universe, which would leave a non-zero cosmological constant $\rho(\phicur(0), 0) - \rho(\phigs(0), 0)$. Conservation of energy is an implicit assumption in this form of the energy density of the Universe. As the transition progresses, regions of the Universe will no longer be in the false vacuum, so using only $\rhof$ instead of a combination of $\rhof$ and $\rhot$ (the subscript $t$ denoting the true vacuum) would appear erroneous. However, energy liberated by expanding true vacuum bubbles reheats the surrounding plasma. Following Ref.\ \cite{Leitao:2015fmj}, we thus assume energy conservation outside of adiabatic cooling.

\begin{figure}
	\centering
	\includegraphics[width=0.6\linewidth]{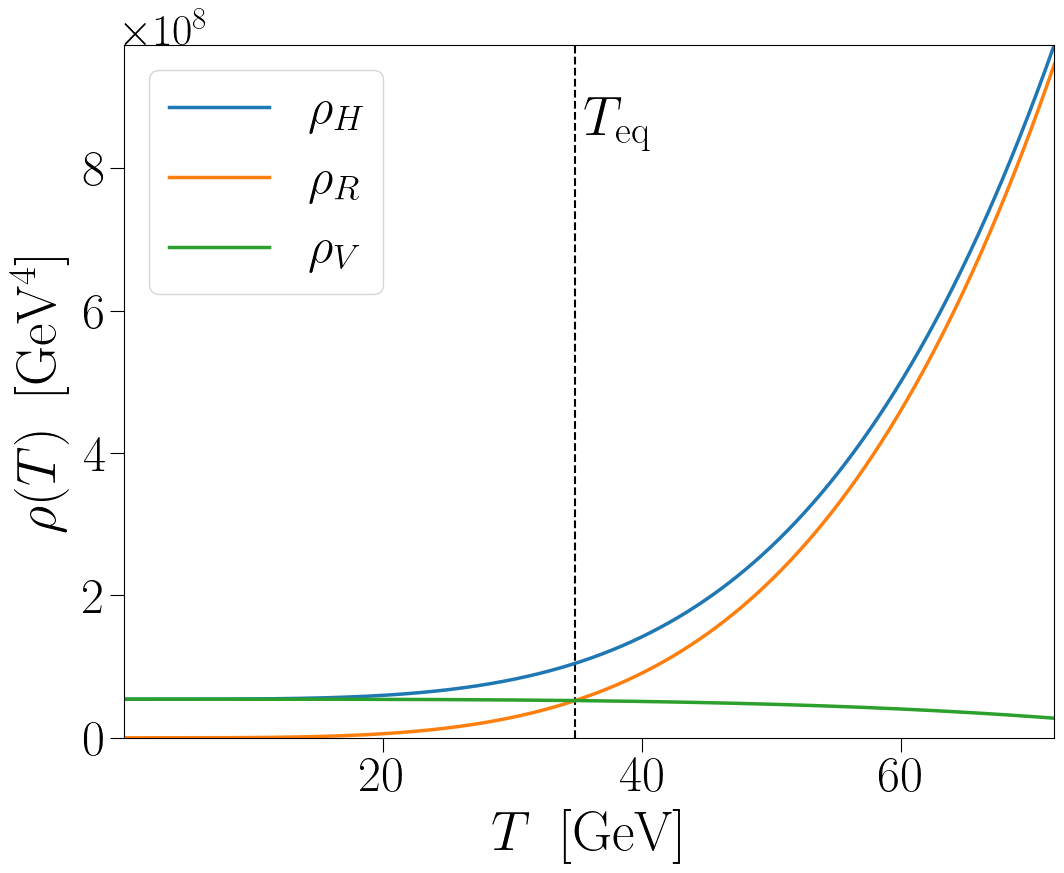}
	\caption{The energy density as a function of temperature for the benchmark point \bench{2}{1} defined in \cref{tab:benchmarks-M2}. The vertical dashed black line corresponds to $\Teq$ (defined in \eqref{eq:Teq}). The energy density is dominated by $\rhov$ at low temperatures and by $\rhor$ at high temperatures.}
	\label{fig:energyDensity}
\end{figure}

It will be convenient to separate out the parts of the potential that scale as radiation (i.e.\ $\propto T^4$), which yields the radiation energy density
\begin{equation}
	\rhor(T) = \frac{\pi^2}{30} g_* T^4 = \frac{T^4}{\xi_g^2}. \label{eq:rhoR}
\end{equation}
Here $g_*$ is the number of relativistic degrees of freedom in the plasma, with a value of $106.75$ at the electroweak scale in the Standard Model, and we introduced $\xi_g^2$ as a shorthand notation for $30/(\pi^2 g_*)$. The remaining terms in the energy density can be considered to be the `vacuum contribution', given by
\begin{equation}
	\rhov(T) = \rhoh(T) - \rhor(T) . \label{eq:rhoV}
\end{equation}
Typically, $\rhov$ freezes to a constant value at low temperatures (see \cref{fig:energyDensity}, or e.g.\ Figure 33 in Ref.\ \cite{Wang:2020jrd}) and is subdominant at high temperatures (e.g.\ towards the critical temperature). Then, a reasonable approximation is
\begin{align}
	\rhoh(T) & \approx \rhor(T) + \rhov(0) \nonumber \\
	& = \rhor(T) + \Vfz - \Vgsz . \label{eq:energyDensityApprox}
\end{align}
One can separate periods of the transition where the energy density was radiation- and vacuum-dominated by defining $\Teq$ such that
\begin{equation}
	\rhov(\Teq) = \rhor(\Teq) . \label{eq:Teq}
\end{equation}
Then we can roughly assume radiation domination when $T > \Teq$ and vacuum domination when $T < \Teq$. We will make use of this concept of distinct eras in our analytic discussion in \cref{sec:methodology}. Note that \eqref{eq:energyDensityApprox} further assumes $\phif$ still exists at $T = 0$, which is not always true. One could instead use $\rhov(T_*)$ instead of $\rhov(0)$, where $T_*$ is the temperature at which the phase $\phif$ disappears.

We argue that a correct description of the expansion rate during a phase transition would ideally reflect a difference in expansion rate inside bubbles, outside bubbles, and near bubble walls. In a thin-wall approximation, one could neglect the latter region, leaving a difference in expansion rate in the two phases. The applicability of such a treatment to the averaged, analytic measure of nucleation and transition progress considered here is not immediately obvious. We reserve an exploration of this marriage of approaches for future work. In line with the averaged analysis for the transition progress, we use \eqref{eq:energyDensityReduced} with the assumption that, due to reheating, it captures the average energy density of the Universe. We use \eqref{eq:energyDensityReduced} in our numerical results and \eqref{eq:energyDensityApprox} in our analytic discussion.

\subsection{Milestone temperatures} \label{sec:transitionTemperatures}

It is useful to define several temperatures corresponding to milestones in the phase transition. The nucleation, percolation and completion temperatures, among others, are often introduced to capture the moment of their respective milestones. Here we discuss these temperatures and their corresponding events.

Significant nucleation does not occur immediately after the critical temperature, because the action diverges at $T_c$. The number of bubbles nucleated within a Hubble volume (of order $H^{-3}$) at temperature $T$ is \cite{Enqvist:1991xw, Guo:2021qcq}
\begin{equation}
	N(T) = \int_{T}^{T_c} \! dT' \, \frac{\Gamma(T') P_f(T')}{T' H^4(T')} , \label{eq:numBubblesProper}
\end{equation}
where \eqref{eq:adiabaticJacobian} has been used to transform the time integral. A factor of $4\pi/3$ could be included in \eqref{eq:numBubblesProper} to consider a spherical Hubble volume, though such a change would usually have little impact due to how rapidly $N$ changes with temperature. We omit this factor to connect with the broader literature. Still, \eqref{eq:numBubblesProper} differs from the common description of the number of nucleated bubbles \cite{Ellis:2018mja},
\begin{equation}
	\Next(T) = \int_{T}^{T_c} \! dT' \, \frac{\Gamma(T')}{T' H^4(T')} , \label{eq:numBubbles}
\end{equation}
in that \eqref{eq:numBubblesProper} does not count phantom bubbles (i.e.\ those nucleated inside other bubbles). Phantom bubbles are important for calculating the extended volume \eqref{eq:extendedVolume}, but should not be counted here because phantom bubbles cannot nucleate. We have labelled the number of bubbles in \eqref{eq:numBubbles} with the superscript ``ext'' to signify the counting of all bubbles in the extended true vacuum volume.

The nucleation temperature, $T_n$, is defined as the temperature when on average one bubble has nucleated in a Hubble volume,
\begin{equation}
	N(T_n) = 1, \label{eq:nucleationTemperatureProper}
\end{equation}
an event we will refer to as ``unit nucleation'' for convenience.%
\footnote{This event is usually referred to as nucleation in the literature. Here we want to differentiate it from the generic nucleation of a bubble.}
The usual definition is
\begin{equation}
	\Next(T_n) = 1 . \label{eq:nucleationTemperature}
\end{equation}
Although approximate, \eqref{eq:nucleationTemperature} will yield a virtually identical nucleation temperature to \eqref{eq:nucleationTemperatureProper}. An exception is when the transition is strongly supercooled, with very few bubbles such that $P_f$ differs from unity appreciably before the nucleation temperature is reached. In \cref{sec:nucleationHeuristic} we will demonstrate cases where the extracted nucleation temperatures from \eqref{eq:nucleationTemperatureProper} and \eqref{eq:nucleationTemperature} differ noticeably.

The nucleation temperature is often used in studies of phase transitions as a characteristic or reference temperature for important processes, such as gravitational wave production, baryogenesis or topological defect formation. It is conveniently simple to determine provided one has a method of computing the action. There are a few approximations used in the literature for the nucleation temperature that further simplify its determination, as discussed in \cref{sec:nucleationHeuristic}. The nucleation temperature is often thought to represent the start of a transition. For a fast transition, the nucleation temperature may lie close to both the critical and completion temperatures. An alternative initial temperature is when non-negligible false vacuum conversion has occurred; for instance, Ref.\ \cite{Megevand:2016lpr} use $P_f(T_I) = 0.99$. For a strongly supercooled transition, the nucleation temperature may be significantly higher than the completion temperature (or not exist at all, as we will show), and is no longer a suitable reference temperature. This brings us to the next milestone temperature of interest.

\begin{figure}
	\centering
	\includegraphics[width=0.7\linewidth]{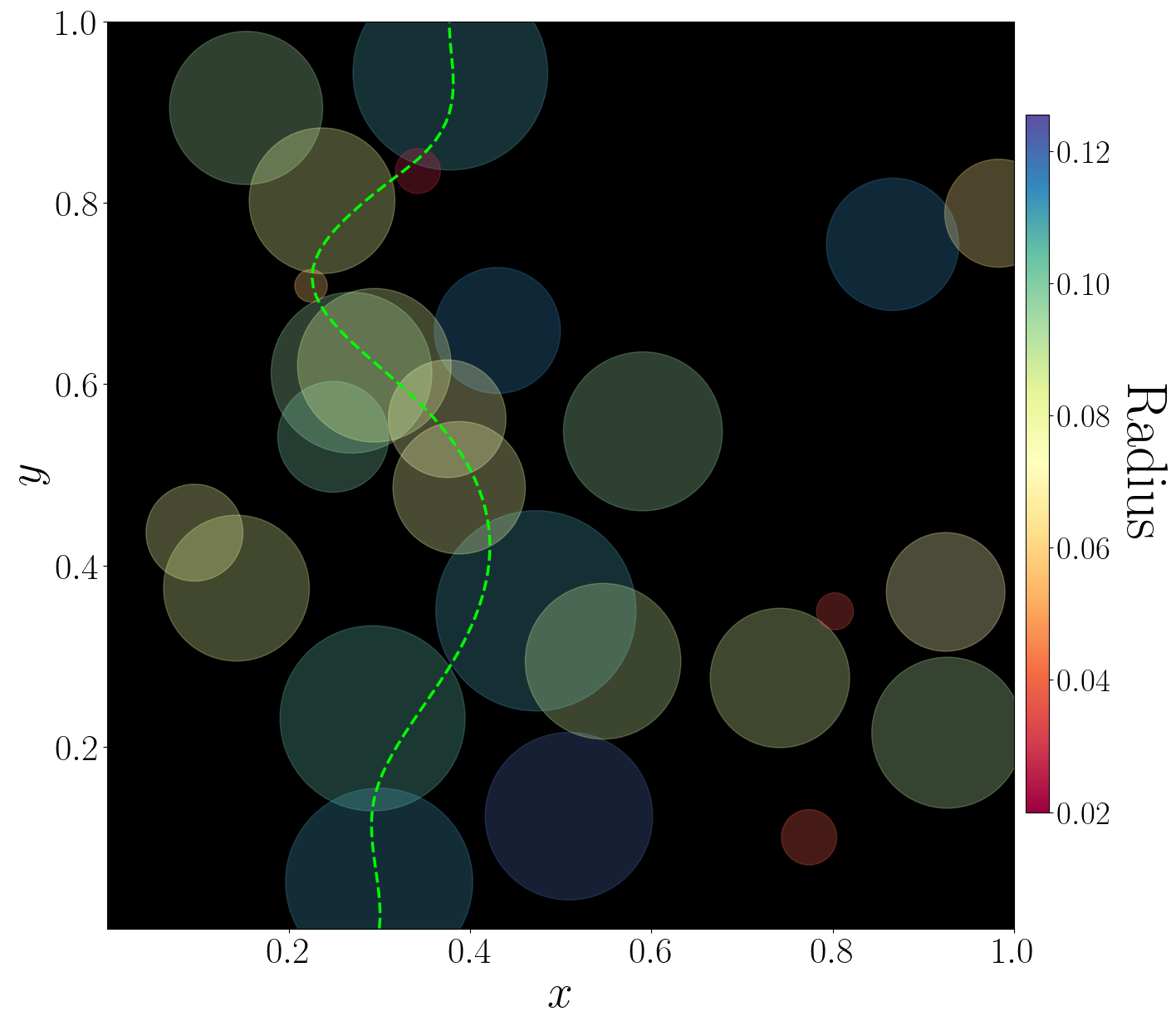}
	\caption{A 2D slice (with $z = 0.422$) of a simple 3D simulation of bubble nucleation in a unit volume. At the moment of plotting, the false vacuum fraction was $P_f \approx 0.7$ suggesting the onset of percolation. Indeed, a path through connected bubbles spanning from $y = 0$ to $y = 1$ is evident (see the green dashed curve), with the small gaps filled at slightly higher $z$ values. In general, the path spanning across the simulation volume may not be clear from a single 2D slice of the volume at the onset of percolation. The colour of a bubble corresponds to its overall radius, which is not equal to the radius of the bubble's cross-section in the $z=0.422$ plane depicted here.}
	\label{fig:percolation}
\end{figure}

The percolation temperature, $T_p$, is defined by
\begin{equation}
	P_f(T_p) \approx 0.71 . \label{eq:percolationTemperature}
\end{equation}
This condition is based on results from percolation threshold studies (see e.g.\ Refs.\ \cite{doi:10.1063/1.1338506, LIN2018299, LI2020112815}). Percolation is defined as having an infinite cluster of --- in our case --- connected bubbles \cite{doi:10.1080/00018737100101261, hunt2014percolation}. For a finite-sized transforming medium, percolation is similarly defined as having a cluster of connected bubbles with size of the order of the medium (as depicted in \cref{fig:percolation}).%
\footnote{In other contexts such as a fluid filtering through a porous medium, the percolation threshold is the maximum connectivity (of the solid regions) in the medium such that the fluid can pass from one side of the medium to the other (i.e.\ there is a path through the entire medium for the fluid to flow). In the present context, the percolation threshold is the minimum fraction of the Universe converted to the true vacuum such that there is a connected cluster of bubbles that span the Universe.}
Percolation coincides with a time when bubbles are colliding (by definition, and as evidenced by \cref{fig:percolation}), marking an important time for gravitational wave production. While the ability to determine the ideal reference temperature for gravitational wave production is currently lacking, the percolation temperature is recommended over the nucleation temperature \cite{Caprini:2019egz, Wang:2020jrd, Guo:2021qcq}.

The completion (or final) temperature, $T_f$, is the temperature at which the transition completes. Ideally, this would be when $P_f(T_f) = 0$. However, \eqref{eq:falseVacuumFraction-temperatureSimplified} only predicts zero false vacuum fraction asymptotically because \eqref{eq:falseVacuumFraction} is a probabilistic treatment of bubble nucleation and growth. Instead, we can define the completion temperature as
\begin{equation}
	P_f(T_f) = \varepsilon ,
\end{equation}
where $\varepsilon \ll 1$. We consider $\varepsilon = 0.01$ in this work. Such a definition ensures that the transition is highly likely to have completed by the time the Universe cools to $T_f$. We remark that $T_f$ varies very weakly with $\varepsilon < 0.01$ except in very strongly supercooled transitions where $P_f$ approaches a non-zero value as $T \rightarrow 0$. Remembering that we do not consider zero-temperature quantum tunnelling in this work, if a transition has $P_f(0) > \varepsilon$ due to thermal fluctuations and finite-temperature quantum tunnelling, we label it as incomplete.

Another criterion that should be satisfied for a successful transition is that the average physical volume of the false vacuum,
\begin{equation}
	\Vphys(t) = a^3(t) P_f(t) ,
	\label{eq:Vphys}
\end{equation}
eventually decreases with time \cite{Turner:1992tz}.%
\footnote{More precisely, \eqref{eq:Vphys} is the average proper volume of the false vacuum contained inside a unit comoving volume, hence the conversion factor $a^3$.}
That is,
\begin{equation}
	\dv{t} \! \left.\left(a^3(t) P_f(t) \right) \right\vert_{t > t_*} < 0 ,
\end{equation}
where $t_*$ is the time at which the physical volume of the false vacuum stops increasing. This leads to the condition
\begin{equation}
	3 H(t) - \dv{\Vext}{t} < 0 
\end{equation}
for some $t > t_*$, where we have used the definition of the Hubble parameter given in \eqref{eq:HubbleParameter}. Or in terms of temperature, using \eqref{eq:adiabaticJacobian},
\begin{equation}
	3 + T \dv{\Vext}{T} < 0 \label{eq:decreasingPhysicalVolume}
\end{equation}
for some $T < T_*$, where $T_*$ corresponds to $t_*$. Ref.\ \cite{Ellis:2018mja} notes that \eqref{eq:decreasingPhysicalVolume} may not be satisfied when $P_f(T) \approx 0.71$, potentially spoiling percolation. For transitions where \eqref{eq:decreasingPhysicalVolume} is only satisfied for temperatures below $T_p$, it is suggested that percolation (and by extension completion) is questionable. To understand this situation, consider two nearby but non-overlapping bubbles. If the growth of the bubbles towards each other is slower than the expansion of the false vacuum between the bubbles, then they will never meet. This situation is possible even as the fraction of space in the false vacuum along the line connecting the centres of the bubbles approaches zero.

\begin{figure}
	\centering
	\begin{subfigure}{.49\textwidth}
		\centering
		\includegraphics[width=1\linewidth]{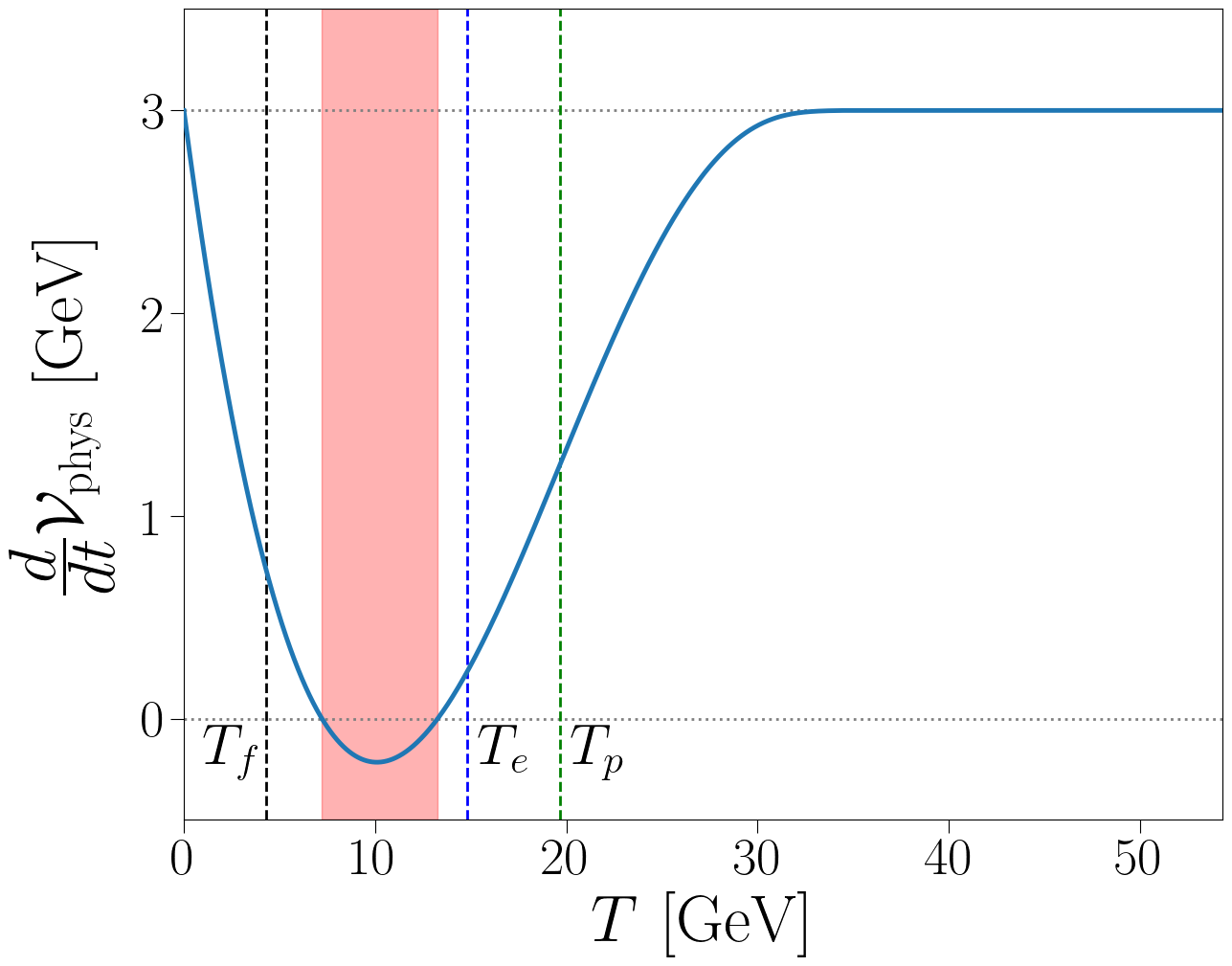}
		\caption{}
		\label{fig:physicalVolume-timeDeriv}
	\end{subfigure}\hspace{0.2cm}%
	\begin{subfigure}{.49\textwidth}
		\centering
		\includegraphics[width=1\linewidth]{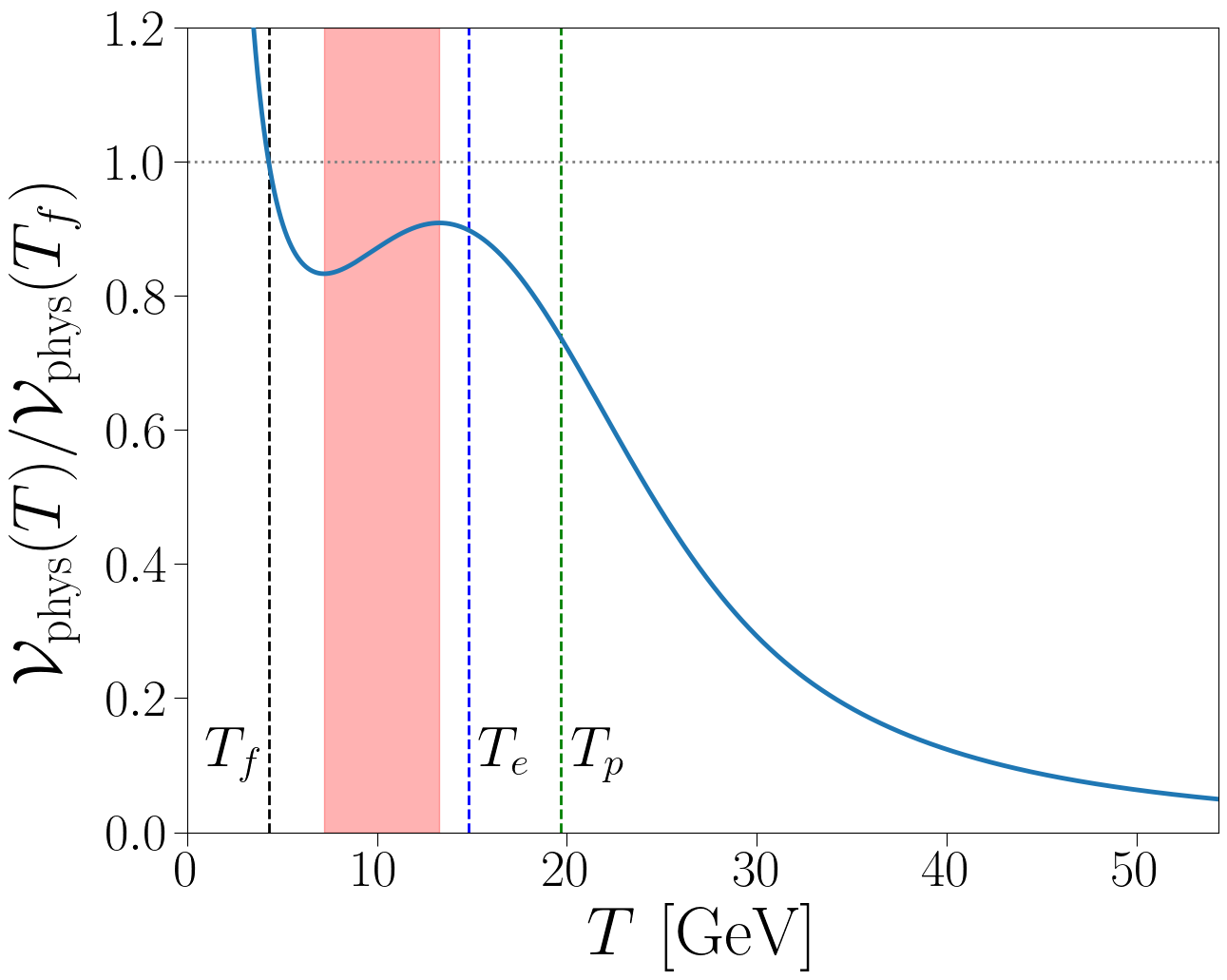}
		\caption{}
		\label{fig:physicalVolume-relative}
	\end{subfigure}
	\vspace{-0.2cm}
	\caption{(a) The rate of change in physical volume of the false vacuum $\Vphys$, and (b) the physical volume of the false vacuum relative to its value at the completion (or final) temperature $\Vphys(T_f)$, for benchmark \bench{1}{4} (defined in \cref{tab:benchmarks-M1}) with $v_w = 1$. The physical volume of the false vacuum is seen to decrease slightly in the small shaded temperature window, yet increases during the rest of the transition, and continues to increase after the transition is complete. We expect finite regions of the Universe to remain in the false vacuum for such a transition, even though \eqref{eq:decreasingPhysicalVolume} is satisfied for a range of temperatures. The vertical dashed lines are, in order from left to right: the completion temperature $T_f$ (black), the temperature $T_e$ for which $\Vext(T) = 1$ (blue), and the percolation temperature $T_p$ (green). The physical volume decreases in the red shaded temperature window and increases everywhere else.
	}
	\label{fig:physicalVolume}
\end{figure}

To be confident in the success of a transition and accuracy of the percolation condition \eqref{eq:percolationTemperature}, ideally we would have \eqref{eq:decreasingPhysicalVolume} satisfied at both $T_p$ and $T_f$. A weaker constraint is that there is a temperature for which $\Vphys$ is decreasing, without regard for the magnitude or duration of this decrease \cite{Ellis:2018mja}. We demonstrate in \cref{fig:physicalVolume} and more generally in \cref{sec:completion} that this constraint is insufficient for the strongly supercooled transitions we consider in this study. Other simple constraints can be considered in lieu of integrating the change in $\Vphys$ over time. One such constraint is that $\Vphys(T_f) < \Vphys(T_p)$, which maps to the condition $T_f \gtrsim 0.24 T_p$. Another possible constraint is that $\Vphys$ is decreasing at $T_f$. The (perhaps model-dependent) hierarchy of these constraints is presented in \cref{sec:completion}.

Another relevant temperature for gravitational wave predictions is the reheating temperature, $\Treh$. As the energy liberated by vacuum conversion thermalises, the plasma reheats. This can have important consequences on the progress of the phase transition if bubbles expand as deflagrations or hybrids (see e.g.\ Refs.\ \cite{Heckler:1994uu, Megevand:2017vtb}), and still affects the gravitational wave signal even for detonations. For instance, the frequency spectrum of gravitational waves is red-shifted due to cooling from $\Treh$ to the present day vacuum temperature, rather than cooling from $T_f$ or $T_p$ \cite{Leitao:2015fmj, Cai:2017tmh}. One can estimate the temperature of the plasma after the phase transition completes on the grounds of energy conservation \cite{Leitao:2015fmj}, by solving
\begin{equation}
	\rho(\phif(T_f), T_f) = \rho(\phit(\Treh), \Treh) , \label{eq:reheat}
\end{equation}
where the solution lies in the range $T_f < \Treh < T_c$. The reheating temperature can also be evaluated at the onset of percolation by replacing $T_f$ with $T_p$ in \eqref{eq:reheat}. The reheating temperature is not directly relevant to this study but will be reported in numerical scans in \cref{sec:nucleationApplicability}.

With all temperatures of interest defined, the remainder of this paper is dedicated to comparing these temperatures to each other and determining their relevance. We will argue that the nucleation temperature based on its standard definition \eqref{eq:nucleationTemperature} is not a fundamental quantity, and that its only use is as a proxy for the more appropriate and well-motivated percolation temperature. A thorough analysis of a phase transition can determine the percolation temperature, rendering the nucleation temperature obsolete, with the slight caveat that the percolation temperature is currently more computationally expensive to determine. In this work, we provide scenarios where the typical use of the nucleation temperature is unsuitable and would lead to incorrect predictions of the dynamics and completion of the transition.

\section{Nucleation, percolation and completion: analytic treatment} \label{sec:methodology}

\subsection{Separating unit nucleation and percolation} \label{sec:separating-events}

\noindent
One of the main purposes of this paper is to carefully examine the separation of two events in a phase transition: unit nucleation \eqref{eq:nucleationTemperatureProper} and percolation \eqref{eq:percolationTemperature}. To do so, we need to compare the conditions for the occurrence of these events. In this subsection, we qualitatively discuss the nucleation and percolation temperatures based on their definitions. In the next subsection, we will show how to compare the events of unit nucleation and percolation quantitatively and gain intuition for the occurrence of these events through an analytic treatment. To begin the qualitative discussion, we consider the definition of these two events.

Unit nucleation (with its corresponding nucleation temperature $T_n$) \eqref{eq:nucleationTemperatureProper} is when an average of one bubble appears per Hubble volume. This may occur very soon after the transition first becomes possible in a fast (or weakly supercooled) transition. In a strongly supercooled transition, this may occur when a significant fraction of the Universe has already been converted to the true vacuum. The few bubbles have sufficient time to expand to sizes that are not achievable in a fast transition. For a discussion of the use of the nucleation temperature, see the text following \eqref{eq:nucleationTemperature}.

The difference between the nucleation and critical temperatures is an indication of the degree of supercooling, or how long the Universe remained in the metastable phase below the critical temperature. A supercooling parameter can be defined as
\begin{equation}
	\scn = \frac{T_c - T_n}{T_c} . \label{eq:supercooling-nucleation}
\end{equation}
This measure of supercooling has two important limitations. Obviously, unit nucleation must occur for this parameter to be defined. Additionally, given a value for $\scn$, one still has no insight as to the progress of the transition.

Percolation (with its corresponding percolation temperature $T_p$) \eqref{eq:percolationTemperature} is when there is a cluster of connected bubbles that spans across the transforming medium. For spherical bubbles, this occurs when approximately $30\%$ of the Universe has been converted to the true vacuum. The event of percolation aligns with when a significant proportion of bubble collisions occur, suggesting its relevance for gravitational wave generation \cite{Kosowsky:1991ua, Kosowsky:1992rz, Kosowsky:1992vn, Kamionkowski:1993fg}. The percolation temperature can deviate significantly from the nucleation temperature in strongly supercooled transitions, as we will show in \cref{sec:nucleationApplicability}. Thus, another supercooling parameter can be defined as
\begin{equation}
	\scp = \frac{T_c - T_p}{T_c} . \label{eq:supercooling-percolation}
\end{equation}
Similarly to $\scn$, this has the limitation that it is only defined if percolation occurs. However, given a value $\scp$, one can determine the approximate temperature window for which the false vacuum fraction deviates significantly from unity.

In this paper we highlight that unit nucleation and percolation are independent events. The occurrence of one does not guarantee the occurrence of the other. Satisfying \eqref{eq:nucleationTemperatureProper} suggests a significant number of bubbles have been nucleated, which naturally assists percolation. However, we will show two interesting scenarios that defy expectation:
\begin{enumerate}[align=left]
	\item[\hspace{0.35cm} Scenario 1:] Unit nucleation without percolation.
	\item[\hspace{0.35cm} Scenario 2:] Percolation without unit nucleation.
\end{enumerate}
These scenarios may also be expressed in terms of completion rather than percolation. Scenario 1 is what one might expect when a transition is tuned to be more strongly supercooled. Bubbles nucleate, but not enough for them to percolate before the Universe cools well below the scale of the transition (e.g.\ the electroweak scale). Slow bubble growth also aids in realising this scenario. Scenario 2 is less obvious. This requires very few bubbles to nucleate, and for those bubbles to grow sufficiently fast to eventually coalesce and encompass the entire Universe. This naturally requires bubbles to expand for a long time and at a significant fraction of the speed of light. A high temperature phase transition aids in realising this scenario, as well as a large and long-lasting potential barrier to suppress bubble nucleation.

The absence of unit nucleation is often used as a guarantee that the transition cannot be successful. We argue that its definition is somewhat misleading. Nucleating less than one bubble per Hubble volume does not preclude the possibility of those bubbles reaching a Hubble volume in which no bubble nucleated. Unit nucleation represents an obvious early stage of a fast transition where the number of bubbles per Hubble volume can be many orders of magnitude above unity. However, unit nucleation fails to represent a consistent milestone for supercooled transitions. This is demonstrated in \cref{sec:nucleationApplicability}.

\subsection{Comparing unit nucleation and percolation} \label{sec:comparing-events}

Eventually, we wish to demonstrate the existence of Scenarios 1 and 2 in a concrete model. First we consider an analytic treatment that will provide insight into the conditions necessary for these scenarios to be realised. Later, we will verify the accuracy of this analytic treatment by cross-checking with a full numerical transition analysis in the models described in \cref{sec:models}.

Scenario 1 requires
\begin{equation}
	N(0) \geq 1 ~~\mathrm{and}~~ P_f(0) > 0.71 , \label{eq:scenario1}
\end{equation}
while Scenario 2 requires
\begin{equation}
	N(0) < 1 ~~\mathrm{and}~~ P_f(0) \leq 0.71 . \label{eq:scenario2}
\end{equation}
Here $N(0)$ is the average number of bubbles nucleated per Hubble volume, and $P_f$ is the fraction of the Universe remaining in the false vacuum. Both of these quantities are evaluated at zero temperature. However, our analysis could be generalised to a transition with a low temperature cutoff.

Our aim is to find threshold values of some parameter for these scenarios. We will use the bubble wall velocity, $v_w$, as this deciding parameter. Consider Scenario 1 (unit nucleation without percolation), and assume unit nucleation occurs while varying the extended volume of true vacuum bubbles. On one side of this threshold value for $v_w$, we have unit nucleation \textit{without} percolation, and on the other side we have unit nucleation \textit{with} percolation. Whether a transition has a bubble wall velocity above or below the threshold value is more important than how far from the threshold value the transition is. This is because we merely want to demonstrate the existence of the scenarios and understand how and when they arise.

\begin{figure}
	\centering
	\includegraphics[width=0.65\linewidth]{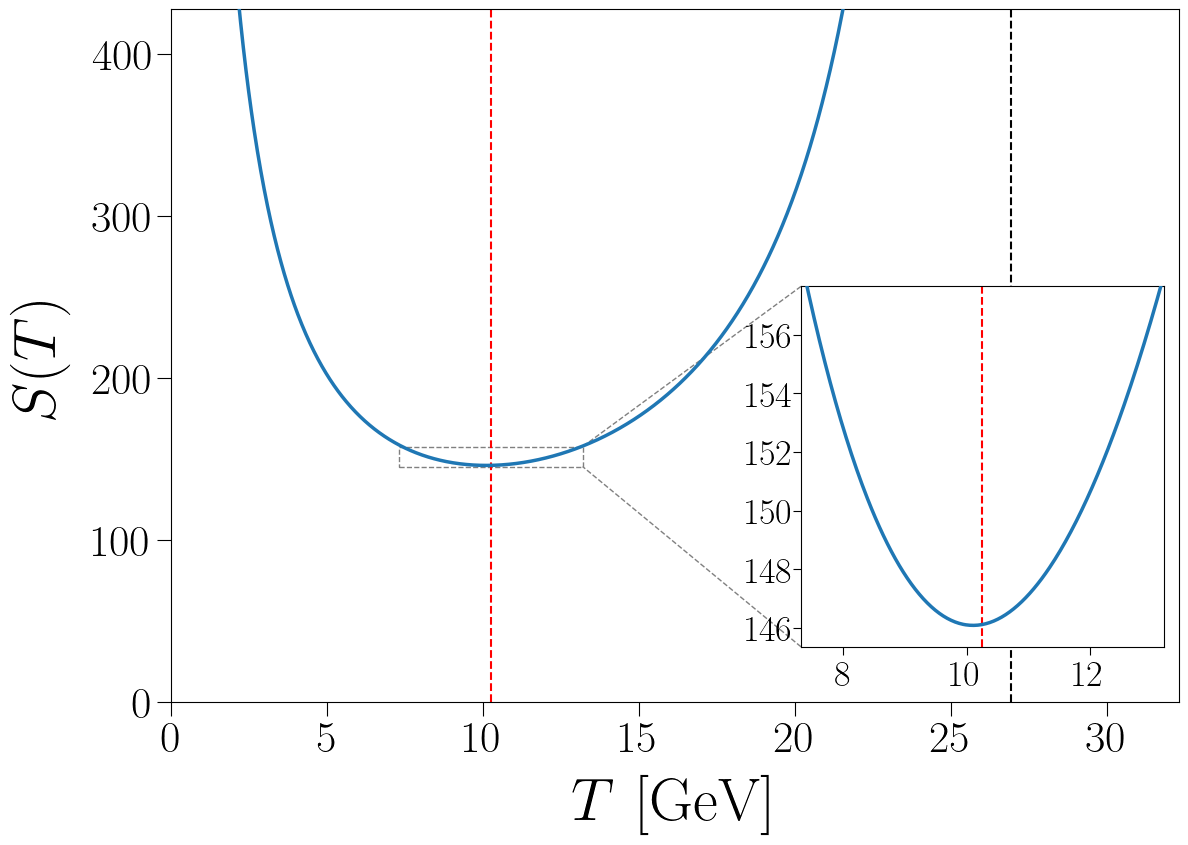}
	\caption{An example of an action curve. This example is from a benchmark, \bench{1}{5}, that we will define later in \cref{sec:M1}. Of our benchmarks in the two models that we investigate in \cref{sec:models}, this benchmark has the largest deviation from a quadratic action around $\Tmax$ (the dashed red line). The maximum nucleation rate occurs at $\Tmax \approx 10.25$ GeV, while the minimum of the action occurs at $\Tmin \approx 10.12$ GeV. Not displayed are the vertical asymptotes at $T=0$ and $T=T_c$ (the dashed black line). This U-shaped action curve is typical of supercooled transitions (see e.g.\ Refs.\ \protect\cite{Megevand:2016lpr, Cai:2017tmh, Kobakhidze:2017mru}).}
	\label{fig:actionCurve}
\end{figure}

Direct integration of \eqref{eq:falseVacuumFraction-temperatureSimplified} and \eqref{eq:numBubblesProper} is not feasible analytically unless many approximations are made. This subsection is dedicated to stating and justifying our assumptions to assist in the comparison of $N(0)$ and $P_f(0)$, and in the next subsection we detail the method of doing so. In fact, we take $N(0)$ as input and compare $P_f(0)$ to this while avoiding direct integration of \eqref{eq:falseVacuumFraction-temperatureSimplified}. While the following assumptions are convenient for an analytic investigation, we later demonstrate numerically through benchmarks --- without these assumptions --- that Scenarios 1 and 2 can occur. The expected action curve, $S(T)$, is shown in \cref{fig:actionCurve} with a minimum at $\Tmin$, and informs the first two assumptions. This shape of the action curve is typical of supercooled transitions, see e.g.\ Refs.\ \cite{Megevand:2016lpr, Cai:2017tmh, Kobakhidze:2017mru}.
\begin{enumerate}[align=left]
	\item[\hspace{0.35cm} Assumption 1:] The action is minimised at temperature $\Tmin$, where $S(\Tmin)$ and $\Tmin$ are both positive. The action is approximately quadratic close to its minimum. Because the nucleation depends exponentially on the action, the nucleation rate is approximately Gaussian, centred near $\Tmin$. We define $\Tmax$ as the temperature that maximises the nucleation rate, about which the Gaussian nucleation rate is centred. Although we expect $\Tmax \approx \Tmin$, we note that $\left.\dv{\Gamma}{T} \right\vert_{\Tmin} = \frac{4\Gamma(\Tmin)}{\Tmin} \neq 0$.%
	\footnote{If reheating occurs during the transition, then we should instead compute $\dv{\Gamma}{t}$ to find the minimum of $\Gamma(T(t))$, which would occur near the \textit{time} that minimises $S(T(t))$ (see e.g.\ Ref.\ \cite{Megevand:2017vtb}).}
	\item[\hspace{0.35cm} Assumption 2:] At zero temperature, the number of nucleated bubbles is not affected significantly by whether or not phantom bubbles are counted; that is, $N(0) \approx \Next(0)$. If we consider Scenario 1, then percolation does not occur so $P_f \approx 1$. If we consider Scenario 2, then very few bubbles nucleate and yet percolation occurs. This suggests that bubbles nucleate early and nucleation is later suppressed while prolonged bubble growth continues. In this case, the bubbles nucleate when $P_f \approx 1$. Thus, we generically expect Assumption 2 to hold in these two scenarios for the models we consider.
	\item[\hspace{0.35cm} Assumption 3:] The energy density \eqref{eq:energyDensityReduced} can be decomposed as a radiation component and a vacuum component. The vacuum component can be approximated as $\rhov(T) \approx \Vfz - \Vgsz$, as in \eqref{eq:energyDensityApprox}. This approximation is reasonable for transitions where the vacuum component away from zero temperature is dominated by the radiation component given by \eqref{eq:rhoR}. As discussed below \eqref{eq:energyDensityApprox}, this decomposition can be generalised to transitions where the false vacuum phase does not persist at zero temperature. We use \eqref{eq:energyDensityReduced} rather than \eqref{eq:energyDensityApprox} for the energy density in our numerical simulations.
\end{enumerate}
In Assumption 1 we defined $\Tmax$ as the temperature that maximises the nucleation rate. For transitions that complete after cooling below $\Tmax$ (i.e.\ $T_f < \Tmax$), we naturally find $\Tmax \approx \Tmin$, where $\Tmin$ is the temperature that minimises the action. The maximum difference between $\Tmin$ and $\Tmax$ in the benchmarks listed in \cref{sec:models} is 0.7\%. Hence, we may use these temperatures interchangeably in the models we consider, which will be important for \cref{sec:nucleationApplicability}.

\subsection{When does a transition percolate?} \label{sec:reduceProblem}

As a reminder, we wish to simultaneously compare $N(0)$ to 1 and $P_f(0)$ to 0.71, which are the number of nucleated bubbles and the false vacuum fraction at zero temperature, respectively. In the following, we decompose $P_f$ in such a way that eventually permits direct comparison of $N(0)$ and $P_f(0)$. Using Assumption 2, we can approximate \eqref{eq:numBubblesProper} as
\begin{equation}
	N(T) \approx \int_{T}^{T_c} \! dT' \, \frac{\Gamma(T')}{T' H^4(T')} = \int_{T}^{T_c} \! dT' \, X(T') ,
\end{equation}
where we have defined
\begin{equation}
	X(T) = \frac{\Gamma(T)}{T H^4(T)} . \label{eq:X}
\end{equation}
To assist in comparing the number of nucleated bubbles to the true vacuum volume, we also want to similarly rewrite $P_f$, so that both $N$ and $P_f$ involve integrals of $X$. Introducing
\begin{equation}
	Y(T', T) = \frac{v_w}{T'} \int_T^{T'} \! dT'' \, \frac{H(T')}{H(T'')} , \label{eq:Y}
\end{equation}
we can decompose $P_f$ as
\begin{equation}
	P_f(T) = \exp(-\frac{4\pi}{3} \int_T^{T'} \! dT' \, X(T') Y^3(T', T)) . \label{eq:falseVacuumFraction-XY}
\end{equation}
One can recognise that 
\begin{equation}
	Y(\Tmax, T) = \frac{r(\Tmax, T)}{r_H(\Tmax)} , \label{eq:Y-radiusRatio} 
\end{equation}
where
\begin{equation}
	r(T', T) = v_w \int_T^{T'} \! dT'' \recip{a(T'') T'' H(T'')}
\end{equation}
is the comoving radius of a bubble nucleated at $\Tmax$ that has grown until $T$, and
\begin{equation}
	r_H(T) = \recip{a(T) H(T)}
\end{equation}
is the comoving Hubble radius. Then $Y$ can be interpreted as the ratio of radii of the most common bubbles to the Hubble volume, in comoving coordinates.%
\footnote{We use comoving radius because the proper radius of bubbles is ill-defined at zero temperature, since $a(T) \propto T^{-1}$.}
The remainder of this subsection is dedicated to an analytic comparison of the number of bubbles nucleated and the true vacuum volume, aided by the use of well-motivated assumptions and approximations.

The conditions for Scenario 1, unit nucleation without percolation, and Scenario 2, percolation without unit nucleation, are given in \eqref{eq:scenario1} and \eqref{eq:scenario2}, respectively. The exclusivity between the conditions for these two scenarios means that once we have found conditions for one scenario, it will be easy to find the conditions for the other. The conditions for Scenario 1 can be expressed as
\begin{equation}
	\int_0^{T_c} \! dT \, X(T) \geq 1 \;\;\; ~\mathrm{and}~ \;\;\; \exp(-\frac{4\pi}{3} \int_0^{T_c} \! dT \, X(T) Y^3(T, 0)) > 0.71 . \label{eq:tempRefToNucleationWithoutPercolation}
\end{equation}
It will be convenient to work with the extended volume of true vacuum bubbles, $\Vext$, which appears inside the exponent of $P_f$ (see \eqref{eq:falseVacuumFraction}), rather than $P_f$ directly. Let $f_i = P_f(T_i)$ denote the false vacuum fraction for some milestone temperature $T_i$. We have $f_p = 0.71$ for percolation and $f_f = 0.01$ for completion, according to our definitions in \cref{sec:transitionTemperatures}. We can express the condition $P_f(0) > 0.71$ as
\begin{equation}
	\int_0^{T_c} \! dT \, X(T) Y^3(T, 0) < - \frac{3 \ln(f_p)}{4\pi} = \vextnFactor{p}^3 , \label{eq:nuclWOPerc-intermediate}
\end{equation}
where we have defined
\begin{equation}
	\vextnFactor{i} = \left(- \frac{3 \ln(f_i)}{4\pi} \right)^{\!\!\frac{1}{3}} . \label{eq:vextn}
\end{equation}
The factor of $3/(4\pi)$ would not be present if a spherical Hubble volume was used in $N$ (see \eqref{eq:numBubblesProper}).

Multiplying the right-hand side of \eqref{eq:nuclWOPerc-intermediate} by a factor of one and using Assumption 2, we have
\begin{equation}
	\int_0^{T_c} \! dT \, X(T) Y^3(T, 0) < \frac{\vextnFactor{p}^3}{N(0)} \int_0^{T_c} \! dT \, X(T) .
	\label{eq:PercCondReexpressed}
\end{equation}
We take $N(0)$ as input in the following analysis. The number of bubbles nucleated throughout the transition requires knowledge of the action as a function of temperature, which we cannot hope to capture in a model-independent way. In fact, our analysis will be completely model-independent, except that the model must admit a phase transition at roughly the electroweak scale, and this transition must satisfy the assumptions detailed in \cref{sec:comparing-events}.

Now consider percolation more generally. The inequality \eqref{eq:PercCondReexpressed} becomes an equation in the general case, which may be written as
\begin{equation}
	\int_0^{T_c} \! dT \, X(T) A(T) = Z , \label{eq:PercCondGeneral}
\end{equation}
where $Z$ is a real number that can take either sign or even vanish, and
\begin{equation}
	A(T) = Y^3(T, 0) - \frac{\vextnFactor{p}^3}{N(0)} . \label{eq:noPercCond-A}
\end{equation}
The sign of $Z$ indicates whether percolation occurs: $Z < 0$, $Z = 0$ and $Z > 0$ are respectively where percolation does not occur, the threshold of percolation ($P_f(0) = 0.71$), and where percolation does occur. That is, $\sign(Z) = \sign(0.71 - P_f(0))$. To classify whether a transition satisfies the conditions of Scenario 1, Scenario 2, or neither, requires only the determination of $\sign(Z)$, along with $N(0)$ which we take as input. However, the integral in \eqref{eq:PercCondGeneral} still cannot be evaluated analytically. We can make one further simplification to avoid the integration in \eqref{eq:PercCondGeneral} altogether.

\begin{figure}
	\centering
	\includegraphics[width=0.7\linewidth]{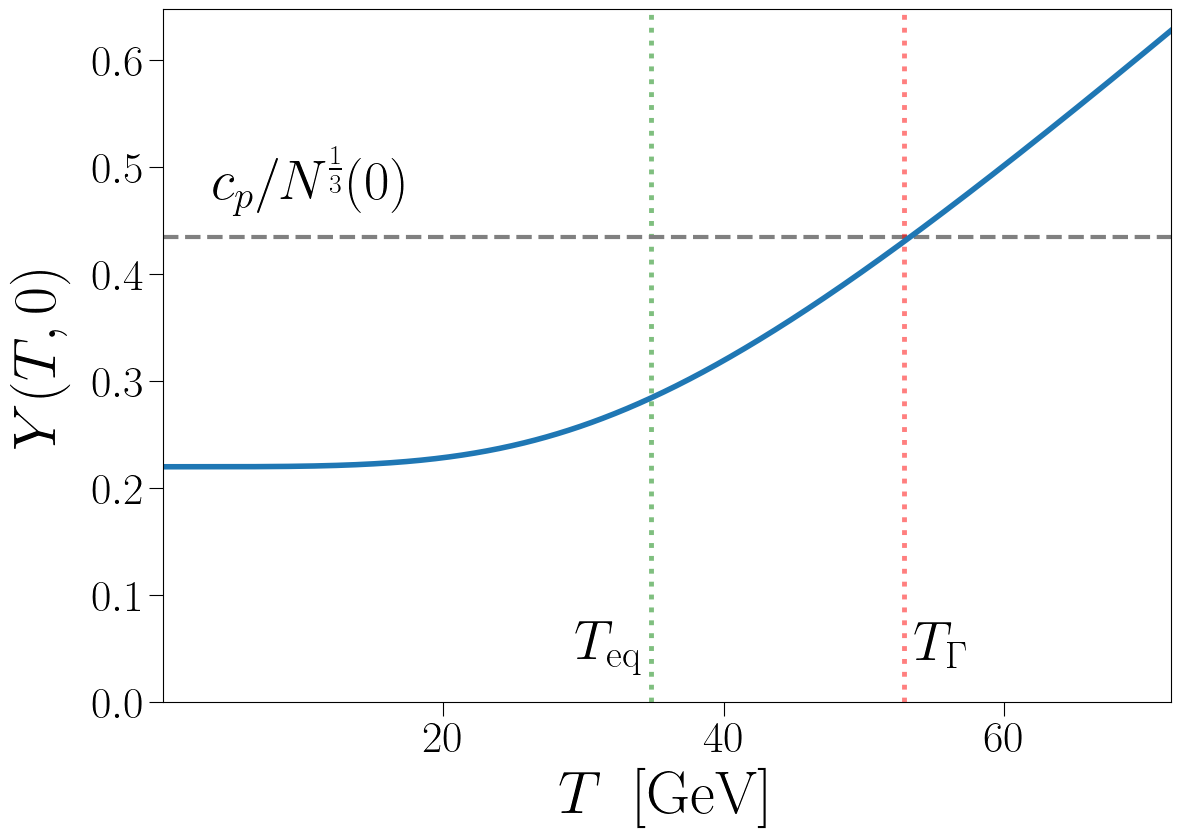}
	\caption{The function $Y(T, 0)$ for benchmark \bench{2}{1} with $v_w = 0.220$. The horizontal dashed line marks the threshold $\vextnFactor{p}/N^{\recip{3}}(0)$, where $N(0) \approx 1$. The vertical dotted lines correspond to $\Teq$ (green) and $\Tmax$ (red). This benchmark has $Y(\Tmax, 0) < \vextnFactor{p}/N^{\recip{3}}(0)$, and hence is predicted to not yield percolation. Indeed, for \bench{2}{1} we numerically find $v_w \approx 0.223$ is required for percolation.}
	\label{fig:Y}
\end{figure}

The form of $Y(T, 0)$ is shown in \cref{fig:Y} for the benchmark \bench{2}{1} (defined in \cref{sec:M2}). This form holds in general and can be understood as follows. During vacuum domination ($T \ll \Teq$), a constant Hubble parameter in \eqref{eq:Y} yields a constant: $Y(T, 0) \approx v_w$. During radiation domination ($T \gg \Teq$), instead $dY(T, 0)/dT$ is approximately constant, which can be seen by using $H(T) \approx T^2/\xi_g$ in \eqref{eq:Y}. Importantly, $Y(T, 0)$ monotonically increases with temperature from zero temperature to the critical temperature, and consequently so does $A(T)$. Because $X(T)$ is strictly positive, we have $\sign(X(T) A(T)) = \sign(A(T))$. Assumption 1 implies that $X(T)$ is an even function (specifically Gaussian) about $\Tmax$. Then we have $Z = 0$ if $A(T)$ is odd about $\Tmax$. Further, we have $\sign(Z) = \sign(c)$ if $A(T) - c$ is odd about $\Tmax$. Because we only need the sign of $Z$ (and not the magnitude) to determine whether percolation occurs, we can reduce the problem of checking \eqref{eq:PercCondReexpressed} to simply computing $\sign(c) = \sign(A(\Tmax))$, provided $A(T) - c$ is odd about $\Tmax$. This property of $A(T)$ can be seen by the fact that it scales as a constant for low temperatures $T \ll \Teq$, and becomes most strongly temperature-dependent at high temperatures $T \gg \Teq$, scaling as $T^3$. Due to the exponential suppression away from $\Tmax$ by $X(T)$, any deviation from the odd function we desire is washed out. This leaves an approximately odd function about $\Tmax$, provided the Gaussian nucleation rate is not broadly peaked.

We numerically confirm that evaluating $\sign(A(\Tmax))$ is a reasonably accurate way to predict whether percolation occurs (with comparison to numerical results shown later in \cref{fig:noPercCases,fig:noCompletionCases}). We expect this to hold in general, unless the energy density of the Universe scales differently with temperature than assumed for the models considered here. Another caveat is that alternate nucleation scenarios can be conceived where there is a prolonged window of significant nucleation, in which case suppression of the nucleation rate away from the maximum is not strong. In either case, the same general principles for realising Scenarios 1 and 2 should still apply, however, the analytic treatment used here would need to be significantly altered.

The percolation condition resulting from the simplification $\sign(Z) = \sign(A(\Tmax))$ is that percolation occurs if
\begin{equation}
	Y(\Tmax, 0) > \frac{\vextnFactor{p}}{N^{\frac13}(0)} . \label{eq:percCond-Y}
\end{equation}
Then, checking \eqref{eq:percCond-Y} and whether $N(0)$ is larger or smaller than unity is sufficient to determine whether Scenarios 1 and 2 occur.

\subsection{Scenario 1: unit nucleation without percolation} \label{sec:scenario1}

In \cref{sec:reduceProblem} we showed that the conditions for unit nucleation without percolation reduce to simply finding when
\begin{equation}
	Y(\Tmax, 0) < \frac{\vextnFactor{p}}{N^{\frac13}(0)} , \label{eq:noPercCond-Y}
\end{equation}
with $N(0) \geq 1$ taken as input. Because $Y$ is an integral of Hubble parameters, it remains to evaluate the Hubble parameter. Using Assumption 3, we can approximate the Hubble parameter by considering a constant vacuum energy density: $\rhov(T) \approx \rhoz$. The Hubble parameter is then
\begin{equation}
	H(T)\approx \sqrt{\frac{8\pi G}{3} \left(\frac{T^4}{\xi_g^2} + \rhoz \right)} . \label{eq:H-approx}
\end{equation}
where $\rhoz = \Vfz - \Vgsz$. We define $\Teq$ as the temperature for which the radiation and vacuum energy densities are equal;
\begin{equation}
	\rhor(\Teq) = \rhov(\Teq) .
\end{equation}
Under the approximation of constant vacuum energy density, we have
\begin{equation}
	\Teq^2 = \xi_g \sqrt{\rhoz} .
\end{equation}
Then we can express \eqref{eq:H-approx} as
\begin{equation}
	H(T) \approx \sqrt{\frac{8\pi G}{3} \rhoz} \sqrt{\left(\frac{\Tmax}{\Teq} \right)^{\!\!4} + 1} .
\end{equation}

This approximation for the Hubble parameter permits an analytic result for the integral of the ratio of Hubble parameters in \eqref{eq:Y}. The condition for Scenario 1 can be expressed as a bound on the bubble wall velocity:
\begin{equation}
	v_w < \frac{\vextnFactor{p}}{N^{\frac13}(0)} \Bigg[\sqrt{\left(\frac{\Tmax}{\Teq} \right)^{\!\!4} + 1} \,\, \hypergeomArgs{-\!\left(\frac{\Tmax}{\Teq} \right)^{\!\!4}} \Bigg]^{-1} , \label{eq:vwBound-scenario1}
\end{equation}
where $\hypergeom$ is a hypergeometric function. Interestingly, the bubble wall velocity required for percolation depends only on two quantities: the number of bubbles nucleated during the transition, $N(0)$; and the temperature at which bubble nucleation is maximised, $\Tmax$, relative to the temperature when the radiation and vacuum energy densities are equal, $\Teq$.

\begin{figure}
	\centering
	\includegraphics[width=0.7\linewidth]{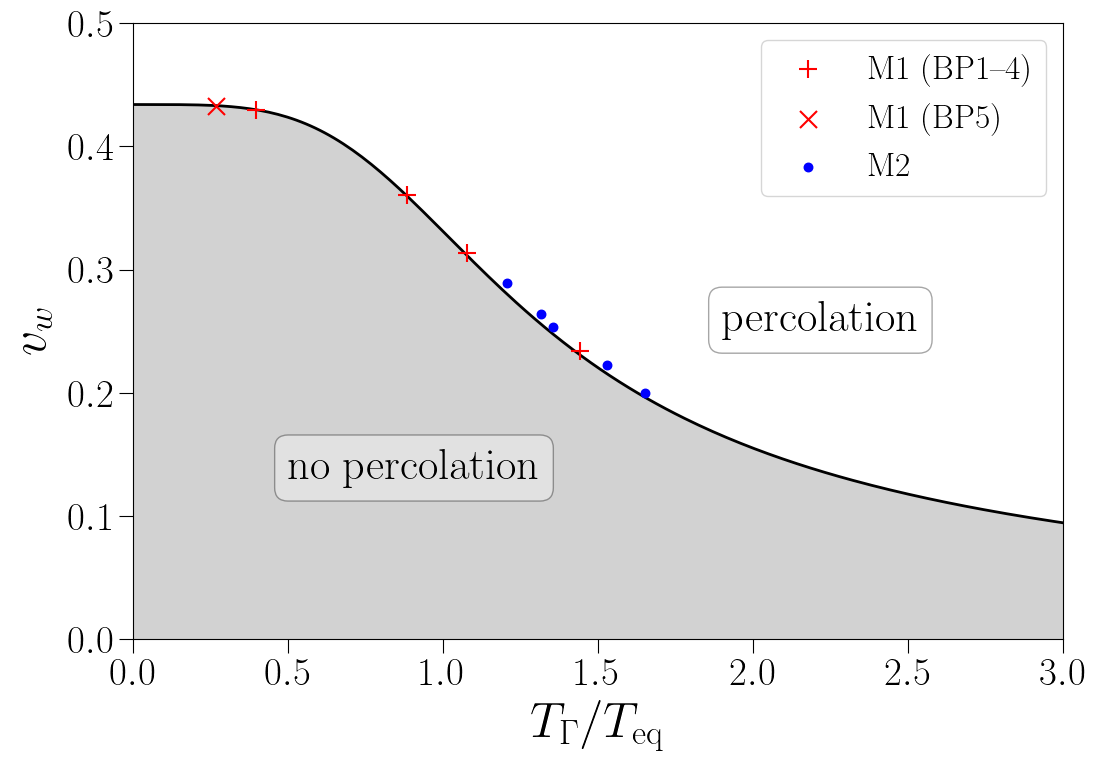}
	\caption{The bounds on the bubble wall velocity for which no percolation is expected, where $N(0) = 1$. The shaded region corresponds to no percolation. For a given value of $\Tmax/\Teq$, if $v_w$ lies above the curve, then percolation is expected. The red pluses are benchmark points for a toy model (M1) (defined in \cref{tab:benchmarks-M1}), and the blue dots are benchmark points for the Standard Model with a scalar singlet (M2) (defined in \cref{tab:benchmarks-M2}). The red cross also belongs to M1 (specifically \bench{1}{5}), but the true vacuum does not exist at $\Teq$. However, $\Teq$ can be obtained analytically for this benchmark as described in \cref{sec:M1}.}
	\label{fig:noPercCases}
\end{figure}

The result for the bubble wall velocity bound \eqref{eq:vwBound-scenario1} is presented in \cref{fig:noPercCases}. To have unit nucleation without percolation (i.e.\ Scenario 1), we must have $N(0) \geq 1$ in addition to satisfying the bound on the bubble wall velocity in \eqref{eq:vwBound-scenario1}. This scenario is realised in the shaded region of \cref{fig:noPercCases}, where $N(0) = 1$ was chosen. The choice of $N(0)$ scales the vertical axis; a value of $N(0)$ above this threshold of unity would permit percolation for lower bubble wall velocities. Intuitively, the results suggest that percolation does not occur if the bubbles expand sufficiently slowly. In a phase transition with slow bubble walls, the transition will complete via nucleation of many small bubbles. Or, if the nucleation rate is suppressed as we consider here, the bubbles will not percolate and the transition will not complete. The reciprocal factor in brackets in \eqref{eq:vwBound-scenario1} has an upper bound of unity as $\Tmax/\Teq \rightarrow 0$. Then for $N(0) < c_p^3$, even $v_w = 1$ is not predicted to yield percolation for $\Tmax \ll \Teq$. This corresponds to the average number of nucleated bubbles (throughout the entire transition) per Hubble volume being very low; specifically $N(0) \lesssim 0.08$, which is an order of magnitude below unit nucleation. Nevertheless, percolation can still occur with even fewer bubbles if $\Tmax \gg \Teq$. For $N(0) \gg 1$, we expect percolation to occur unless bubbles grow very slowly during adiabatic expansion. Although we take $v_w$ to be a free parameter, bubbles that expand as deflagrations or hybrids cause reheating in the false vacuum whereas detonations cause reheating in the true vacuum. This is an important distinction \cite{Heckler:1994uu, Megevand:2016lpr, Megevand:2017vtb} that we do not capture in this study, with our analysis being most suitable for detonations.%
\footnote{Modelling of the false vacuum fraction evolution in the presence of reheating from deflagrations, as well as the effects of reheating on the gravitational wave signal, was recently performed in Ref.\ \cite{Ajmi:2022nmq}.}
Reheating from bubbles expanding as deflagrations or hybrids could significantly delay percolation and may even prevent percolation for strongly supercooled transitions.%
\footnote{However, with sufficiently strong supercooling, deflagrations may not be a stable hydrodynamic solution \cite{Megevand:2013yua, Megevand:2014dua}. The instability is expected to accelerate the bubble walls, perhaps somewhat mitigating the hindrance to percolation from reheating. The realisation and analysis of deflagrations in the presence of strong supercooling may be an interesting avenue for future investigation.}

We now discuss the implications of the obtained bubble wall velocity bounds. In \cref{fig:noPercCases} we see that percolation is hardest to achieve when bubbles are predominantly nucleated during vacuum domination ($\Tmax \ll \Teq$); in that a larger bubble wall velocity is required for percolation compared to cases where bubbles nucleate when the radiation energy density is still significant ($\Tmax \gtrsim \Teq$). This result is expected because the bubbles that nucleate during radiation domination have both the radiation- and vacuum-dominated eras during which to grow, whereas bubbles that nucleate during the vacuum-dominated era can only grow during vacuum domination. Thus, the ratio $\Tmax/\Teq$ indicates how long bubbles can grow for. For Scenario 1, unit nucleation without percolation, we must have $N(0) \geq 1$, otherwise unit nucleation does not occur. Then the largest bubble wall velocity for which percolation does not occur is $v_w = c_p \approx 0.43$, according to the bound for $v_w$ with $\Tmax \ll \Teq$ and $N(0) = 1$. Thus, percolation is guaranteed to occur if unit nucleation occurs, provided the bubbles expand as detonations or hybrids, which have $v_w > c_s \simeq 0.58$ for a perfect fluid in the MIT bag equation of state. Strictly speaking, Scenario 1 lies outside the regime of validity of our analysis because we neglect reheating, and therefore only model detonations correctly.
However, in \cref{app:reheating} we discuss how reheating affects Scenarios 1 and 2 by hindering percolation, and we reason that reheating should only strengthen the case for the existence of Scenario 1 for deflagrations. Additionally, reheating may allow Scenario 1 to be realised for bubbles growing as hybrids. Although we generically expect detonations for strong supercooling, hybrids may be possible if friction from the plasma is sufficient. We then predict that Scenario 1 is possible for strongly supercooled transitions (with $N(0) \approx 1$) that proceed with bubbles growing as deflagrations or hybrids. Conversely, we predict that unit nucleation implies percolation for transitions that are not strongly supercooled or where bubbles grow as detonations.

Formulating the condition for Scenario 1 as a bound on the bubble wall velocity is quite intuitive. Percolation occurs once a large enough fraction of the Universe is contained in true vacuum bubbles. A low bubble wall velocity results in bubbles expanding slower, delaying and perhaps precluding the onset of percolation. One can further connect this to the bubble radius using \eqref{eq:Y-radiusRatio} in \eqref{eq:noPercCond-Y} to give the condition
\begin{equation}
	\frac{r(\Tmax, 0)}{r_H(\Tmax)} < \frac{\vextnFactor{p}}{N^{\recip{3}}(0)} \label{eq:noPercCond-RH}
\end{equation}
for Scenario 1, again with $N(0) \geq 1$. The left-hand side is the bubble's comoving radius as a fraction of the comoving Hubble radius at the time of nucleation. For percolation to occur, this fractional radius $\mathcal{R}(T) = r(\Tmax, T)/r_H(\Tmax)$ must to be at least $c_p/N^{\recip{3}}(0)$ as $T \rightarrow 0$. When considering a spherical Hubble volume (i.e.\ including the factor of $4\pi/3$ in \eqref{eq:numBubblesProper}) and using $f_p = 0.71$, we have the constraint $\mathcal{R}(0) \gtrsim 0.70 / N^{\recip{3}}(0)$ for percolation to occur. Replacing $\vextnFactor{p}$ with $\vextnFactor{f}$ in \eqref{eq:noPercCond-RH} and using $f_f = 0.01$, we have the constraint $\mathcal{R}(0) \gtrsim 1.7 / N^{\recip{3}}(0)$ for completion to occur.

\subsection{Scenario 2: percolation without unit nucleation} \label{sec:scenario2}

The analysis of Scenario 2 follows from that of Scenario 1. The condition for Scenario 2 equivalent to \eqref{eq:noPercCond-Y} is given in \eqref{eq:percCond-Y}, and we must have $N(0) < 1$. Fortunately, we can reuse all results from Scenario 1 simply by reversing the inequalities. Scenario 2 occurs in the non-shaded region above the curve in \cref{fig:noPercCases}.%
\footnote{This second usage of \cref{fig:noPercCases} is possible because we consider $N(0) = 1$; the threshold of unit nucleation. Reducing $N(0)$ infinitesimally in \cref{fig:noPercCases} will not affect the partitions in the plot, but will then be considering the case of no unit nucleation.}
Even with a low number density of nucleated bubbles, percolation and completion are still possible, provided the bubbles expand quickly and have enough time to expand (that is, they nucleate at a high enough temperature relative to $\Teq$). Percolation is more difficult to achieve if bubbles predominantly nucleate during the vacuum-dominated era ($\Tmax \ll \Teq$) because the bubbles have less time to grow. As long as $N(0) > \vextnFactor{p}^3 \approx 0.08$, percolation without unit nucleation remains possible even if $\Tmax \ll \Teq$, provided the bubbles grow as detonations. Superluminal bubble wall velocities are required for percolation if $N(0) < \vextnFactor{p}^3$ and bubbles nucleate during the vacuum-dominated era. If $N(0) \sim 1$ and bubbles grow as deflagrations or hybrids, reheating in front of the bubble walls may spoil percolation (as discussed in \cref{app:reheating}). For instance, the bubble wall velocity bound when bubbles nucleate predominantly in the radiation-dominated era ($\Tmax \gtrsim \Teq$) suggests that bubble wall velocities even in the range $v_w \sim$ 0.1--\hspace{1pt}0.2 could yield percolation. However, percolation is hindered by the suppression of bubble nucleation and growth due to reheating in deflagrations and hybrids, so larger bubble wall velocities than predicted are required for percolation. Thus, the obtained bubble wall velocity bounds are necessary but not sufficient conditions for Scenario 2 outside of the detonation regime. Besides, a constant bubble wall velocity throughout the transition is not a reasonable approximation when reheating occurs \cite{Heckler:1994uu, Megevand:2017vtb}. Quantitative predictions for Scenarios 1 and 2 in the case of reheating are beyond the scope of this study. 

The obtained bubble wall velocity bounds provide no insight of the time ordering of unit nucleation and percolation events. A natural extension to Scenario 2 is percolation before unit nucleation. A similar analytic treatment to that used for Scenarios 1 and 2 could be applied for this scenario, with some difficulty and much tedium. We argue that such a scenario is largely unimportant, with this assessment following from our findings in \cref{sec:nucleationApplicability} that the event of unit nucleation is an arbitrary milestone. We note that Ref.\ \cite{Wang:2020jrd} demonstrate strongly supercooled cases where $T_n$ can become close to $T_p$, and also cases where $T_n < \Teq$. Ref.\ \cite{Kobakhidze:2017mru} consider cases where $T_p < \Teq < T_n$,%
\footnote{Although Ref.\ \cite{Kobakhidze:2017mru} stated the ordering of event times, we have neglected reheating, so the ordering of temperatures is the reverse ordering of times.}
and mention that scale-invariant models can have $T_p < T_n < \Teq$. We find both cases in our benchmarks (stated in \cref{sec:models}), but, like Refs.\ \cite{Kobakhidze:2017mru, Wang:2020jrd}, we do not find any benchmarks where $0 < T_n < T_p$.

\subsection{A note on completion} \label{sec:completion}

\begin{figure}
	\centering
	\includegraphics[width=0.7\linewidth]{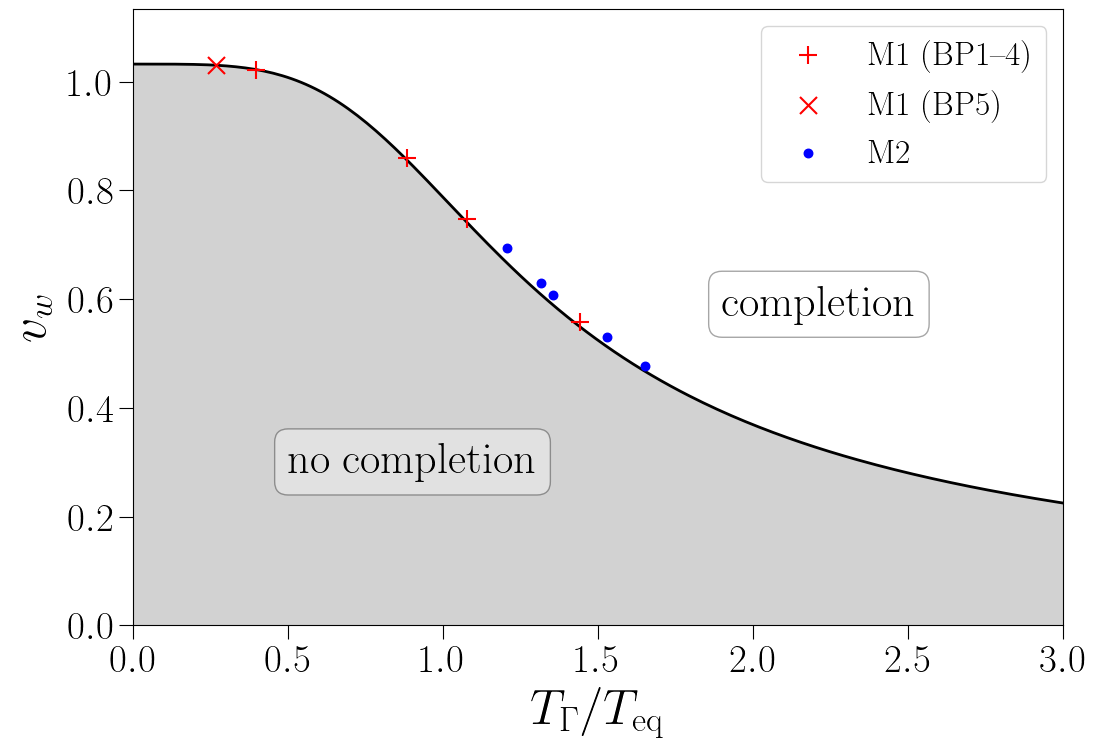}
	\caption{The same as \cref{fig:noPercCases} except for completion rather than percolation. The shaded region now corresponds to where completion is not expected. Since $v_w > 1$ is invalid, completion without unit nucleation is not possible for sufficiently small $\Tmax/\Teq$, even if $N(0) \sim 1$.}
	\label{fig:noCompletionCases}
\end{figure}

The bubble wall velocity bound obtained for Scenarios 1 and 2 in \cref{sec:scenario1,sec:scenario2} can also be applied to unit nucleation without completion and completion without unit nucleation, respectively, simply by replacing $\vextnFactor{p}$ with $\vextnFactor{f}$ (see \eqref{eq:vextn}). The results are shown in \cref{fig:noCompletionCases}. We see that completion is no longer possible for $\Tmax \lesssim \Teq/2$ at the threshold of unit nucleation, even with $v_w = 1$. Additionally, completion is not possible (even neglecting reheating) for deflagrations at the threshold of unit nucleation, unless $\Tmax \gtrsim 1.38 \Teq$. However, one may na{\"\i}vely expect that percolation can still occur in some of the cases where completion does not. We will demonstrate that this is not necessarily true.

Even if the fraction of the Universe in the false vacuum decreases to some completion threshold $P_f(T_f) = \epsilon \ll 1$ and beyond, finite pockets of the false vacuum can persist if the rate of false vacuum conversion does not exceed the rate of expansion of the false vacuum. At the end of \cref{sec:transitionTemperatures} we mentioned five completion criteria (CC):
\begin{enumerate}[align=left]
	\item[\hspace{0.35cm} \textbf{\cc{1}} \,] The false vacuum fraction becomes sufficiently small.
	\item[\hspace{0.35cm} \textbf{\cc{2}} \,] The physical volume of the false vacuum, $\Vphys$, decreases for some temperature $T > T_f$.
	\item[\hspace{0.35cm} \textbf{\cc{3}} \,] $\Vphys$ decreases on average from percolation to completion: $\Vphys(T_f) < \Vphys(T_p)$. This constraint conveniently maps to the condition $T_f \gtrsim 0.24 T_p$.
	\item[\hspace{0.35cm} \textbf{\cc{4}} \,] $\Vphys$ is decreasing at $T_f$: $\left. \displaystyle \dv{\Vphys}{t} \right\vert_{T_f} < 0$.
	\item[\hspace{0.35cm} \textbf{\cc{5}} \,] $\Vphys$ is decreasing at $T_p$: $\left. \displaystyle \dv{\Vphys}{t} \right\vert_{T_p} < 0$.
\end{enumerate}
In this study we do not determine simple, model-independent analytic conditions for these constraints equivalent to the conditions found in \cref{sec:scenario1,sec:scenario2} for Scenarios 1 and 2. Instead we numerically find the bubble wall velocity, $v_w$, for which each of these constraints are satisfied. We do this for each of the benchmarks stated in \cref{sec:models}, which we use in \cref{fig:noPercCases,fig:noCompletionCases}. We consistently find that the value of $v_w$ for which \cc{2}, \cc{3}, \cc{4} and \cc{5} are satisfied in our benchmarks are respectively 10-15\%, 18-21\%, 21-23\% and 70-123\% above the value of $v_w$ required for completion (\cc{1}), where the ranges show the full variation in results amongst our benchmarks. This holds for each benchmark, with the required bubble wall velocity being progressively larger in the order of \cc{1-5}. A significant variation between benchmarks is only observed for \cc{5}. We tentatively propose the more general bound (analogous to \eqref{eq:percCond-Y}) for a successful transition,
\begin{equation}
	Y^3(\Tmax, 0) \geq \frac{\ccFactor{i} \vextnFactor{f}^3}{N(0)} , \label{eq:completionBounds}
\end{equation}
with $\ccFactor{1} = 1$, $\ccFactor{2} = 1.15$, $\ccFactor{3} = 1.21$, $\ccFactor{4} = 1.23$ and $\ccFactor{5} = 2.23$. Recalling the situation depicted in \cref{fig:physicalVolume}, we recommend the use of \cc{5}, or at least \cc{4}, for a conservative estimate of transition completion. If one wishes to rule out the possibility of transition completion, then failure to satisfy \cc{1} is sufficient. In general, one should be cautious if any of the success criteria are not satisfied.

Although these results suggest that a transition completes (i.e.\ $T_f > 0$) if \cc{5} is satisfied, it remains to be proven. We have provided analytic arguments that support the claim that a transition completes if and only if \eqref{eq:completionBounds} is satisfied, using $\ccFactor{1}$. No such analytic arguments have been made for the physical volume of the false vacuum and we are not aware of any existing proof that $\dv{t} \Vphys (T_p) < 0 \implies P_f(0) \rightarrow 0$.

To further investigate the ordering of $\ccFactor{i}$ seen in our benchmarks, we consider a simple transition model that allows for analytic investigation of $\Vphys$ and $P_f$. We refer to this as the simultaneous nucleation model. We assume instantaneous nucleation of all bubbles at some temperature $T_*$ such that $\Gamma(T) = n_B \delta(T - T_*)$, with $n_B$ being the number density of bubbles. We also assume vacuum domination throughout the transition, with the further approximation $H(T) = H_*$. Here, $T_*$ and $H_*$ are both constant. With these simplifications, the extended volume in \eqref{eq:falseVacuumFraction-temperatureSimplified} is
\begin{equation}
	\Vext(T < T_*) = \frac{4\pi}{3} b^3 \! \left(1 - \frac{T}{T_*} \right) ,
\end{equation}
where we have defined
\begin{equation}
	b = v_w \! \left(\frac{n_B}{T_* H_*^4} \right)^{\!\!\recip{3}} = v_w N^{\frac13}(0) . \label{eq:b-new}
\end{equation}
The last equality in \eqref{eq:b-new} is found using \eqref{eq:numBubblesProper}, with $N(0)$ being the number of bubbles per Hubble volume nucleated throughout the entire transition. The percolation and final temperatures are defined through $P_f(T_i) = f_i$, or equivalently $\Vext(T_i) = -\ln(f_i)$. In this simple model, the percolation and final temperatures are given by
\begin{equation}
	T_i = T_* \! \left(1 - \frac{\vextnFactor{i}}{b} \right) , \label{eq:Ti-new}
\end{equation}
where $c_i$ is as defined in \eqref{eq:vextn}. Additionally, we can determine whether the physical volume of the false vacuum, $\Vphys$, is decreasing with time. This condition can be checked by finding $d\Vext/dT$, as in \eqref{eq:decreasingPhysicalVolume}. This is simple in the simultaneous nucleation model under consideration, because
\begin{equation}
	\dv{\Vext}{T} = -\frac{4\pi b^3}{T_*} \! \left(1 - \frac{T}{T_*} \right)^{\!\!2} .
\end{equation}
Then \eqref{eq:decreasingPhysicalVolume} becomes
\begin{equation}
	3 - 4\pi b^3 \frac{T_i}{T_*} \! \left(1 - \frac{T_i}{T_*} \right)^{\!\!2} < 0 . \label{eq:decreasingPhysicalVolume-simple-new}
\end{equation}
We can use \eqref{eq:Ti-new} in \eqref{eq:decreasingPhysicalVolume-simple-new} to check whether $\Vphys$ is decreasing at the percolation or completion temperature. This yields the constraint
\begin{equation}
	b > \vextnFactor{i} + \frac{3}{4\pi \vextnFactor{i}^2} . \label{eq:decreasingPhysicalVolume-simple-new-Ti}
\end{equation}

It is possible to determine analytic bounds on $b = v_w N^{\frac13}(0)$ for all completion criteria in the simultaneous nucleation model, using \eqref{eq:Ti-new}--\eqref{eq:decreasingPhysicalVolume-simple-new-Ti}. The bounds are
\begin{enumerate}[align=left]
	\item[\hspace{0.35cm} \textbf{\cc{1}} \,] $T_f > 0$: \; $b > \vextnFactor{f} \approx 1.032$.
	\item[\hspace{0.35cm} \textbf{\cc{2}} \,] $\displaystyle \dv{\Vphys}{t} < 0$ for some $T_f < T(t) < T_*$: \; $\displaystyle b > \left(\frac{81}{16\pi} \right)^{\!\!\frac13} \approx 1.172$.
	\item[\hspace{0.35cm} \textbf{\cc{3}} \,] $\Vphys(T_f) < \Vphys(T_f)$: \; $\displaystyle b > \left[\left(\frac{f_p}{f_f} \right)^{\!\!\frac13} \vextnFactor{f} - \vextnFactor{p} \right] \left[\left(\frac{f_p}{f_f} \right)^{\!\!\frac13} - 1 \right]^{-1} \approx 1.223$.
	\item[\hspace{0.35cm} \textbf{\cc{4}} \,] $\displaystyle \left. \dv{\Vphys}{t} \right\vert_{T_f} < 0$: \; $\displaystyle b > \vextnFactor{f} + \frac{3}{4\pi \vextnFactor{f}^2} \approx 1.256$.
	\item[\hspace{0.35cm} \textbf{\cc{5}} \,] $\displaystyle \left. \dv{\Vphys}{t} \right\vert_{T_p} < 0$: \; $\displaystyle b > \vextnFactor{p} + \frac{3}{4\pi \vextnFactor{p}^2} \approx 1.701$.
\end{enumerate}
In determining the bound for \cc{2}, one can recognise that $T_i/T_* = 1/3$ minimises \eqref{eq:decreasingPhysicalVolume-simple-new}. The condition for percolation is $b > \vextnFactor{p}$. Thus, the condition for unit nucleation without percolation is $v_w < \vextnFactor{p} / N^{\frac13}(0)$ and $N(0) \geq 1$, just as in the vacuum-dominated limit ($\Tmax \ll \Teq$) of the model-independent analysis in \cref{sec:scenario1}.

We now compare these results for completion criteria for the simultaneous nucleation model to the empirical bounds on the bubble wall velocity for the toy model, M1 (see \cref{sec:M1}), and the real scalar singlet model, M2 (see \cref{sec:M2}). The latter empirical bounds are encoded in the correction factors $\ccFactor{i}$ listed below \eqref{eq:completionBounds}. Since $\ccFactor{1} = 1$ by definition (see \eqref{eq:completionBounds}), we can view the remaining correction factors $\ccFactor{2-4}$ as ratios to $\ccFactor{1}$. For a fixed value of $N(0)$, a bubble wall velocity $\ccFactor{i}$ times larger is required to satisfy \cc{i} compared to \cc{1}. We can similarly determine correction factors for the simultaneous nucleation model, and interpret these as ratios of $v_w$ with fixed $N(0)$ because $b = v_w N^{\frac13}(0)$. The correction factors are $\ccFactor{1} = 1$, $\ccFactor{2} \approx 1.136$, $\ccFactor{3} \approx 1.185$, $\ccFactor{4} \approx 1.217$ and $\ccFactor{5} \approx 1.648$. These correction factors agree closely with those found empirically in M1 and M2 (see \eqref{eq:completionBounds}). The largest discrepancy is with the empirical $\ccFactor{5}$ value, which would be 1.70 if we instead took the lowest value in the observed range of 70-123\% above $\ccFactor{1}$. Additionally, the ordering of the correction factors is the same in all models. The agreement supports the utility of the empirical bounds on the bubble wall velocity for the various completion criteria considered in this study.

However, we note that the $\ccFactor{i}$ values are model dependent, and our numerical exploration of the benchmarks does not preclude more variation or different orderings amongst the $\ccFactor{i}$ values. Furthermore, the simultaneous nucleation model considered here may not accurately capture the evolution of the fraction and physical volume of the false vacuum in other models with strong supercooling. Thus a more general commentary is necessary. By definition, $\ccFactor{2} < \ccFactor{3\text{--}5}$, because \cc{2} requires only that $\Vphys$ is decreasing at some time, whereas \cc{3--5} require $\Vphys$ to decrease at specific times. Consequently, \cc{2} imposes a weaker constraint on the bubble wall velocity because it is a weaker condition. By definition, $\ccFactor{1} < \ccFactor{3,4}$, because \cc{1} is the condition that there is a completion temperature, and \cc{3,4} rely on the completion temperature. However, we are not aware of any definitive arguments for why the other hierarchies must hold in general.

We now briefly consider fast transitions: the extreme case of transitions opposite to that of strong supercooling. In fast transitions, $v_w$ is many orders of magnitude above what is required for completion because $N(0)$ is many orders of magnitude above unity. This guarantees the transition completes sufficiently fast such that the expansion of space can be neglected. Then $\Vphys$ decreases throughout the entire duration of the transition, including well before percolation (as discussed in Ref.\ \cite{Turner:1992tz}). Clearly this reasoning does not apply to the strongly supercooled transitions considered in this study, where the expansion of space plays an important role in transition progress. Our observed ordering of $\ccFactor{i}$ values in strongly supercooled transitions suggests that $\Vphys$ decreases somewhere between $T_p$ and $T_f$, or not at all. This is expected from \eqref{eq:Vphys} since $P_f$ changes most rapidly in this range of temperatures. In contrast, we expect $\Vphys$ to decrease for all temperatures during a fast phase transition. As the level of supercooling decreases, the temperature window for which $\Vphys$ decreases expands out from a range somewhere between $T_p$ and $T_f$. Then for some sufficiently small level of supercooling, $\Vphys$ is decreasing at both $T_p$ and $T_f$ such that all of the completion criteria are satisfied.

From our results, we conclude that an analysis of unit nucleation alone does not indicate the completion or non-completion of a transition. Many studies of cosmological first-order phase transitions assume the lack of unit nucleation guarantees the transition cannot complete, or that the occurrence of unit nucleation indicates a successful transition. We have demonstrated the contrary in \cref{sec:scenario1}. Additionally, in \cref{app:bubbleRadius} we show that the volume of a bubble can exceed the Hubble volume, especially during vacuum domination, intuitively allowing completion when $N(0) < 1$.

\section{Nucleation, percolation and completion: numerical treatment} \label{sec:models}

In this section we demonstrate realisations of the two main scenarios considered in \cref{sec:methodology} using our full numerical treatment in two different models. These scenarios are unit nucleation without percolation (Scenario 1), and percolation without unit nucleation (Scenario 2). We present five benchmarks for each model. The first model (M1) is a simple model obtained through a high-temperature expansion of the Standard Model with a zero-temperature potential barrier, which we treat as a toy model in this study. The second model (M2) is a minimal real scalar singlet extension of the Standard Model commonly used in studies of gravitational waves from a first-order electroweak phase transition \cite{Ashoorioon:2009nf, Huang:2016cjm, Hashino:2016xoj, Huang:2017jws, Alves:2018oct, Alves:2018jsw, Gould:2019qek, Bian:2019bsn, Alanne:2019bsm, Zhou:2019uzq, Alves:2020bpi, Liu:2021jyc, Guo:2021qcq, Ellis:2022lft}. Our perturbative treatment of M2 is known to suffer from gauge and renormalisation scale dependence, and IR divergences \cite{Chiang:2018gsn, Croon:2020cgk, Athron:2022jyi}. Thus, the benchmarks we present for M2 yield the desired scenarios subject to our choice of gauge parameter and renormalisation scale, as well as our treatment of daisy resummation and the Goldstone catastrophe. Additionally, it has recently been shown that two-loop effects may be large, particularly for strongly first-order transitions \cite{Niemi:2021qvp}.

\subsection{Toy model (M1)} \label{sec:M1}

First we consider the toy model (M1) with the effective potential%
\footnote{The last field-independent term does not affect the phase structure of the potential, yet is important for the analysis of phase transitions. It affects the pressure and its temperature derivative, thus it affects the energy density and consequently cosmic expansion (see \eqref{eq:energyDensityPressure} and \eqref{eq:HubbleParameter}). For gravitational wave studies, this can also affect the speed of sound in the plasma \cite{Giese:2020znk}.}
\begin{equation}
	V(\phi, T) = D (T^2 - T_0^2) \phi^2 - (ET + A) \phi^3 + \recip{4} \lambda \phi^4 - \frac{\pi^2}{90} g_* T^4 , \label{eq:M1}
\end{equation}
where $\phi$ is a scalar field. Neglecting the $A\phi^3$ term, \eqref{eq:M1} comes from a high-temperature expansion of the Standard Model \cite{Anderson:1991zb, Dine:1992wr, Megevand:2003tg}, while the $A\phi^3$ term can arise from multi-scalar field potentials (see e.g.\ Ref.\ \cite{Profumo:2007wc}). However, we do not attempt to choose realistic values of the parameters, instead treating \eqref{eq:M1} as a toy model. Following Ref.\ \cite{Megevand:2016lpr} we parameterise this potential by the ratio $\aonv = A/v$, where $v$ is the field value of the global minimum at zero temperature. As long as $T_0^2>0$, the first term in the potential is negative for $T < T_0$, leading to a maximum at the origin and (electroweak) symmetry breaking through the Higgs mechanism. At higher temperatures, this quadratic field coefficient is positive and symmetry can be restored. However, there are also temperatures where the quadratic field term is positive while the cubic field term is negative, resulting in a potential barrier separating two phases. This can lead to a first-order phase transition, which proceeds through bubble nucleation. The barrier dissolves at $T_0$, so a phase transition would no longer occur through bubble nucleation below $T_0$. On the other hand, if $T_0^2<0$ the cubic term allows for a potential barrier that persists at $T=0$, which will be the scenario we consider here.

Fixing $v = 250 \gev$ (as done in Ref.\ \cite{Megevand:2016lpr}), we can solve for $T_0$ by using the usual electroweak symmetry breaking condition for the potential at zero temperature. This is the condition that $v$ is a stationary point of the zero-temperature potential, and gives
\begin{equation}
	T_0^2 = \frac{\lambda - 3 \aonv}{2D} v^2 . \label{eq:T0Sq}
\end{equation}
The zero-temperature scalar mass $m_{\phi}$ is given by
\begin{equation}
	m_\phi^2 = \left. \pdv[2]{V}{\phi} \right\vert_{(v, 0)} = (2\lambda - 3\aonv) v^2, 
\end{equation}
where we used \eqref{eq:T0Sq}.
Following Ref.\ \cite{Megevand:2016lpr}, we choose to fix the scalar mass to be $m_{\phi} = v/2$, leading to the constraint
\begin{equation}
	\lambda = \recip{8} + \frac{3}{2} \aonv , \label{eq:lambda}
\end{equation}
and we set the number of effective degrees of freedom in the plasma to $g_* = 100$.

The energy density in this model is conveniently simple because the false vacuum remains fixed at the origin (i.e.\ $\phi_f(T) = 0$). This leaves $\rho(\phi_f, T) = \rhor(T)$. Hence, we find that $\rhov = -\rho(v, 0)$ is independent of temperature, by using \eqref{eq:energyDensityGeneral} and \eqref{eq:rhoV}. Then, using \eqref{eq:T0Sq} and \eqref{eq:lambda}, we have
\begin{equation}
	\rhov = \recip{8} \! \left(\recip{4} - \aonv \right) \! v^4 .
\end{equation}
This allows for the exact determination of $\Teq$ through \eqref{eq:Teq}, giving
\begin{equation}
	\Teq = \left[\frac{15}{4 \pi^2 g_*} \! \left(\recip{4} - \aonv \right) \right]^{\!\recip{4}} \! v. \label{eq:Teq-toy}
\end{equation}

We use M1 for probing cases where bubbles predominantly nucleate during the vacuum-domi\-nated era, and connecting to previous studies \cite{Megevand:2016lpr, Wang:2020jrd} of strong supercooling.\footnote{The Standard Model with a dimension-six operator, as used in e.g.\ Refs.\ \cite{Leitao:2015fmj, Cai:2017tmh, Ellis:2018mja, Wang:2020jrd}, would also serve the latter purpose.}
We present the benchmarks for this model in \cref{tab:benchmarks-M1}. These benchmarks demonstrate that the scenarios considered in \cref{sec:methodology} (unit nucleation without percolation, and percolation without unit nucleation) can be realised without the approximations used in the analytic treatment. Each benchmark is designed to lie at the threshold of unit nucleation being possible, which we ensure by tuning $\aonv$ such that $\Next(0) = 1$, for a fixed choice of the other input parameters. Perturbing the input parameters slightly then considers transitions both with and without unit nucleation. Whether percolation and completion occur then depends on the bubble wall velocity, $v_w$, and a bound on $v_w$ for these scenarios can be obtained numerically. For the purpose of matching to the bounds on $v_w$ obtained in \cref{sec:scenario1}, we treat $v_w$ as a free parameter and scan over it. As we discuss later and in \cref{app:reheating}, reheating occurs for sufficiently small bubble wall velocities corresponding to deflagrations and hybrids. Our model for the progress of the transition breaks down in such cases. However, reheating would only affect the $v_w$ thresholds for each scenario, rather than preventing the scenarios from being possible.

Note that we consider $\Next$ rather than $N$ here due to convenience for scanning. The approximate form, $\Next(0)$, does not depend on the bubble wall velocity if no reheating occurs during the transition. On the other hand, $N(0)$ depends on the false vacuum fraction (which we calculate by numerically evaluating \eqref{eq:falseVacuumFraction-temperatureSimplified}), and by extension depends on the bubble wall velocity. Additionally, we find that the difference between $N(0)$ and $\Next(0)$ is negligible in our benchmarks. The use of $\Next(0)$ allows the tuning of $\Next(0) = 1$ to be independent of the bubble wall velocity.

In fact, all quantities reported in \cref{tab:benchmarks-M1} are independent of the bubble wall velocity, except for the bounds on the bubble wall velocity itself, $v_w^{p,f}$. We do not report the percolation or completion temperatures in \cref{tab:benchmarks-M1} because the false vacuum fraction depends on the bubble wall velocity. We do report the temperature that maximises the nucleation rate, $\Tmax$, the presence of which indicates a minimum in the action. A non-zero value for $\Tmax$ is common to all benchmarks, and is necessary for studying transitions with $N(0) \sim 1$. We also report the temperature at which the radiation and vacuum energy densities are equal, $\Teq$. The numerically determined values of $\Teq$ precisely match the predicted values from \eqref{eq:Teq-toy}. The ratio $\Tmax/\Teq$ sets the horizontal position of the benchmarks in \cref{fig:noPercCases,fig:noCompletionCases}. Bubbles nucleate during radiation domination for $\Tmax \gg \Teq$, or during vacuum domination for $\Tmax \ll \Teq$. For $\Tmax \sim \Teq$, bubbles nucleate during an era where the radiation and vacuum energy densities are similar. The benchmark \bench{1}{5} does not have a numerical solution for $\Teq$ because the Universe is vacuum dominated when the true vacuum first appears (at $T \approx 29.88$ GeV). However we can still obtain $\Teq \approx 38.37$ GeV analytically using \eqref{eq:Teq-toy}.

\begin{table}
{\footnotesize
\setlength{\tabcolsep}{3.5pt}
\begin{center}
\begin{tabular}{|l||l|l|l|l||l|l|l|l|l|l|l|}
	\hline
	$\vphantom{1^{1^1}}$ & \multicolumn{1}{|c|}{$\aonv$} & \multicolumn{1}{|c|}{$v$} & \multicolumn{1}{|c|}{$D$} & \multicolumn{1}{|c||}{$E$} & \multicolumn{1}{|c|}{$\Delta N$} & \multicolumn{1}{|c|}{$T_c$} & \multicolumn{1}{|c|}{$\Tmax$} & \multicolumn{1}{|c|}{$\Teq$} & \multicolumn{1}{|c|}{$T_n$} & \multicolumn{1}{|c|}{$v_w^p$} & \multicolumn{1}{|c|}{$v_w^f$} \\\hline
	\bench{1}{1} $\vphantom{1^{1^1}}$ & $0.129064$ & $250$ & $0.442484$ & $0.0625$	& $1\!\times\!10^{-3}$ & $67.90$ & $32.30$ & $36.60$ & $27.09$ & $0.361$ & $0.859$ \\[0.1em]\hline
	\bench{1}{2} $\vphantom{1^{1^1}}$ & $0.110220$ & $250$ & $1.76993$  & $0.125$	& $1\!\times\!10^{-3}$ & $35.83$ & $15.08$ & $37.94$ & $12.55$ & $0.430$ & $1.02$ \\[0.1em]\hline
	\bench{1}{3} $\vphantom{1^{1^1}}$ & $0.144762$ & $250$ & $0.2$      & $0.05$	& $8\!\times\!10^{-4}$ & $102.8$ & $50.96$ & $35.35$ & $42.45$ & $0.234$ & $0.558$ \\[0.1em]\hline
	\bench{1}{4} $\vphantom{1^{1^1}}$ & $0.134623$ & $250$ & $0.3$      & $0.05$	& $1\!\times\!10^{-3}$ & $80.35$ & $39.00$ & $36.17$ & $32.69$ & $0.314$ & $0.747$ \\[0.1em]\hline
	\bench{1}{5} $\vphantom{1^{1^1}}$ & $0.104005$ & $250$ & $3.5$      & $0.2$		& $1\!\times\!10^{-3}$ & $26.92$ & $10.28$ & $38.37$ & $8.507$ & $0.433$ & $1.03$ \\[0.1em]\hline
\end{tabular}
\end{center}}
\vspace{-0.5cm}
\caption{Benchmark points for the toy model, M1. The first four columns are the input parameters that define the potential (see \cref{sec:M1}). The vacuum expectation value, $v$, and the temperatures have units of GeV. We have defined $\Delta N = \Next(0) - 1$ to indicate the deviation from the threshold of unit nucleation. The velocities $v_w^{p,f}$ denote the bubble wall velocity required for percolation and completion, respectively, when $\Next(0) = 1$ exactly. These velocities were numerically determined by varying $v_w$ until percolation and completion ceased. The combination $(\Tmax/\Teq, v_w^{p,f})$ defines the coordinate for each benchmark in \cref{fig:noPercCases,fig:noCompletionCases}.}
\label{tab:benchmarks-M1}
\end{table}

The benchmarks were selected to span a wide range of $\Tmax/\Teq$. They were generated by fixing values for $D$, $E$ and $v$, then scanning over $\aonv$ to find where $\Next(0) = 1$. No benchmarks for M1 were rejected from this study. The agreement between our model-independent predictions for the threshold bubble wall velocity values for Scenarios 1 and 2 and the numerically determined thresholds, $v_w^{p,f}$, indicate the validity of our assumptions and approximations used in \cref{sec:methodology}. The software pipeline we used for analysing each benchmark consists of \PT\ (version 1.1.0) \cite{Athron:2020sbe} to track the evolution of each phase and the in-development code, \TS\, to analyse all relevant transitions and determine the phase history. We made use of a modified version of \CT\ (version 2.0.6) \cite{Wainwright:2011kj} to evaluate the action. See \cref{app:action} for details of the modifications.

Strong supercooling typically requires that the potential barrier separating the phases must persist at low temperatures. A phase transition no longer occurs through bubble nucleation when the potential barrier vanishes. Thus, we focus on cases where there is a potential barrier at zero temperature, which corresponds to the condition $T_0^2 < 0$ in M1. This condition can be expressed as a constraint on $\aonv$, giving the lower bound $\aonv > \recip{12}$. Additionally, we can obtain the upper bound $\aonv < \recip{4}$ by requiring $\Teq > 0$ (or equivalently $\rhov > 0$). Each of the benchmarks listed in \cref{tab:benchmarks-M1} satisfy $\recip{12} < \aonv < \recip{4}$.

Benchmarks \bench{1}{2} and \bench{1}{5} have bubbles nucleating in the strongly vacuum-dominated era. Consider \bench{1}{2}, which has the higher value for $\Tmax/\Teq$ of the two benchmarks. We found the maximal nucleation rate occurs at $\Tmax \approx 0.40 \Teq$. Because the radiation energy density scales as $T^4$, we have $\rhor(\Tmax) \approx 0.026 \rhor(\Teq)$. By definition, $\rhor(\Teq) = \rhov(\Teq)$. Meanwhile, $\rhov$ should vary much slower between $\Tmax$ and $\Teq$; in fact, in this model $\rhov$ is constant. Thus, vacuum domination at $\Tmax$ is a good assumption for this benchmark, with $\rhov$ comprising 98\% of the total energy density at $\Tmax$. Vacuum domination is even stronger at $\Tmax$ in \bench{1}{5}.

In \cref{tab:benchmarks-M1} we state bubble wall velocities which we numerically found to define the boundary of Scenarios 1 and 2 at the threshold of unit nucleation. For instance, $v_w^p$ corresponds to the maximal bubble wall velocity for which percolation does not occur, when $N(0) = 1$. The threshold for completion is similarly marked by $v_w^f$. Consider \bench{1}{2} again. This benchmark --- like all others in this study --- lies at the threshold of unit nucleation. We numerically find that percolation does not occur for $v_w \lesssim 0.430$, while a larger wall velocity would yield percolation (if reheating can be ignored). Meanwhile, completion does not occur if $v_w \lesssim 1.03$, indicating that completion is not possible at or below the threshold of unit nucleation. We have argued that vacuum domination at $\Tmax$ is a good approximation for \bench{1}{2}. The numerically determined $v_w^{p,f}$ agree closely with the prediction \eqref{eq:vwBound-scenario1} in \cref{sec:scenario1} for the vacuum-dominated limit $\Tmax/\Teq \ll 1$, which corresponds to $v_w^p \approx 0.434$ and $v_w^f \approx 1.03$. For an analytic computation of $v_w^f$ we replace $\vextnFactor{p}$ with $\vextnFactor{f}$ in \eqref{eq:vwBound-scenario1}. The predicted bubble wall velocity bounds are shown in \cref{fig:noPercCases,fig:noCompletionCases}.

Benchmarks \bench{1}{1} and \bench{1}{4} both have bubbles predominantly nucleating when the radiation and vacuum energy densities are similar ($\Tmax \sim \Teq$). In this situation, we match $v_w^{p,f}$ to our analytic predictions for bubbles nucleating in an era where neither energy density contribution dominates, and the bubbles continue to grow during vacuum domination. Meanwhile, \bench{1}{3} has bubbles predominantly nucleating during radiation domination and continuing to grow during vacuum domination. In \cref{fig:noPercCases,fig:noCompletionCases} we see that the analytic predictions for the bubble wall velocity at the threshold of percolation and completion agree very closely with the numerical results, even away from the vacuum-dominated limit. The agreement supports the validity of the assumptions made in \cref{sec:comparing-events} and the simplification made in \cref{sec:reduceProblem} (discussed below \eqref{eq:PercCondGeneral}), at least for this model.

We numerically confirm that bubbles grow as detonations in each of our benchmarks,%
\footnote{We follow the methods described in Refs.\ \cite{Ellis:2019oqb, Ellis:2020nnr} to estimate the terminal bubble wall velocity. We consistently find that $v_w$ is larger than the Chapman-Jouguet velocity, which marks the lower bound on $v_w$ for detonations. This holds for all benchmarks (in both M1 and M2) using friction models with $\gamma^1$ and $\gamma^2$ scaling, where $\gamma = (1-v_w^2)^{-1/2}$ is the Lorentz factor of the bubble wall in the plasma frame.}
which suggests that reheating can be ignored. In \cref{sec:scenario1} we predicted that percolation is guaranteed if unit nucleation occurs, provided bubbles grow as detonations. Thus, none of our benchmarks demonstrate unit nucleation without percolation (Scenario 1) given the assumed particle content and current friction models. Other extensions to the Standard Model may allow for Scenario 1 to be realised more readily, with non-detonations occurring even with strong supercooling. However, we can realise percolation without unit nucleation (Scenario 2) by reducing $N(0)$ below unity. This is easily achieved by slightly increasing $\aonv$ for any benchmark listed in \cref{tab:benchmarks-M1}, because $N(0)$ decreases as $\aonv$ increases. Increasing $\aonv$ strengthens the potential barrier, and hence suppresses bubble nucleation further. Completion without unit nucleation is not possible for \bench{1}{2} and \bench{1}{5} because $v_w^f > 1$, but is possible for the remaining benchmarks, where the estimated bubble wall velocity is always larger than $v_w^f$. Our numerical results for benchmarks \bench{1}{2} and \bench{1}{5} agree with our analytic predictions that completion without unit nucleation is not possible if bubbles are predominantly nucleated during vacuum domination --- specifically if $\Tmax \lesssim \Teq/2$. Additionally, benchmarks \bench{1}{1,3,4} support our analytic prediction that completion is guaranteed even at the threshold of unit nucleation, provided the bubbles grow as detonations and nucleate before vacuum domination.

\subsection{Real scalar singlet model (M2)} \label{sec:M2}

\noindent We now consider a realistic and very popular extension of the Standard Model of particle physics: the (non-$\mathbb{Z}_2$-symmetric) real scalar singlet model (M2), which extends the Standard Model with an additional real scalar singlet \cite{Ham:2004cf, OConnell:2006rsp, Ahriche:2007jp, Profumo:2007wc, Barger:2007im, Espinosa:2011ax, No:2013wsa, Fuyuto:2014yia, Profumo:2014opa, Chen:2014ask, Sannino:2015wka, Kozaczuk:2015owa, Ghosh:2015apa, Kotwal:2016tex, Kanemura:2016lkz, Huang:2017jws, Lewis:2017dme, Chen:2017qcz, Li:2019tfd, Niemi:2021qvp, Ghorbani:2021rgs, Huang:2022him}. We will demonstrate in this model that the scenarios we have considered through analytic approximations and in the toy model may also exist in realistic models for new physics.

The tree-level potential in this model is \cite{Profumo:2007wc}
\begin{equation}
	\VTL(H, S) = \mu_h^2 (H^\dagger \! H) + \lambda_h (H^\dagger \! H)^2 + \half \mu_s^2 S^2 + \recip{4} \lambda_s S^4 + \half \lambda_{hs} (H^\dagger \! H) S^2 + \half \kappa_{hhs} (H^\dagger \! H) S + \recip{3} \kappa_{sss} S^3 ,
\end{equation}
where 
\begin{equation}
	H = \begin{pmatrix} G^+ \\ \recip{\sqrt{2}} (h + i G^0) \end{pmatrix} 
\end{equation}
is the Higgs doublet and $S$ is a real scalar singlet. Choosing the vacuum expectation value (VEV) of $H$ to lie along the real axis, and being interested in the vacuum structure, we can set $h \to \phi_h$, $G^{\pm,0} \to 0$, and $S \rightarrow \phi_s$, giving
\begin{equation}
	\VTLd = \half \mu_h^2 \phi_h^2 + \recip{4} \lambda_h \phi_h^4 + \half \mu_s^2 \phi_s^2 + \recip{4} \lambda_s \phi_s^4 + \recip{4} \lambda_{hs} \phi_h^2 \phi_s^2 + \recip{4} \kappa_{hhs} \phi_h^2 \phi_s + \recip{3} \kappa_{sss} \phi_s^3 ,
\end{equation}
where $\field = (\phi_h, \phi_s)$. At zero temperature, the singlet has a non-zero VEV $v_s \neq 0$, and the Higgs field has a VEV at $v_h \sim 246 \gev$.

The one-loop perturbative corrections to the tree-level potential at zero temperature are given by the Coleman-Weinberg potential \cite{Coleman:1973jx}. In the \msbar\ scheme and Landau gauge, the corrections are \cite{Quiros:1999jp}
\begin{equation}
	V_{CW}(\field,T) = \recip{64\pi^2} \!\! \sum_{i \in \{b\},\{f\}} \!\!\!\! (-1)^{2s_i} n_i m_i^4(\field,T) \left(\ln(\frac{m_i^2(\field,T)}{\mu_R^2}) - c_i \right) .
\end{equation}
Here $i$ runs over the mass spectrum with $\{b\}$ and $\{f\}$ being the set of bosons and fermions; $m_i$, $n_i$, and $s_i$ are the mass, number of degrees of freedom and spin, respectively, of each particle; $\mu_R = m_Z$ is the renormalisation scale; and $c_i = 5/6$ for gauge bosons and $c_i = 3/2$ for fermions and scalars. The thermal corrections to the masses are included in the Coleman-Weinberg potential according to the Parwani method of thermal resummation \cite{Parwani:1991gq}. See \cref{app:M2} for details of the mass spectrum.

The finite-temperature corrections to the potential include terms of the form \cite{Dolan:1973qd}
\begin{equation}
	V_T(\field,T) = \frac{T^4}{2 \pi^2} \! \Bigg[\sum_{i \in \{b\}} \! n_i J_b\!\left(\frac{m_i^2(\field,T)}{T^2} \right) + \sum_{i \in \{f\}} \! n_i J_f\!\left(\frac{m_i^2(\field,T)}{T^2} \right) \!\! \Bigg] , \label{VT}
\end{equation}
where
\begin{equation}
	J_{b,f}(z^2) = \pm \int_0^\infty \! dx \, x^2 \ln(1 \mp \exp(-\sqrt{x^2 + z^2})) .
\end{equation}
The full effective potential is the sum of these contributions, as well as thermal corrections from light particles we ignore in the mass spectrum;
\begin{equation}
	\Vd = \VTLd + \VCWd + \VFTd - \frac{\pi^2}{90} g_*' T^4 . \label{eq:ssm-eff_pot}
\end{equation}
Low mass particles are omitted from the one-loop corrections because they have a negligible effect on the potential and corresponding phase structure. However, they still contribute significant $T^4$ terms in $\VFT$, which have an important effect on the energy density \eqref{eq:energyDensityReduced}. We account for these $T^4$ terms by including the last term of \eqref{eq:ssm-eff_pot}. The real scalar singlet model has $g_* = 107.75$ effective degrees of freedom. However, we already explicitly include $27.5$ of them in the effective potential. The other $80.25$ effective degrees of freedom come from particles that are omitted from the one-loop corrections because they have small masses and contribute negligibly to all field-dependent terms. Thus, we set $g_*' = 80.25$ so that no $T^4$ terms are omitted from the potential.

Our benchmarks for this model are listed in \cref{tab:benchmarks-M2}. The generation process for these benchmarks was similar to that used for M1. However, due to the dimensionality of the parameter space and complicated phase structure of M2, we performed a parameter space scan to identify benchmark candidates that yield a first-order phase transition. From that subset of points in the parameter space scan, we selected five points with sufficiently distinct parameter values. For each of these five points, we selected a parameter to scan over to find where $\Next(0) = 1$. Benchmarks for M2 were rejected from this study only if the potential was non-perturbative, had no first-order phase transition, or a nearby point satisfying $\Next(0) = 1$ could not be found.

\begin{table}
{\footnotesize
\setlength{\tabcolsep}{2.5pt}
\begin{center}
\begin{tabular}{|c||c|c|c|c|c||r|c|c|c|c|c|c|}
	\hline
	$\vphantom{1^{1^1}}$ & $\kappa_{hhs}$ & $\kappa_{sss}$ & $\theta$ & $v_s$ & $m_s$ & \multicolumn{1}{|c|}{$\Delta N$} & $T_c$ & $\Tmax$ & $\Teq$ & $T_n$ & $v_w^p$ & $v_w^f$ \\\hline
	\bench{2}{1}\!\! $\vphantom{1^{1^1}}$ & $\!-1450.76$ & $-342.609$ & $0.298703$ & $699.930$ & $425.628$ & $-2\!\times\!10^{-3}$	& $99.84$	& $53.33$	& $34.85$ & --		& $0.223$ & $0.531$ \\[0.1em]\hline
	\bench{2}{2}\!\! $\vphantom{1^{1^1}}$ & $\!-1259.83$ & $-272.907$ & $0.261558$ & $663.745$ & $351.183$ & $7\!\times\!10^{-3}$		& $106.1$	& $59.95$	& $35.98$ & $52.63$	& $0.200$ & $0.478$ \\[0.1em]\hline
	\bench{2}{3}\!\! $\vphantom{1^{1^1}}$ & $\!-2002.02$ & $-1971.27$ & $0.273396$ & $700.189$ & $896.659$ & $1\!\times\!10^{-3}$		& $87.53$	& $40.73$	& $33.33$ & $33.80$	& $0.291$ & $0.694$ \\[0.1em]\hline
	\bench{2}{4}\!\! $\vphantom{1^{1^1}}$ & $\!-1389.99$ & $-1122.98$ & $0.289510$ & $521.645$ & $723.960$ & $3\!\times\!10^{-4}$		& $91.52$	& $45.08$	& $33.87$ & $36.47$	& $0.264$ & $0.630$ \\[0.1em]\hline
	\bench{2}{5}\!\! $\vphantom{1^{1^1}}$ & $\!-1263.96$ & $-656.768$ & $0.281675$ & $440.431$ & $645.128$ & $-6\!\times\!10^{-4}$	& $92.26$	& $46.47$	& $33.95$ & --		& $0.255$ & $0.608$ \\[0.1em]\hline
\end{tabular}
\end{center}}
\vspace{-0.5cm}
\caption{Benchmark points for the real scalar singlet model, M2. The first five columns are the input parameters that define the potential (see \cref{app:M2}). The remaining columns and strategy for benchmark selection are as in \cref{tab:benchmarks-M1}, and all dimensionful quantities have units of GeV. The benchmarks with no reported $T_n$ have $\Next(0)$ just below unity. We ensure that all benchmarks have $\lambda_{h} < \frac{2}{3} \pi$, $\lambda_s < \frac{2}{3} \pi$ and $\lambda_{hs} < 4\pi$, thus satisfying perturbativity constraints \cite{Lerner:2009xg}. We do not check whether these benchmarks violate constraints from colliders or other experiments. The percolation and completion threshold velocities $v_w^{p,f}$ are defined in \cref{tab:benchmarks-M1}.}
\label{tab:benchmarks-M2}
\end{table}

Each benchmark in this model has dominant bubble nucleation occurring during radiation domination. The numerically determined bubble wall velocity bounds for which percolation and completion occur agree closely with our predictions in \cref{sec:methodology}, as seen in \cref{fig:noPercCases,fig:noCompletionCases}. While none of the benchmarks in this model have bubbles nucleating after radiation domination ($\Tmax \lesssim \Teq$), it may be possible for significant bubble nucleation to be delayed until the vacuum-dominated era in some regions of the parameter space. Our benchmarks are in no way representative of the entire viable parameter space of the real scalar singlet model, nor the subset of the parameter space that permits strong supercooling.

Mixing between the scalar singlet and the Higgs boson modifies the electroweak symmetry breaking, making it possible to have a first-order phase transition rather than a crossover like in the Standard Model. Therefore, each benchmark has a significant mixing angle. We do not apply constraints from Higgs couplings%
\footnote{Our benchmarks satisfy the 95\% confidence limit set by collider experiments \cite{ATLAS:2016neq, Carena:2018vpt}, though these values may now be in tension with the latest data \cite{CMS:2020gsy-new, ATLAS:2022vkf}. A high mixing angle is not required to achieve supercooling, so it should be possible to find similar scenarios with a lower mixing angle.}
or other phenomenological constraints, such as those from electroweak fits. Our main aim is to demonstrate that Scenarios 1 and 2 can occur in popular and viable models, not to perform a detailed phenomenological analysis of the model.

We require a minimum in the action to suppress bubble nucleation for these scenarios to occur. Such a minimum implies the existence of a potential barrier between the phases at zero temperature, as was argued in \cref{sec:M1}. A zero-temperature potential barrier can be generated by a large and negative cubic term in the potential, so it is not surprising that all benchmarks have a large negative $\kappa_{hhs}$, and most have a large negative $\kappa_{sss}$ as well.

The phase history for each benchmark apart from \bench{2}{2} consists of a single first-order phase transition from the high temperature phase $\field_1$ with $\phi_h = 0, \phi_s \neq 0$, to the electroweak symmetry-broken phase $\field_2$ with $\phi_h\neq0, \phi_s \neq 0$. The phase structure for these benchmarks consists only of these two phases. \bench{2}{2} had an additional phase $\field_3$ along the $\phi_h = 0$ axis, and the same transition $\field_1 \rightarrow \field_2$. However, this time the transition competed with another transition: $\field_1 \rightarrow \field_3$. The potential barrier separating the phases $\field_1$ and $\field_3$ prevented the corresponding transition from occurring. Both transitions had similar critical temperatures: $T_c^{1\rightarrow2} = 106.1$ GeV and $T_c^{1\rightarrow3} = 101.8$ GeV. If the transition $\field_1 \rightarrow \field_3$ had occurred, the subsequent transition $\field_3 \rightarrow \field_2$ may have been possible because $\field_3$ had higher free energy density than $\field_2$, leading to a multi-step transition or potentially simultaneous transitions. Transitions occurring simultaneously \cite{Croon:2018new, Morais:2018uou, Morais:2019fnm} or in multiple steps \cite{Angelescu:2018dkk, Zhao:2022cnn} have already been considered in this model and similar models. A careful determination of the transition progress is necessary in such studies. The task of correctly identifying the phase history in models with richer phase structures is currently an open problem with interesting ramifications.

The real scalar singlet model is often used in studies of gravitational waves from first-order phase transitions \cite{Ashoorioon:2009nf, Huang:2016cjm, Hashino:2016xoj, Huang:2017jws, Alves:2018oct, Alves:2018jsw, Gould:2019qek, Bian:2019bsn, Alanne:2019bsm, Zhou:2019uzq, Alves:2020bpi, Liu:2021jyc, Guo:2021qcq}. The detectability of gravitational waves increases with their amplitude, which in turn increases with supercooling. Strong supercooling can increase the energy available for gravitational wave production, although it may also shift the peak frequency of the signal away from a given detector's sensitivity window. The appropriate choice of reference temperature where gravitational waves are produced is important for accurate predictions. As we will show in the next section, the nucleation temperature is not an appropriate choice in strongly supercooled transitions. While performing parameter tuning scans for strong supercooling in this model, we observed that the percolation and nucleation temperatures can differ by more than 10 GeV for $v_w = 1$, before unit nucleation ceases altogether. The difference would increase for smaller bubble wall velocities, and could become 50 GeV or more. We find (for $v_w = 1$) that the characteristic length scale of the transition --- as measured by the average bubble radius, mean bubble separation or volume-weighted bubble radius --- can change by an order of magnitude between the nucleation and completion temperatures. A larger characteristic length scale could enhance the gravitational wave signal \cite{Ajmi:2022nmq}. Additionally, the fluid kinetic efficiency coefficient (determined using the pseudo-trace \cite{Giese:2020rtr}) was found to double from unit nucleation to percolation. We reserve further investigation of the implications of strong supercooling on gravitational wave predictions for future work.

\section{Applicability of the nucleation temperature} \label{sec:nucleationApplicability}

\begin{figure}[h!]
	\centering
	\begin{subfigure}{.508\textwidth}
		\centering
		\includegraphics[width=1\linewidth]{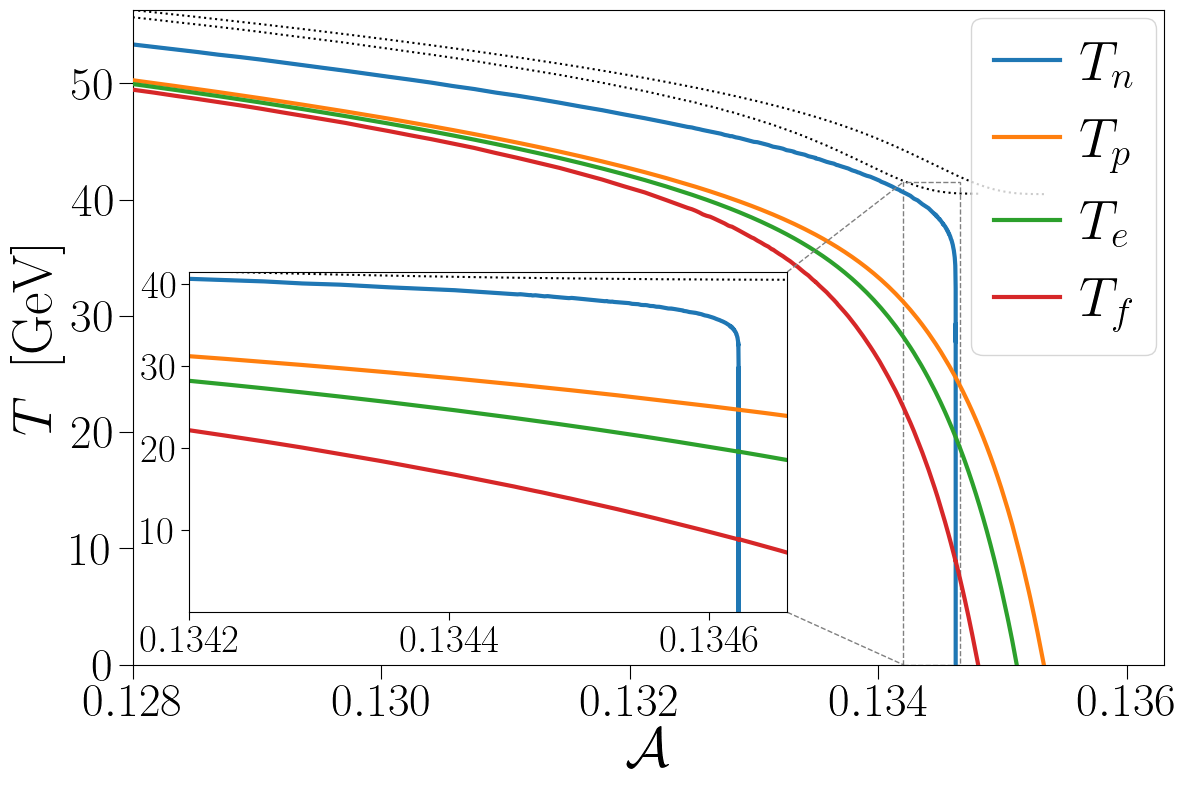}
		\caption{}
		\label{fig:t-vs-aonv-M1-BP4}
	\end{subfigure}%
	\begin{subfigure}{.492\textwidth}
		\centering
		\includegraphics[width=1\linewidth]{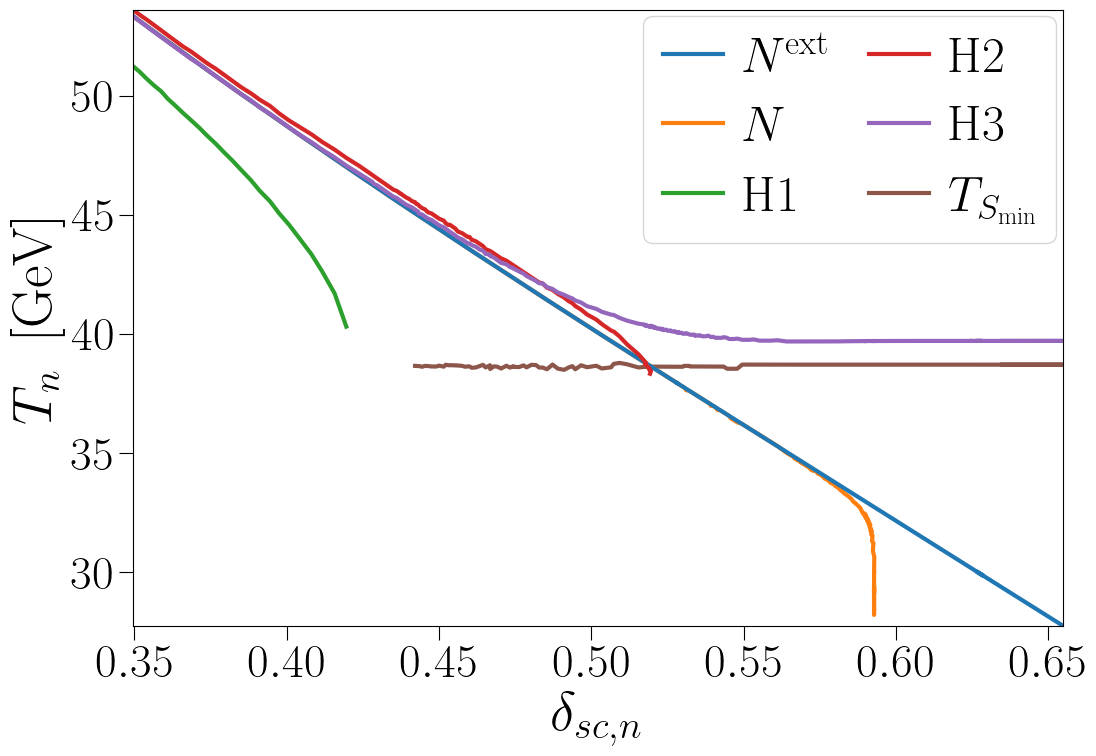}
		\caption{}
		\label{fig:tn-vs-dsc-M1-BP4}
	\end{subfigure}
	\begin{subfigure}{.505\textwidth}
		\centering
		\includegraphics[width=1\linewidth]{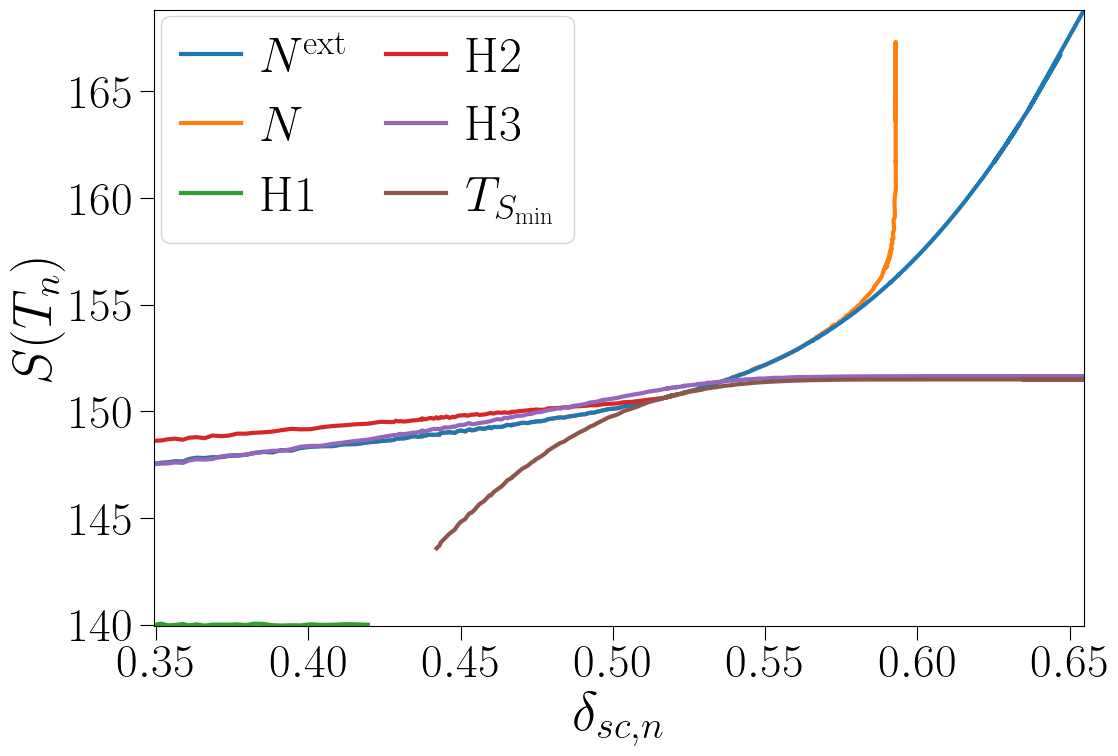}
		\caption{}
		\label{fig:stn-vs-dsc-M1-BP4}
	\end{subfigure}%
	\begin{subfigure}{.495\textwidth}
		\centering
		\includegraphics[width=1\linewidth]{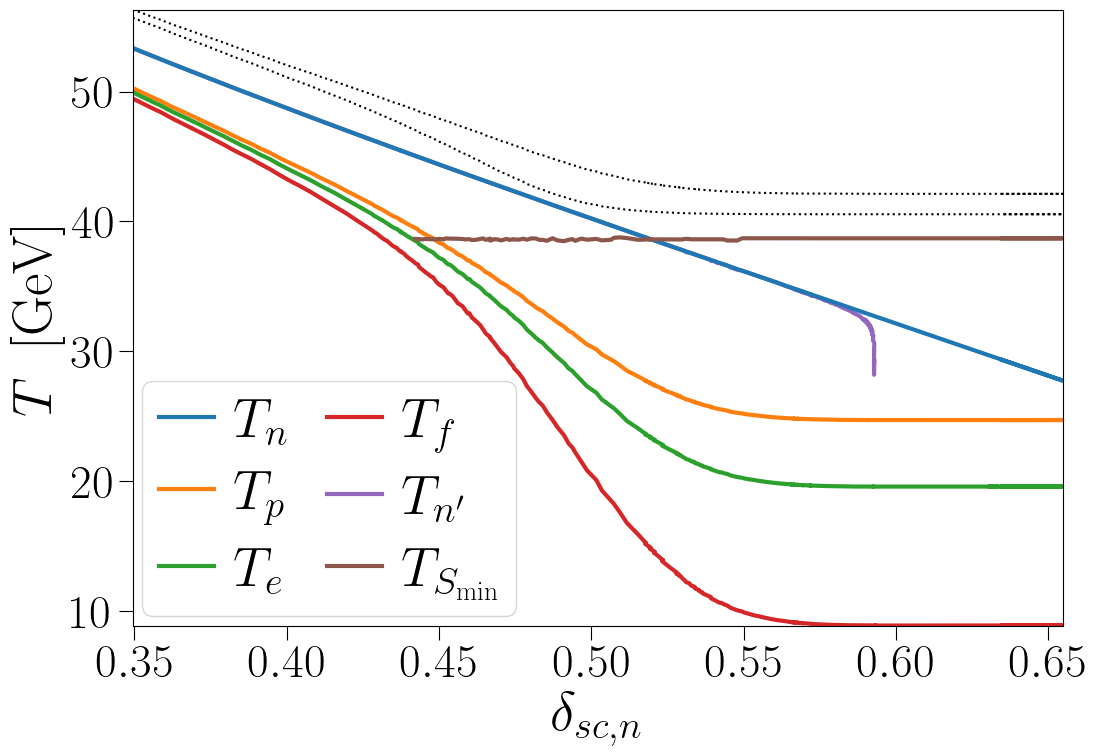}
		\caption{}
		\label{fig:t-vs-dsc-M1-BP4}
	\end{subfigure}
	\vspace{-0.2cm}
	\caption{Results of a scan over $\aonv = A/v$ with $v_w = 1$, taking all other inputs of the toy model as in \bench{1}{4} (see \cref{tab:benchmarks-M1}). This scan yields a range of levels of supercooling, parameterised by $\scn$ (see \eqref{eq:supercooling-nucleation}). We have not plotted results for smaller values of $\aonv$, where supercooling is weaker and the milestone temperatures have less separation and less dependence on $\aonv$. (a) Key transition temperatures as a function of $\aonv$, where $T_e$ is defined as $P_f(T_e) = 1/e$ and $T_n$ is calculated using \eqref{eq:nucleationTemperature}. These temperatures drop sharply near a critical value of $\aonv$. (b) The nucleation temperature obtained both numerically and through nucleation heuristics, as a function of supercooling. Here, $N$ and $\Next$ respectively refer to the nucleation temperatures numerically extracted from \eqref{eq:nucleationTemperatureProper} and \eqref{eq:nucleationTemperature}. (c) The action at the nucleation temperature (obtained with different approximations) as a function of supercooling. (d) Key transition temperatures as a function of supercooling, where $T_{n'}$ is extracted from \eqref{eq:nucleationTemperatureProper} (i.e.\ phantom bubbles are not counted). The temperatures related to the false vacuum fraction are seen to plateau for $\scn \gtrsim 0.55$. The supercooling parameter is calculated using $T_n$ rather than $T_{n'}$ to avoid dependence on the bubble wall velocity. The higher and lower dotted black curves in (a) and (d) correspond to the reheating temperature, $\Treh$, evaluated at $T_p$ and $T_f$, respectively. The qualitative features of these plots hold for all benchmarks considered in this study.}
	\label{fig:supercoolingScan}
\end{figure}

In this section we investigate unit nucleation in cases of strong supercooling, particularly when the transition completes after the peak of bubble nucleation is reached ($T_f < \Tmax$). First, in \cref{sec:nucleationHeuristic}, we assess the accuracy of various approximations to the condition $N(T_n) = 1$ and the necessity of nucleation heuristics. Then, we discuss the importance of unit nucleation in transition analyses in \cref{sec:relevanceUnitNucleation}. Prior to those discussions, we briefly describe the data presented in \cref{fig:t-vs-aonv-M1-BP4}, which is used to generate the results in the remainder of \cref{fig:supercoolingScan}. While \bench{1}{4} was used to generate the results displayed in \cref{fig:supercoolingScan}, all other benchmarks give the same qualitative features.

In \cref{fig:t-vs-aonv-M1-BP4}, various milestone temperatures are shown to rapidly vanish as $\aonv = A/v$ approaches some critical value, with all other potential inputs fixed to their values in \bench{1}{4}. Well above this critical value of $\aonv$, the temperatures are seen to vary more slowly. This behaviour is well known as shown in e.g.\ Refs.\ \cite{Cai:2017tmh, Ellis:2018mja, Megevand:2016lpr}, and can be explained as follows. The false vacuum fraction and number of bubbles nucleated depend exponentially on the action. The action itself varies approximately exponentially with temperature away from its minimum (if one exists). This leads to the milestone temperatures (and transition dynamics more generally) depending logarithmically on the action. Hence the dependence of the milestone temperatures on the Lagrangian parameters affecting the action is washed out. However, if the transition dynamics depend on the action at or near its minimum, a small change in the action curve results in an exponential change in the milestone temperatures and transition dynamics; notably the nucleation of bubbles. For large enough $\aonv$, the transition does not complete before reaching the temperature that minimises the action. Increasing $\aonv$ further, the exponential sensitivity suppresses the transition to the point where none of the milestone temperatures are reached before zero temperature. Although the milestone temperatures rapidly decrease as $\aonv$ approaches various critical values, the same is not true for the reheating temperature, $\Treh$, as seen in \cref{fig:t-vs-aonv-M1-BP4,fig:t-vs-dsc-M1-BP4}. The reheating temperature is comparatively insensitive to $\aonv$ because $\rho_H$ is approximately constant for low temperatures, leading to only a small variation in the solution of \eqref{eq:reheat}.

A numerical determination of transition dynamics (including the milestone temperatures defined in \cref{sec:transitionTemperatures}) is consequently sensitive to numerical errors in the determination of the action for strongly supercooled transitions. Indeed, the fluctuations in results presented in \cref{fig:supercoolingScan} are due to numerical uncertainties from \CT\ \cite{Wainwright:2011kj}; the bounce solver used in this study. We discuss these numerical errors and our attempts to minimise them in \cref{app:action}.

\subsection{Accuracy of nucleation heuristics} \label{sec:nucleationHeuristic}

Studies of gravitational waves from first-order phase transitions typically use gravitational wave fits from hydrodynamic simulations or models (for reviews, see Refs.\ \cite{Caprini:2015zlo, Weir:2017wfa, Caprini:2019egz}). These fits require a reference temperature at which to evaluate the gravitational wave signal. The nucleation temperature, $T_n$, is often used for simplicity. One requires only the action $S(T) = S_3(T)/T$ and Hubble parameter $H(T)$ as functions of temperature to determine $T_n$, as seen in \eqref{eq:nucleationRate} and \eqref{eq:numBubbles}. Tools \cite{Wainwright:2011kj, Masoumi:2016wot, Athron:2019nbd, Sato:2019wpo, Guada:2020xnz} exist to determine the action at a given temperature, and the Hubble parameter is typically expected to receive a dominant contribution from the radiation energy density, $\rhor$ \eqref{eq:rhoR}. Yet, it is common to simplify the problem further by employing the use of a nucleation heuristic; that is, an approximation for finding $T_n$ without evaluating the integral in \eqref{eq:numBubbles}.

The simplest and most commonly used heuristic for the nucleation temperature is
\begin{equation}
	\frac{S_3(T_n)}{T_n} \sim 140 , \label{eq:nucleation-140}
\end{equation}
which is used implicitly in studies that use the default implementation of \CT\ \cite{Wainwright:2011kj} for transition analysis. If one can sample the action as a function of temperature, which is made easy through the use of tools, a root-finding method can be employed to find $T_n$ from \eqref{eq:nucleation-140}. We also consider two other heuristics for the nucleation temperature, noting that many exist. The second heuristic (see Appendix B of Ref.\ \cite{No:2009thesis}) is
\begin{equation}
	\frac{S_3(T)}{T_n} \simeq 143 + 4 \log(\frac{\Delta \phi(T)}{T_c(1 - \scn)}) + \log(\frac{\scn}{1 - \scn}) , \label{eq:nucleation-NoRedondo}
\end{equation}
where $\Delta \phi(T) = \abs{\phif(T) - \phit(T)}$ is the distance between minima, and $\scn$ is defined in \eqref{eq:supercooling-nucleation}. Radiation domination up to the time of unit nucleation is assumed in the derivation of \eqref{eq:nucleation-NoRedondo}. The third heuristic (featured in the recent review, Ref.\ \cite{Caprini:2019egz}) is \cite{Huber:2007vva}
\begin{equation}
	\frac{S_3(T_n)}{T_n} \simeq 141.4 - 4 \log(\frac{T_n}{100 \gev}) - \log(\frac{\beta(T_n)}{100}) , \label{eq:nucleation-HuberKonstandin}
\end{equation}
where $\beta$ is the characteristic inverse duration of the transition, given by
\begin{equation}
	\beta(T) = \dv{t} \! \left(\frac{S_3}{T} \right) = T H(T) \dv{T} \! \left(\frac{S_3}{T} \right) .
\end{equation}

We will refer to \eqref{eq:nucleation-140}, \eqref{eq:nucleation-NoRedondo} and \eqref{eq:nucleation-HuberKonstandin} as nucleation heuristics H1, H2 and H3, respectively. All mentioned heuristics perform a low-order Taylor expansion of the action around the nucleation temperature. This expansion breaks down if the nucleation temperature lies near or below the minimum of the action, $\Tmin$. Heuristic H1 suggests unit nucleation cannot occur if $S(\Tmin) > 140$; yet we see in \cref{fig:tn-vs-dsc-M1-BP4,fig:stn-vs-dsc-M1-BP4} that unit nucleation can still occur in this case. We find that H2 predicts an action at unit nucleation smaller than $S(\Tmin)$ for $T_n \lesssim \Tmin$, for the benchmark considered in \cref{fig:supercoolingScan}. Heuristic H2 cannot predict unit nucleation in such cases (see \cref{fig:stn-vs-dsc-M1-BP4}). However, for about half of the benchmarks in both models, the form of the predicted nucleation temperature and action from H2 follow that of H3. The form of predictions from H3 was found to be consistent in both models. The inverse duration $\beta$ is identically zero at $\Tmin$ and becomes negative below $\Tmin$, so $\log(\beta)$ diverges at $\Tmin$. Consequently, H3 predicts an action at unit nucleation that is very sensitive to temperature near $\Tmin$. Additionally, H3 cannot predict $T_n < \Tmin$ because $\log(\beta)$ becomes complex-valued. In general, these heuristics suffer from missing solutions where $T_n < \Tmin$, and instead either finding an incorrect solution with a higher temperature above $\Tmin$ or finding no unit nucleation at all.

In \cref{fig:tn-vs-dsc-M1-BP4}, the nucleation heuristics H2 and H3 are seen to hold well for moderate supercooling (e.g.\ for $\scn \lesssim 0.4$). For many other benchmarks we found the agreement to only hold up to $\scn \lesssim 0.2$, where even H1 is a reasonable approximation. The roughest approximation, H1, breaks down much sooner than H2 and H3. These instead break down where $T_n$ approaches $\Tmin$, which corresponds to $\scn \gtrsim 0.5$ for \bench{1}{4}. This is also the point where the nucleation temperature decouples from the transition progress (i.e.\ the percolation and completion temperatures), as seen in \cref{fig:t-vs-dsc-M1-BP4}. We will return to this feature in the next subsection. In \cref{fig:stn-vs-dsc-M1-BP4}, we see that the action at the nucleation temperature can be arbitrarily high for strong supercooling. However, it is very difficult to precisely resolve the point at which $N(T) = 1$ for strong supercooling. This is because nucleation becomes strongly suppressed if the nucleation temperature is not reached before or near $\Tmin$. Then reaching the unit nucleation threshold for low temperatures becomes a highly fine-tuned problem. In fact, numerical errors in the bounce solver used (i.e.\ \CT) prevent precise determination of $N(T) = 1$, which limits our ability to probe $\scn \gtrsim 0.7$ in M1, and $\scn \gtrsim 0.5$ in M2. These numerical issues are discussed further in \cref{app:action}. Finally, we remark that, as expected, the nucleation heuristics yield a reasonable estimation of the nucleation temperature for weakly supercooled (or fast) transitions where radiation domination holds. However, for sufficiently fast transitions, any finite value of $S$ would yield a reasonable estimation of the nucleation temperature, because the action decreases extremely quickly below the critical temperature. We also see an $\mathcal{O}(10\%)$ difference between the nucleation temperatures extracted from \eqref{eq:nucleationTemperatureProper} and \eqref{eq:nucleationTemperature} for strong supercooling. Although our scans did not target this feature, it is evident that this difference would reach $\mathcal{O}(100\%)$ for a very small range of $\aonv$. In other models, the difference in nucleation temperatures from \eqref{eq:nucleationTemperatureProper} and \eqref{eq:nucleationTemperature} may become significant with less fine-tuning.

Now that we have evaluated the accuracy of these nucleation heuristics, we pause to comment on their necessity. Admittedly, H1 is invitingly convenient. One requires only a bounce solver and a root-finding algorithm, the latter of which is trivial given the shape of $S(T)$. Heuristics H2 and H3 require little extra effort to use. We note that one must determine the Hubble parameter to use H3. However, if one has the action and the Hubble parameter, then the integration in \eqref{eq:numBubbles} can be performed numerically. That is, one could avoid the use of a heuristic to find $T_n$ altogether. Using \eqref{eq:numBubbles} has the added benefit of being valid even with strong supercooling. The difference in method of computation is that the root-finding algorithm should be replaced with an integration scheme with an adaptive temperature sampler. The sampling should be such that the action is densely sampled around $T_n$ (or $\Tmin$ if $T_n < \Tmin$) with higher temperatures being less important.

\subsection{Relevance of unit nucleation} \label{sec:relevanceUnitNucleation}

It is often argued that the percolation temperature, and the corresponding reheating temperature, are more appropriate reference temperatures for gravitational wave predictions than the nucleation temperature (see e.g.\ Ref.\ \cite{Ellis:2018mja}). The nucleation temperature can be used provided it is close to the percolation temperature, which holds for fast transitions. Regardless, the nucleation temperature typically heralds the beginning of the transition, with significant nucleation occurring from then on. The usual expectation is that the transition cannot complete if unit nucleation does not occur. However, we have shown that it is possible for a transition to complete without the occurrence of unit nucleation. In \cref{fig:t-vs-aonv-M1-BP4}, a small range of the parameter $\aonv$ perturbed about \bench{1}{4} yields a phase transition with percolation but no unit nucleation. Our benchmarks listed in \cref{tab:benchmarks-M1,tab:benchmarks-M2} demonstrate that percolation and completion are still possible when $T_n < \Tmin \approx \Tmax$ (where $\Tmax$ maximises the nucleation rate), as shown in \cref{fig:supercoolingScan}. Our results support the finding of Ref.\ \cite{Wang:2020jrd} that the occurrence of a temperature window where $\Vphys$ is decreasing disappears roughly when $T_n < \Tmin$, although we find a small window of the parameter space that contradicts the heuristic that the success of a transition is questionable when $T_n \lesssim \Tmin$. We have also shown that the nucleation temperature (if it exists) may correspond to an arbitrary point during the phase transition. This is evident in \cref{fig:t-vs-dsc-M1-BP4}, where the nucleation temperature is seen to decouple from the transition progress (e.g.\ the percolation temperature) for strong supercooling ($\scn \gtrsim 0.5$). In a fast transition, the nucleation temperature corresponds to a time when there is no significant progress in the transition. In a strongly supercooled transition, the nucleation temperature may correspond to a time when an arbitrarily large fraction of the Universe has already been converted to the true vacuum --- or it may not be defined at all.

One may then wonder what the significance of the nucleation temperature is. We argue that unit nucleation does not necessarily correspond to a significant milestone in a cosmological phase transition. It is related to only the number of bubbles nucleated and not their volume, and so cannot be used to estimate the temperature at which (for example) bubble surface area or collisions are maximised for a general phase transition. Thus, the nucleation temperature will not provide accurate predictions for phenomenology from phase transitions such as gravitational waves.

However, we note that the nucleation temperature is typically more convenient to use. Its determination only requires knowledge of the action and the Hubble parameter. Often radiation domination can be assumed, and tools exist to calculate the action for a given model. This leaves little additional work to obtain a rough estimate of the gravitational wave signal. To use the percolation temperature, the false vacuum fraction and consequently the bubble growth must be tracked. Owing to the double exponentiation of the action in the false vacuum fraction (see \eqref{eq:nucleationRate} and \eqref{eq:falseVacuumFraction-temperatureSimplified}), the temperature integral in \eqref{eq:falseVacuumFraction-temperatureSimplified} must be discretised very finely. Let $n$ be the number of samples of the integrand used in the numerical integration. The evaluation of the nested integral in \eqref{eq:falseVacuumFraction} naively has a time complexity of $\mathcal{O}(n^2)$. If one was to track the false vacuum as the Universe cools (i.e.\ take $n$ samples of $P_f$), then the time complexity would become $\mathcal{O}(n^3)$. We typically use $\mathcal{O}(10^3-10^4)$ temperature samples in our results for high precision, so a brute-force evaluation of $P_f(T)$ from $T_c$ to $T_f$ would be expected to require at least $\mathcal{O}(10^{10})$ operations. This approach is then too slow for large-scale parameter space scans without sacrificing precision. One could approximate the bubble growth to avoid the nested integration, leaving $\mathcal{O}(n)$ time for a single evaluation of $P_f$. However, we have developed an optimisation (without approximation) in the form of a recurrence relation: $P_f(T_i) = P_f(T_{i-1}) + \Delta_i$, with $\Delta_i$ requiring $\mathcal{O}(1)$ time to evaluate. Thus, the false vacuum fraction can be tracked from $T_c$ to $T_f$ in $\mathcal{O}(n)$ time without approximation beyond that of the integration scheme and the approximations introduced in \eqref{eq:falseVacuumFraction-temperatureSimplified}.

We conclude this section by noting that while we argue that the nucleation temperature is not a relevant quantity, it is a convenient and often reasonable proxy for a more appropriate reference temperature for gravitational wave predictions. Presently, a thorough transition analysis and accurate prediction of gravitational waves requires significant effort for any given model. This limits the feasibility of this approach for many studies, particularly large-scale parameter space scans where additional running time is problematic. We aim to address this issue with the upcoming release of our code, \TS.

\section{Conclusions} \label{sec:theend}

We have demonstrated that unit nucleation is not an important event in a cosmological phase transition and that its use as a characteristic temperature for gravitational wave production is not appropriate for strongly supercooled transitions. For studies where tracking the false vacuum fraction is not feasible, we provide simple conditions to determine whether a transition percolates and completes, but not when. Specifically, the conditions are presented in \cref{sec:scenario1}. The main condition \eqref{eq:noPercCond-Y} is expressed as a bound on the bubble wall velocity in \eqref{eq:vwBound-scenario1} or as a bound on the comoving radius of bubbles in \eqref{eq:noPercCond-RH}. If a reference temperature is required for baryogenesis or gravitational wave production, a full analysis of the transition is essential for strongly supercooled phase transitions, particularly if bubble nucleation is suppressed at low temperature. We have attempted to treat the transition analysis as rigorously as our analytic methods allow. Although reheating from non-detonations could affect the accuracy of these conditions for percolation and completion (as discussed in \cref{app:reheating}), it does not invalidate the analysis.

We have investigated two scenarios: unit nucleation without percolation, and percolation without unit nucleation. We showed that both scenarios are possible by using the analytic treatment in \cref{sec:methodology} and a full numerical transition analysis in a toy model and the real scalar singlet model in \cref{sec:models}. Next-to-leading-order corrections in the nucleation rate would change the number of bubbles nucleated throughout the transition, and gauge and renormalisation scale dependence of M2 could also significantly affect the transition. Nonetheless, the benchmarks along with the analytic treatment in \cref{sec:methodology} provide an important proof of principle that the scenarios of unit nucleation without percolation, and percolation without unit nucleation, can occur in realistic models.

An interesting feature we did not capture in our analysis is how concrete particle physics models can realise deflagrations or hybrids during strongly supercooled transitions. Additionally, our analysis applies to transitions that remain possible at arbitrarily low temperatures. Even within M2 there are possible phase histories involving transitions between phases that do not persist at zero temperature. However, our analytic treatment in \cref{sec:methodology} can easily be adapted to transitions that are halted at some finite temperature, which may arise for example when there is no tree-level barrier in M2.

We also performed a detailed examination of the conditions for a phase transition to complete. The condition that the false vacuum fraction becomes negligible is necessary but not sufficient for a phase transition to successfully convert the entire Universe to the true vacuum. Even if the false vacuum fraction vanishes asymptotically, the physical volume of the false vacuum, $\Vphys$, may not decrease. To address this issue, we provide correction factors $\ccFactor{i}$ (see \eqref{eq:completionBounds} and surrounding text) to the bubble wall velocity bounds for completion. These correction factors guarantee that $\Vphys$ decreases for all benchmarks considered. Yet even a successful transition with $N(0) \lesssim 1$ may be at odds with the observed isotropy and homogeneity of the cosmic microwave background \cite{Guth:1982pn, Hawking:1982ga, Turner:1992tz}. If only a few large bubbles were responsible for converting the observable Universe into the electroweak symmetry-broken vacuum, it is not clear that the latent heat would be guaranteed to thermalise. This issue is left for future work.

Despite this caveat, our completion criteria provide an important test of extensions to the Standard Model of particle physics. For example, when a first-order phase transition to our current vacuum is essential (e.g.\ for baryogenesis or gravitational waves),%
\footnote{However, gravitational wave production merely requires that bubbles and their sound shells collide, not that the transition completes. Thus, there may be scenarios where a phase transition does not complete but leaves a detectable gravitational wave signal. To our knowledge, such a possibility has not been considered before in the literature.}
the phase transition must complete for that new physics scenario to be realised in nature. The most promising parameter points in terms of gravitational wave detectability typically involve strongly supercooled transitions. Therefore, any regions of parameter space that are constrained using less stringent completion criteria (such as merely requiring unit nucleation) may be excluded when using our completion criteria. Plus, particularly promising parameter points potentially providing phase transitions that are strongly supercooled might be too hastily discarded, due to the assumption that completion requires unit nucleation. This may lead to incorrect bounds being placed on models, and potentially missing a scenario that is actually realised in nature. This necessitates an increase in the rigour in which transition analyses are performed. We hope this paper helps advance that rigour.

\acknowledgments

The work of P.A.\ in this paper has been supported by the National Natural Science Foundation of China (NNSFC) Research Fund for International Excellent Young Scientists, grant No.\ 12150610460, the  National Foreign Bureau Supporting Fund for Foreign Experts grant No.\ wgxz2022021L and the Australian Research Council Future Fellowship grant FT160100274. The work of P.A. and C.B. was also supported by the Australian Research Council Discovery Project grants DP180102209 and DP210101636. The work of L.M.\ was supported by an Australian Government Research Training Program (RTP) Scholarship and a Monash Graduate Excellence Scholarship (MGES). L.M.\ would like to thank Ariel M\'egevand for clarifying notation for the energy density used throughout the literature, and illuminating the difference in assumptions for $\rhoh$ used in Refs.\ \cite{Leitao:2015fmj,Megevand:2007sv}. We thank Andrew Fowlie for discussions regarding the derivation of the false vacuum fraction and the bubble nucleation rate. We also thank Yang Zhang for discussions regarding \PT.

\appendix

\section{Treatment of the real scalar singlet model} \label{app:M2}

\subsection{Symmetry breaking conditions at one-loop}

We impose two constraints on the zero-temperature, one-loop potential: there is a stationary point at $\vv$, and the eigenvalues of the (squared) mass matrix are the (squared) pole masses $m_h^2$ and $m_s^2$. The first constraint reads
\begin{align}
	\atvev{\pdv{V}{\phi_h}} & = 0 , \label{eq:EWSBh} \\
	\atvev{\pdv{V}{\phi_s}} & = 0 , \label{eq:EWSBs}
\end{align}
which respectively allow for the elimination of $\mu_h^2$ and $\mu_s^2$ from the list of free parameters.

For the latter constraint we introduce a mixing angle, $\theta$, to diagonalise the mass matrix, and we have
\begin{align}
	\atvev{\pdv[2]{V}{\phi_h}} & = m_h^2 \cos[2](\theta) + m_s^2 \sin[2](\theta) , \\
	\atvev{\pdv[2]{V}{\phi_s}} & = m_h^2 \sin[2](\theta) + m_s^2 \cos[2](\theta) , \\
	\atvev{\frac{\partial^2 V}{\partial \phi_h \partial \phi_s}} & = \half \! \left(m_h^2 - m_s^2 \right) \sin(2\theta) . \label{eq:mass_hs}
\end{align}
This allows the elimination of $\lambda_h$, $\lambda_s$ and $\lambda_{hs}$, leaving $\kappa_{hhs}, \kappa_{sss}, \theta, v_s$ and $m_s$ as free parameters, with $m_h$ and $v_h$ fixed by experimental observations as described in \cref{app:massSpectrum}. We employ an iterative method to numerically extract the $\mu^2_i$ and $\lambda_i$ Lagrangian parameters from the system of implicit equations described by \eqref{eq:EWSBh}-\eqref{eq:mass_hs}.

\subsection{Mass spectrum, couplings and renormalisation scale} \label{app:massSpectrum}

We use tree-level masses within the Coleman-Weinberg potential. We start with $\alpha = 1 / 137.036$, $m_W = 80.379 \gev$, $m_Z = 91.1876 \gev$ and $m_t = 162.5 \gev$, where the $W$ and $Z$ pole masses and the top running mass have been used \cite{ParticleDataGroup:2020ssz}. From this we calculate the Weinberg angle to be $\theta_W = 0.4918$, and $g = 0.6412$, $g' = 0.3435$ and $y_t = 0.9167$. We take the Higgs mass to be $m_h = 125.1 \gev$ and set the renormalisation scale to $\mu_R = m_Z$ (i.e.\ the scale at which the couplings have been extracted). These couplings and the renormalisation scale are held fixed, even though the masses are field- and temperature-dependent. The Higgs VEV is then $v_h \approx 250.7 \gev$.

The masses included in $\VCW$ and $V_T$ are $m_t$, $m_{W^\pm}$, $m_{Z}$, $m_{\gamma}$, $m_{GB}$, $m_h$ and $m_s$. The scalar masses $m_h$, $m_s$ and $m_{GB}$ are extracted at tree-level, hence using $\mu_{h,0}^2$ and $\mu_{s,0}^2$ obtained through \eqref{eq:EWSBh} and \eqref{eq:EWSBs} with the replacement $V \rightarrow V_0$. The degrees of freedom are $n_t = 12$, $n_{W^\pm} = 6$, $n_Z = n_\gamma = n_{GB} = 3$ and $n_h = n_s = 1$. The thermal mass corrections as well as the explicit forms for these masses (in a $\mathbb{Z}_2$-symmetric variant of this model) can be found in Appendix C of Ref.\ \cite{Athron:2022jyi}.

Lastly, we note that an infrared divergence appears in $V''$ due to terms in the Coleman-Weinberg potential of the form
\begin{equation}
	\atvev{\frac{\partial^2 V_{CW}}{\partial \phi_j \partial \phi_k}} \supset \atvev{\pdv{\mitwo}{\phi_j}} \atvev{\pdv{\mitwo}{\phi_k}} \ln(\frac{\mitwov}{\renorm}) . \label{eq:IRdivergence}
\end{equation}
A vanishing mass in the logarithm causes a divergence if the derivative of the mass at the VEV is non-zero. There are two particles in our mass spectrum that vanish at the VEV at zero temperature; the photon and the Goldstone boson. The photon term in \eqref{eq:IRdivergence} vanishes because the photon is field-independent at zero temperature. However, the Goldstone boson is field-dependent, so the Goldstone term in \eqref{eq:IRdivergence} diverges.

We adopt the following prescription to avoid this IR divergence \cite{Elias-Miro:2014pca, Martin:2014bca}. We shift the Goldstone mass by its one-loop self-energy evaluated at $p^2 = 0$; that is, $m_{GB}^2(\field, T) \rightarrow m_{GB}^2(\field, T) + \kappa \Pi_{GB}(\field)$ where $\kappa = 1 / (16 \pi^2)$. This self-energy is calculated as
\begin{equation}
	\kappa \Pi_{GB}(\field) = \recip{h} \pdv{V'_{CW}}{h} ,
\end{equation}
where the prime denotes the removal of the Goldstone boson contribution to the Coleman-Weinberg potential. Thus,
\begin{equation}
	\Pi_{GB}(\field) = \recip{2 h} \sum_{i \neq GB} (-1)^{2s_i} n_i \pdv{\mitwo}{h} \mitwop \! \left[\log(\frac{\mitwop}{\renorm}) - c_i + \half \right] .
\end{equation}
This mass shift prevents the IR divergence in the second derivative of the Coleman-Weinberg potential, because the Goldstone mass no longer vanishes in the VEV.

\section{Time-temperature relation} \label{app:timeTemperatureRelation}

Under the assumption that the Universe expands adiabatically during the phase transition,%
\footnote{Reheating does not occur in the false vacuum if bubbles grow as detonations, so reheating does not affect the dynamics of the phase transition. However, temperature-dependent relics such as gravitational waves are affected by this reheating.}
we have
\begin{equation}
	\dv{t} \! \left(s(t) a^3(t) \right) = 0 ,
\end{equation}
where $s$ is the entropy density of the plasma and $a$ is the scale factor of the Universe. This can be expressed as
\begin{equation}
	\dv{s}{t} = -3 H(t) s(t). \label{eq:dsdt}
\end{equation}
The entropy density is given by
\begin{equation}
	s(T) = \pdv{p}{T} = -\pdv{V}{T} . \label{eq:entropyDensity}
\end{equation}
Writing \eqref{eq:dsdt} in terms of temperature and using \eqref{eq:entropyDensity}, we have
\begin{equation}
	\dv{T}{t} \pdv[2]{V}{T} = -3 H(T) \pdv{V}{T} ,
\end{equation}
from which we can extract
\begin{equation}
	\dv{T}{t} = -3 H(T) \frac{\partial_T V}{\partial_{TT} V} .
\end{equation}
Here we have used the notation $\partial_{T} V = \pdv{V}{T}$ and $\partial_{TT} V = \pdv[2]{V}{T}$.
Because we assume the energy density is approximately homogeneous even in the presence of bubbles due to energy conservation, we evaluate the derivatives of the potential at the false vacuum. That is,
\begin{equation}
	\dv{T}{t} = -3 H(T) \frac{\partial_T V(\phif(T), T)}{\partial_{TT} V(\phif(T), T)} . \label{eq:dTdt-general}
\end{equation}
The typical next step is to assume an equation of state to simplify \eqref{eq:dTdt-general} and the transition analysis in general. A common choice is the MIT bag equation of state \cite{Chodos:1974je}, where the potential in the false vacuum is modelled as
\begin{equation}
	V(T) = a T^4 + b ,
\end{equation}
with $a$ and $b$ constant in temperature. Under this approximation, \eqref{eq:dTdt-general} reduces to the familiar form of
\begin{equation}
	\dv{T}{t} = -T H(T) . \label{eq:dTdt-simple}
\end{equation}
This simple form is essential for analytic comparison of unit nucleation and percolation in \cref{sec:methodology}. We have numerically compared \eqref{eq:dTdt-general} and \eqref{eq:dTdt-simple} and found very good agreement in each of our benchmarks. The most significant deviation occurs for $T < 10 \gev$, where sub-quartic thermal terms in the potential become significant. That is, where the terms proportional to $T$, $T^2$ and $T^3$ are no longer dominated by the terms proportional to $T^4$. The former are neglected in the high-temperature approximation used in the bag equation of state.

\section{Ratio of scale factors} \label{app:scaleFactorRatio}

The ratio of scale factors at different times and at different temperatures respectively appear in the false vacuum fraction in \eqref{eq:falseVacuumFraction-fullTime} and \eqref{eq:falseVacuumFraction-withScaleFactorRatio}. Here we show how to evaluate these ratios. A particularly simple form arises when using temperature.

The definition of the Hubble parameter,
\begin{equation}
	H(t) = \frac{\dot{a}(t)}{a(t)} = \recip{a(t)} \dv{a}{t} ,
\end{equation}
is a separable ordinary differential equation. Integrating, we have
\begin{equation}
	\int_{t_0}^t \! dt' \, H(t') = \int_{a(t_0)}^{a(t)} \! da \, \recip{a(t)} ,
\end{equation}
where $t_0 < t$ is some reference time. We will see that the reference time is not important here since it cancels when taking the ratio of scale factors. Solving the integral equation for $a(t)$, we find
\begin{equation}
	a(t) = a(t_0) \exp(\int_{t_0}^t \! dt' \, H(t')) .
\end{equation}
The ratio of scale factors at times $t_1$ and $t_2$ is then
\begin{equation}
	\frac{a(t_1)}{a(t_2)} = \exp(\int_{t_2}^{t_1} \! dt' \, H(t')) , \label{eq:scaleFactorRatio-time}
\end{equation}
where all dependence on the reference time $t_0$ has disappeared.

Now consider the ratio of scale factors at temperatures $T_1$ and $T_2$. Using the adiabatic time-temperature relation \eqref{eq:dTdt-simple} derived in \cref{app:timeTemperatureRelation}, the ratio in \eqref{eq:scaleFactorRatio-time} can be expressed in terms of temperature as
\begin{equation}
	\frac{a(T_1)}{a(T_2)} = \exp(\int_{T_1}^{T_2} \! dT' \, \recip{T'}) = \frac{T_2}{T_1} .
\end{equation}
This simple form is exact if the Universe expands adiabatically and the MIT bag equation of state holds.

\section{The effects of reheating} \label{app:reheating}

Reheating occurs when the energy density accumulated in and near bubble walls is redistributed to the surrounding plasma \cite{Heckler:1994uu, Megevand:2017vtb}. This has the effect of increasing the temperature of the surrounding plasma. For bubbles expanding as detonations, this reheating occurs inside the bubble, while for deflagrations it occurs outside the bubble. For a hybrid expansion mode, reheating occurs on both sides of the bubble wall. See Ref.\ \cite{Espinosa:2010hh} for a description of the three bubble expansion modes. If a bubble propagates as a deflagration or hybrid, then the reheating slows the expansion of the bubble and suppresses the bubble nucleation rate nearby.%
\footnote{The suppression of bubble nucleation due to reheating assumes that the nucleation rate is lower at higher temperatures. However, if $T < \Tmax$, where $\Tmax$ maximises the nucleation rate, then reheating towards $\Tmax$ may bring about another era of significant bubble nucleation.}
Then it would appear that our treatment of the true vacuum volume is an upper bound on the correct result that incorporates reheating. We now explain that this does not invalidate our demonstration of the existence of Scenarios 1 and 2 (see \cref{sec:methodology}) when the bubbles expand as deflagrations or hybrids. Reheating for bubbles expanding as detonations is not expected to affect bubble expansion or the nucleation rate.

First, consider Scenario 1: unit nucleation without percolation. We potentially slightly overestimate bubble nucleation, encoded in $X(T)$ (see \eqref{eq:X}). However, $X(T)$ should be largely unaffected because the fraction of the Universe in the reheated shell surrounding true vacuum bubbles must be small. This is because $P_f(T) > 0.71$ for all temperatures in Scenario 1. Regardless, $X(T)$ equally affects the number of bubbles and the true vacuum volume and thus cannot affect their comparison. This realisation follows from the simplification made below \eqref{eq:PercCondGeneral}. The bubble wall velocity will be reduced to an unknown extent. A lower bubble wall velocity would reduce the true vacuum volume, hence further delaying --- perhaps inhibiting --- the onset of percolation. Hence, Scenario 1 only becomes easier to realise with the incorporation of reheating effects. The upper bound on bubble wall velocity for which Scenario 1 is possible then becomes an overly strong constraint. It is possible for Scenario 1 to be realised for bubble wall velocities higher than the predicted upper bound.

Next, consider Scenario 2: percolation without unit nucleation. Again, we overestimate $X(T)$, but perhaps substantially this time as a significant fraction of the Universe may be contained in the reheated shells surrounding bubbles. The case of no unit nucleation is then easier to realise. However, for this scenario we still need percolation to occur. With a slowing of the bubble propagation and less bubbles nucleated, percolation is delayed just as discussed for Scenario 1. The lower bound on bubble wall velocity for which Scenario 2 is possible then becomes an insufficient constraint. It is possible for Scenario 2 to not be realised for bubble wall velocities higher than the predicted lower bound. However, by our arguments, at the very least detonations still allow for the possibility of Scenario 2.

From these general principles, we then expect Scenario 1 to be easier to realise than predicted, and Scenario 2 to still be possible because reheating from detonations occurs only inside the bubbles and should not affect the transition progress.

\section{Comparing bubble and Hubble volumes} \label{app:bubbleRadius}


Here we prove that bubbles can be larger than the Hubble volume provided sufficient supercooling occurs. We need to show that
\begin{equation}
	r(T', T) > r_H(T) , \label{eq:outgrowHubble-comoving}
\end{equation}
where $r(T', T)$ is the comoving radius of a bubble that nucleated at temperature $T'$ and has grown until temperature $T$, and $r_H(T)$ is the Hubble volume at temperature $T$, given by
\begin{equation}
	r_H(T) = \recip{a(T) H(T)} .
\end{equation}

The largest bubble possible at any given temperature is one that nucleated very close to the critical temperature. The initial comoving bubble radius, $r_0$, at the critical temperature is not well-defined. Instead we consider the bubble to have nucleated slightly below this temperature and ignore $r_0$, with the expectation that $r(T_c, T) \gg r_0$. Further assuming a constant wall velocity, as justified in \cref{sec:transitionAnalysis}, the comoving bubble radius (in terms of time) is
\begin{align}
	r(t', t) = v_w \int_{t'}^t \! dt'' \recip{a(t'')} . \label{eq:comovingBubbleRadius}
\end{align}
Using the time-temperature relation derived in \cref{app:timeTemperatureRelation}, \eqref{eq:comovingBubbleRadius} can be expressed in terms of temperature as
\begin{equation}
	r(T', T) = v_w \int_T^{T'} \! dT'' \recip{T'' H(T'') a(T'')} . \label{eq:bubbleRadius-radiation}
\end{equation}
Multiplying by the scale factor at an arbitrary temperature $T_* > 0$ and using the ratio of scale factors derived in \cref{app:scaleFactorRatio}, \eqref{eq:bubbleRadius-radiation} simplifies to
\begin{equation}
	a(T_*) r(T', T) = \frac{v_w}{T_*} \int_T^{T'} \! dT'' \recip{H(T'')} .
\end{equation}
Thus, we have
\begin{equation}
	a(T_*) \left(r(T_c, T) - r_H(T) \right) = \frac{v_w}{T_*} \int_T^{T_c} \! dT' \recip{H(T')} - \frac{T}{T_* H(T)} , \label{eq:compareRadii-final}
\end{equation}
which we require to be positive to satisfy \eqref{eq:outgrowHubble-comoving}.

\begin{figure}
	\centering
	\includegraphics[width=0.7\linewidth]{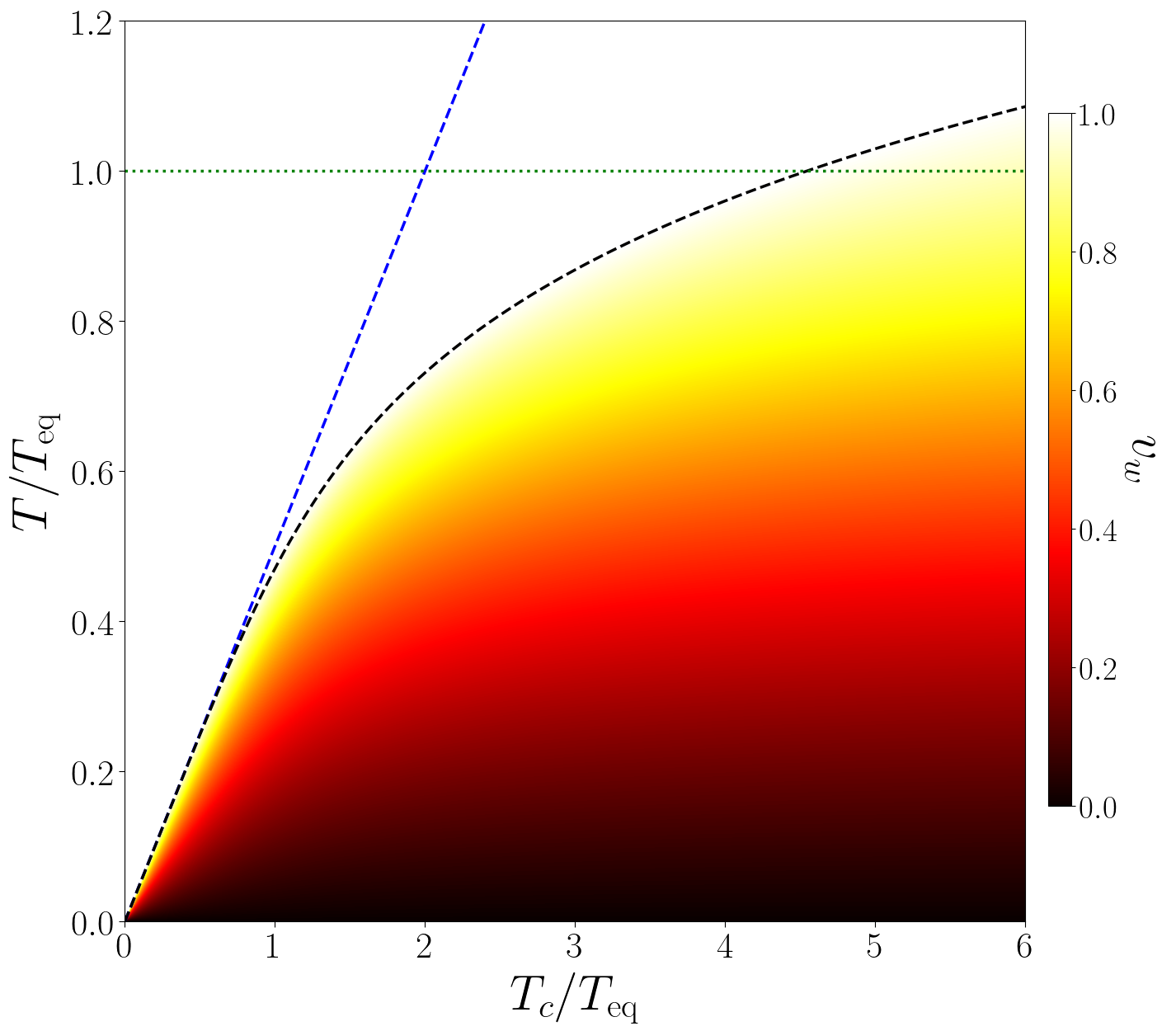}
	\caption{The bubble wall velocity bound for which the volume of bubbles nucleated at the critical temperature can exceed the Hubble volume. The dashed black line corresponds to the contour where a bubble wall velocity of $v_w = 1$ is required. The bubble volume can only exceed the Hubble volume below this contour (in the coloured region). The dashed blue line corresponds to the vacuum-dominated limit: $T < T_c/2$. The horizontal dotted green line corresponds to $T = \Teq$, marking the boundary between the radiation- and vacuum-dominated eras.}
	\label{fig:bubbleVsHubbleVolume}
\end{figure}

We proceed by using the model-independent approximation for the Hubble parameter defined in \eqref{eq:H-approx}. Pulling out a positive factor $1/(T_* H(T))$, we can obtain a bound on the bubble wall velocity for which \eqref{eq:outgrowHubble-comoving} is satisfied,
\begin{equation}
	v_w > \left(\sqrt{\left(\frac{T}{\Teq} \right)^{\!\!4} + 1} \left[\frac{T_c}{T} \, \hypergeomArgs{-\!\left(\frac{T_c}{\Teq} \right)^{\!\!4}} - \hypergeomArgs{-\!\left(\frac{T}{\Teq} \right)^{\!\!4}} \right] \right)^{-1} , \label{eq:vwBound-outgrowHubble}
\end{equation}
independent of the choice of $T_*$. In the vacuum-dominated limit $T_c \ll \Teq$, the hypergeometric functions $\hypergeom$ approach unity, leaving the bound
\begin{equation}
	T < T_c \! \left(1 + \recip{v_w} \right)^{\!\!-1} . \label{eq:vwBound-outgrowHubble-vacDom}
\end{equation}
This bound reduces to $T < T_c/2$ when $v_w = 1$. One can also obtain \eqref{eq:vwBound-outgrowHubble-vacDom} by using a constant Hubble parameter in \eqref{eq:compareRadii-final}.

When the right-hand side of \eqref{eq:vwBound-outgrowHubble} is larger than unity, even the largest bubbles possible are smaller than the Hubble volume. The result \eqref{eq:vwBound-outgrowHubble} is shown in \cref{fig:bubbleVsHubbleVolume}, along with the $(T, T_c)$ contour for which the right-hand side of \eqref{eq:vwBound-outgrowHubble} is equal to unity. We plot the bound $T < T_c/2$ obtained in the vacuum-dominated limit, where subluminal solutions exist. In \cref{fig:bubbleVsHubbleVolume}, we see that the largest bubble volume is smaller than the Hubble volume at high temperature but may exceed the Hubble volume if permitted to grow for sufficiently long. One could replace $T_c$ with $\Tmax$ in \eqref{eq:vwBound-outgrowHubble} and \cref{fig:bubbleVsHubbleVolume} to compare the volume of the most common bubbles to the Hubble volume. One could replace $T$ with $T_f$ to determine whether bubbles can outgrow the Hubble volume before the phase transition ends. For sufficiently large $T_c/\Teq$ (specifically $T_c \gtrsim 4.55\Teq$), we find that bubbles can be larger than the Hubble volume even during the radiation-dominated era. However, the ratio $T/T_c$ for which \eqref{eq:vwBound-outgrowHubble} is satisfied decreases with $T_c$. This suggests that a phase transition that first becomes possible at a high energy scale compared to $\Teq$ requires more supercooling before bubbles can outgrow the Hubble volume.

This investigation demonstrates that it is possible for the volume of a single bubble to exceed that of the Hubble volume if the phase transition is sufficiently supercooled. Then the expectation of having one bubble per Hubble volume is overly restrictive. However, even with bubbles smaller than the Hubble volume, it is possible for these bubbles to percolate without having one in every Hubble volume (assuming some form of Hubble sphere packing). In our simulations with $N(0) \lesssim 1$ (e.g.\ the benchmarks listed in \cref{tab:benchmarks-M1,tab:benchmarks-M2}) we found that the average bubble radius was below $R_H$ at the percolation temperature and above $R_H$ at the completion temperature, where $R_H(T) = 1/H(T)$ is the proper Hubble radius. Ref.\ \cite{Levi:2022bzt} also recently demonstrated the possibility of bubbles exceeding the Hubble volume in Figure 9.


\section{Numerical uncertainty of the bounce action} \label{app:action}

Although tools exist to determine the action, the underlying problem remains computationally difficult and expensive. There is naturally a trade-off between running time and accuracy which becomes particularly relevant for studies that require some combination of parameter space scanning, precise predictions, and strong supercooling. We used \CT\ (version 2.0.6) \cite{Wainwright:2011kj} to compute the action. We found it necessary to make a few key changes to improve the accuracy and precision of our results. The following is a list of changes we made:
\begin{itemize}
	\item We tightened the tolerances \texttt{xtol} and \texttt{phitol} input to \texttt{tunneling1D.findProfile}, from $10^{-4}$ to $10^{-6}$. Further tightening was not found to consistently improve precision, yet incurred substantial penalties to running time.
	\item We resolved many underflow and overflow errors that commonly appear in thin-walled cases and sometimes elsewhere.
	\item We introduced a \TWE\ exception class to differentiate general failure of the path deformation algorithm from failures that occur specifically for thin-walled cases.
	\item Phases were traced using \PT\ (version 1.1.0) \cite{Athron:2020sbe} rather than with \CT.
	\item Transitions were analysed using \TS\ rather than with \CT.
\end{itemize}
Together these changes prevent the otherwise frequent occurrence of crashes and exceptions (outside of expected \TWEs), and suppress sawtooth patterns in the action curve arising from loose tolerances. \TWEs\ are raised typically for $T \approx T_c$ where nucleation is negligible, and importantly provide a distinction between this expected case of failure and more general failure due to unhandled exceptions. We find the \TWE\ to be consistently raised above a threshold temperature near $T_c$, with all action evaluations successful below this threshold temperature. Due to fixing underflow and overflow errors, the more general failure of the action evaluation no longer occurs.

Despite these improvements, we find an $\mathcal{O}(0.1\%)$ discontinuity at the minimum of the action (i.e.\ at $\Tmin$), particularly in M2. Additionally, the numerically determined action curve oscillates slightly near $\Tmin$ due to negligible perturbations of the Lagrangian parameters. The discontinuity and oscillations lead to noticeable uncertainty in quantities directly related to $S_{\text{min}}$, as seen in \cref{fig:supercoolingScan}, and ultimately limits our ability to resolve $\scn \rightarrow 1$. This uncertainty is slightly amplified by the discrete sampling of the action. Nevertheless, our changes alleviate issues presented in the appendix of Ref.\ \cite{Freitas:2021yng} (and possibly the issues that motivated Appendix A of Ref.\ \cite{Azatov:2022tii}) which we also experienced. We suspect issues associated with evaluating the action are more prevalent than the lack of such reports would suggest.
	
\bibliography{References}

\providecommand{\href}[2]{#2}\begingroup\raggedright\begin{thebibliography}{100}

\bibitem{Dolan:1973qd}
L.A.~Dolan and R.W.~Jackiw, \emph{{Symmetry Behavior at Finite Temperature}},
  \href{https://doi.org/10.1103/PhysRevD.9.3320}{\emph{Phys. Rev. D} {\bfseries
  9} (1974) 3320}.

\bibitem{Weinberg:1974hy}
S.~Weinberg, \emph{{Gauge and Global Symmetries at High Temperature}},
  \href{https://doi.org/10.1103/PhysRevD.9.3357}{\emph{Phys. Rev. D} {\bfseries
  9} (1974) 3357}.

\bibitem{Kirzhnits:1974as}
D.A.~Kirzhnits and A.D.~Linde, \emph{{A Relativistic phase transition}},
  {\emph{Zh. Eksp. Teor. Fiz.} {\bfseries 67} (1974) 1263}.

\bibitem{Kirzhnits:1976ts}
D.A.~Kirzhnits and A.D.~Linde, \emph{{Symmetry Behavior in Gauge Theories}},
  \href{https://doi.org/10.1016/0003-4916(76)90279-7}{\emph{Annals Phys.}
  {\bfseries 101} (1976) 195}.

\bibitem{Dine:1992wr}
M.~Dine, R.G.~Leigh, P.Y.~Huet, A.D.~Linde and D.A.~Linde, \emph{{Towards the
  theory of the electroweak phase transition}},
  \href{https://doi.org/10.1103/PhysRevD.46.550}{\emph{Phys. Rev. D} {\bfseries
  46} (1992) 550} [\href{https://arxiv.org/abs/hep-ph/9203203}{{\ttfamily
  hep-ph/9203203}}].

\bibitem{Aoki:2006br}
Y.~Aoki, Z.~Fodor, S.D.~Katz and K.K.~Szab\'o, \emph{{The QCD transition
  temperature: Results with physical masses in the continuum limit}},
  \href{https://doi.org/10.1016/j.physletb.2006.10.021}{\emph{Phys. Lett. B}
  {\bfseries 643} (2006) 46}
  [\href{https://arxiv.org/abs/hep-lat/0609068}{{\ttfamily hep-lat/0609068}}].

\bibitem{Aoki:2006we}
Y.~Aoki, G.~Endr\H{o}di, Z.~Fodor, S.D.~Katz and K.K.~Szab\'o, \emph{{The Order
  of the quantum chromodynamics transition predicted by the standard model of
  particle physics}}, \href{https://doi.org/10.1038/nature05120}{\emph{Nature}
  {\bfseries 443} (2006) 675}
  [\href{https://arxiv.org/abs/hep-lat/0611014}{{\ttfamily hep-lat/0611014}}].

\bibitem{Coleman:1977py}
S.R.~Coleman, \emph{{The Fate of the False Vacuum. 1. Semiclassical Theory}},
  \href{https://doi.org/10.1103/PhysRevD.16.1248}{\emph{Phys. Rev. D}
  {\bfseries 15} (1977) 2929}.

\bibitem{Callan:1977pt}
C.G.~Callan, Jr. and S.R.~Coleman, \emph{{The Fate of the False Vacuum. 2.
  First Quantum Corrections}},
  \href{https://doi.org/10.1103/PhysRevD.16.1762}{\emph{Phys. Rev. D}
  {\bfseries 16} (1977) 1762}.

\bibitem{Linde:1980tt}
A.D.~Linde, \emph{{Fate of the False Vacuum at Finite Temperature: Theory and
  Applications}},
  \href{https://doi.org/10.1016/0370-2693(81)90281-1}{\emph{Phys. Lett. B}
  {\bfseries 100} (1981) 37}.

\bibitem{Linde:1981zj}
A.D.~Linde, \emph{{Decay of the False Vacuum at Finite Temperature}},
  \href{https://doi.org/10.1016/0550-3213(83)90293-6}{\emph{Nucl. Phys. B}
  {\bfseries 216} (1983) 421}.

\bibitem{Trodden:1998ym}
M.~Trodden, \emph{{Electroweak baryogenesis}},
  \href{https://doi.org/10.1103/RevModPhys.71.1463}{\emph{Rev. Mod. Phys.}
  {\bfseries 71} (1999) 1463}
  [\href{https://arxiv.org/abs/hep-ph/9803479}{{\ttfamily hep-ph/9803479}}].

\bibitem{Cline:2006ts}
J.M.~Cline, \emph{{Baryogenesis}},  in \emph{{Les Houches Summer School -
  Session 86: Particle Physics and Cosmology: The Fabric of Spacetime}}, 9,
  2006 [\href{https://arxiv.org/abs/hep-ph/0609145}{{\ttfamily
  hep-ph/0609145}}].

\bibitem{Morrissey:2012db}
D.E.~Morrissey and M.J.~Ramsey-Musolf, \emph{{Electroweak baryogenesis}},
  \href{https://doi.org/10.1088/1367-2630/14/12/125003}{\emph{New J. Phys.}
  {\bfseries 14} (2012) 125003}
  [\href{https://arxiv.org/abs/1206.2942}{{\ttfamily 1206.2942}}].

\bibitem{White:2016nbo}
{White, Graham Albert}, \emph{{A Pedagogical Introduction to Electroweak
  Baryogenesis}}, 2053-2571, Morgan \& Claypool Publishers (2016),
  \href{https://doi.org/10.1088/978-1-6817-4457-5}{10.1088/978-1-6817-4457-5}.

\bibitem{Bertone:2019irm}
G.~Bertone et~al., \emph{{Gravitational wave probes of dark matter: challenges
  and opportunities}},
  \href{https://doi.org/10.21468/SciPostPhysCore.3.2.007}{\emph{SciPost Phys.
  Core} {\bfseries 3} (2020) 007}
  [\href{https://arxiv.org/abs/1907.10610}{{\ttfamily 1907.10610}}].

\bibitem{Vilenkin:1981bx}
A.~Vilenkin, \emph{{Gravitational radiation from cosmic strings}},
  \href{https://doi.org/10.1016/0370-2693(81)91144-8}{\emph{Phys. Lett. B}
  {\bfseries 107} (1981) 47}.

\bibitem{Vilenkin:1984ib}
A.~Vilenkin, \emph{{Cosmic Strings and Domain Walls}},
  \href{https://doi.org/10.1016/0370-1573(85)90033-X}{\emph{Phys. Rept.}
  {\bfseries 121} (1985) 263}.

\bibitem{Kawasaki:2011vv}
M.~Kawasaki and K.~Saikawa, \emph{{Study of gravitational radiation from cosmic
  domain walls}},
  \href{https://doi.org/10.1088/1475-7516/2011/09/008}{\emph{JCAP} {\bfseries
  09} (2011) 008} [\href{https://arxiv.org/abs/1102.5628}{{\ttfamily
  1102.5628}}].

\bibitem{Figueroa:2012kw}
D.G.~Figueroa, M.~Hindmarsh and J.~Urrestilla, \emph{{Exact Scale-Invariant
  Background of Gravitational Waves from Cosmic Defects}},
  \href{https://doi.org/10.1103/PhysRevLett.110.101302}{\emph{Phys. Rev. Lett.}
  {\bfseries 110} (2013) 101302}
  [\href{https://arxiv.org/abs/1212.5458}{{\ttfamily 1212.5458}}].

\bibitem{Witten:1984rs}
E.~Witten, \emph{{Cosmic Separation of Phases}},
  \href{https://doi.org/10.1103/PhysRevD.30.272}{\emph{Phys. Rev. D} {\bfseries
  30} (1984) 272}.

\bibitem{Hogan:1986qda}
C.J.~Hogan, \emph{{Gravitational radiation from cosmological phase
  transitions}}, \href{https://doi.org/10.1093/mnras/218.4.629}{\emph{Mon. Not.
  Roy. Astron. Soc.} {\bfseries 218} (1986) 629}.

\bibitem{Kosowsky:1991ua}
A.~Kosowsky, M.S.~Turner and R.~Watkins, \emph{{Gravitational radiation from
  colliding vacuum bubbles}},
  \href{https://doi.org/10.1103/PhysRevD.45.4514}{\emph{Phys. Rev. D}
  {\bfseries 45} (1992) 4514}.

\bibitem{Kosowsky:1992rz}
A.~Kosowsky, M.S.~Turner and R.~Watkins, \emph{{Gravitational waves from first
  order cosmological phase transitions}},
  \href{https://doi.org/10.1103/PhysRevLett.69.2026}{\emph{Phys. Rev. Lett.}
  {\bfseries 69} (1992) 2026}.

\bibitem{Kajantie:1996mn}
K.~Kajantie, M.~Laine, K.~Rummukainen and M.E.~Shaposhnikov, \emph{{Is there a~
  hot electroweak phase transition at $m_H \gtrsim m_W$?}},
  \href{https://doi.org/10.1103/PhysRevLett.77.2887}{\emph{Phys. Rev. Lett.}
  {\bfseries 77} (1996) 2887}
  [\href{https://arxiv.org/abs/hep-ph/9605288}{{\ttfamily hep-ph/9605288}}].

\bibitem{Rummukainen:1998as}
K.~Rummukainen, M.~Tsypin, K.~Kajantie, M.~Laine and M.E.~Shaposhnikov,
  \emph{{The Universality class of the electroweak theory}},
  \href{https://doi.org/10.1016/S0550-3213(98)00494-5}{\emph{Nucl. Phys. B}
  {\bfseries 532} (1998) 283}
  [\href{https://arxiv.org/abs/hep-lat/9805013}{{\ttfamily hep-lat/9805013}}].

\bibitem{Csikor:1998eu}
F.~Csikor, Z.~Fodor and J.~Heitger, \emph{{Endpoint of the hot electroweak
  phase transition}},
  \href{https://doi.org/10.1103/PhysRevLett.82.21}{\emph{Phys. Rev. Lett.}
  {\bfseries 82} (1999) 21}
  [\href{https://arxiv.org/abs/hep-ph/9809291}{{\ttfamily hep-ph/9809291}}].

\bibitem{Beniwal:2017eik}
A.~Beniwal, M.~Lewicki, J.D.~Wells, M.~White and A.G.~Williams,
  \emph{{Gravitational wave, collider and dark matter signals from a scalar
  singlet electroweak baryogenesis}},
  \href{https://doi.org/10.1007/JHEP08(2017)108}{\emph{JHEP} {\bfseries 08}
  (2017) 108} [\href{https://arxiv.org/abs/1702.06124}{{\ttfamily
  1702.06124}}].

\bibitem{Kurup:2017dzf}
G.~Kurup and M.~Perelstein, \emph{{Dynamics of Electroweak Phase Transition In
  Singlet-Scalar Extension of the Standard Model}},
  \href{https://doi.org/10.1103/PhysRevD.96.015036}{\emph{Phys. Rev. D}
  {\bfseries 96} (2017) 015036}
  [\href{https://arxiv.org/abs/1704.03381}{{\ttfamily 1704.03381}}].

\bibitem{Chiang:2017nmu}
C.-W.~Chiang, M.J.~Ramsey-Musolf and E.~Senaha, \emph{{Standard Model with a
  Complex Scalar Singlet: Cosmological Implications and Theoretical
  Considerations}},
  \href{https://doi.org/10.1103/PhysRevD.97.015005}{\emph{Phys. Rev. D}
  {\bfseries 97} (2018) 015005}
  [\href{https://arxiv.org/abs/1707.09960}{{\ttfamily 1707.09960}}].

\bibitem{Bernon:2017jgv}
J.~Bernon, L.~Bian and Y.~Jiang, \emph{{A new insight into the phase transition
  in the early Universe with two Higgs doublets}},
  \href{https://doi.org/10.1007/JHEP05(2018)151}{\emph{JHEP} {\bfseries 05}
  (2018) 151} [\href{https://arxiv.org/abs/1712.08430}{{\ttfamily
  1712.08430}}].

\bibitem{Alves:2018jsw}
A.~Alves, T.~Ghosh, H.-K.~Guo, K.~Sinha and D.~Vagie, \emph{{Collider and
  Gravitational Wave Complementarity in Exploring the Singlet Extension of the
  Standard Model}}, \href{https://doi.org/10.1007/JHEP04(2019)052}{\emph{JHEP}
  {\bfseries 04} (2019) 052}
  [\href{https://arxiv.org/abs/1812.09333}{{\ttfamily 1812.09333}}].

\bibitem{Li:2019tfd}
H.-L.~Li, M.J.~Ramsey-Musolf and S.~Willocq, \emph{{Probing a scalar
  singlet-catalyzed electroweak phase transition with resonant di-Higgs boson
  production in the $4b$ channel}},
  \href{https://doi.org/10.1103/PhysRevD.100.075035}{\emph{Phys. Rev. D}
  {\bfseries 100} (2019) 075035}
  [\href{https://arxiv.org/abs/1906.05289}{{\ttfamily 1906.05289}}].

\bibitem{Athron:2019teq}
P.~Athron, C.~Bal\'azs, A.~Fowlie, G.~Pozzo, G.~White and Y.~Zhang,
  \emph{{Strong first-order phase transitions in the NMSSM \textemdash{} a
  comprehensive survey}},
  \href{https://doi.org/10.1007/JHEP11(2019)151}{\emph{JHEP} {\bfseries 11}
  (2019) 151} [\href{https://arxiv.org/abs/1908.11847}{{\ttfamily
  1908.11847}}].

\bibitem{Wang:2019pet}
X.~Wang, F.P.~Huang and X.~Zhang, \emph{{Gravitational wave and collider
  signals in complex two-Higgs doublet model with dynamical CP-violation at
  finite temperature}},
  \href{https://doi.org/10.1103/PhysRevD.101.015015}{\emph{Phys. Rev. D}
  {\bfseries 101} (2020) 015015}
  [\href{https://arxiv.org/abs/1909.02978}{{\ttfamily 1909.02978}}].

\bibitem{Demidov:2021lyo}
S.V.~Demidov, D.S.~Gorbunov and E.~Kriukova, \emph{{Gravitational waves from
  first-order electroweak phase transition in a model with light sgoldstinos}},
  \href{https://doi.org/10.1007/JHEP07(2022)061}{\emph{JHEP} {\bfseries 07}
  (2022) 061} [\href{https://arxiv.org/abs/2112.06083}{{\ttfamily
  2112.06083}}].

\bibitem{Schettler:2010dp}
S.~Schettler, T.~Boeckel and J.~Schaffner-Bielich, \emph{{Imprints of the QCD
  Phase Transition on the Spectrum of Gravitational Waves}},
  \href{https://doi.org/10.1103/PhysRevD.83.064030}{\emph{Phys. Rev. D}
  {\bfseries 83} (2011) 064030}
  [\href{https://arxiv.org/abs/1010.4857}{{\ttfamily 1010.4857}}].

\bibitem{Iso:2017uuu}
S.~Iso, P.D.~Serpico and K.~Shimada, \emph{{QCD-Electroweak First-Order Phase
  Transition in a Supercooled Universe}},
  \href{https://doi.org/10.1103/PhysRevLett.119.141301}{\emph{Phys. Rev. Lett.}
  {\bfseries 119} (2017) 141301}
  [\href{https://arxiv.org/abs/1704.04955}{{\ttfamily 1704.04955}}].

\bibitem{Rezapour:2020mvi}
S.~Rezapour, K.~Bitaghsir~Fadafan and M.~Ahmadvand, \emph{{Gravitational waves
  of a first-order QCD phase transition at finite coupling from holography}},
  \href{https://doi.org/10.1016/j.aop.2021.168731}{\emph{Annals Phys.}
  {\bfseries 437} (2022) 168731}
  [\href{https://arxiv.org/abs/2006.04265}{{\ttfamily 2006.04265}}].

\bibitem{Daniel:1980xn}
M.~Daniel and C.E.~Vayonakis, \emph{{The Phase Transition in the SU(5) Model at
  High Temperatures}},
  \href{https://doi.org/10.1016/0550-3213(81)90422-3}{\emph{Nucl. Phys. B}
  {\bfseries 180} (1981) 301}.

\bibitem{Dev:2016feu}
P.S.B.~Dev and A.~Mazumdar, \emph{{Probing the Scale of New Physics by Advanced
  LIGO/VIRGO}}, \href{https://doi.org/10.1103/PhysRevD.93.104001}{\emph{Phys.
  Rev. D} {\bfseries 93} (2016) 104001}
  [\href{https://arxiv.org/abs/1602.04203}{{\ttfamily 1602.04203}}].

\bibitem{Balazs:2016tbi}
C.~Bal\'azs, A.~Fowlie, A.~Mazumdar and G.~White, \emph{{Gravitational waves at
  aLIGO and vacuum stability with a scalar singlet extension of the Standard
  Model}}, \href{https://doi.org/10.1103/PhysRevD.95.043505}{\emph{Phys. Rev.
  D} {\bfseries 95} (2017) 043505}
  [\href{https://arxiv.org/abs/1611.01617}{{\ttfamily 1611.01617}}].

\bibitem{Croon:2018kqn}
D.~Croon, T.E.~Gonzalo and G.~White, \emph{{Gravitational Waves from a
  Pati-Salam Phase Transition}},
  \href{https://doi.org/10.1007/JHEP02(2019)083}{\emph{JHEP} {\bfseries 02}
  (2019) 083} [\href{https://arxiv.org/abs/1812.02747}{{\ttfamily
  1812.02747}}].

\bibitem{Huang:2020bbe}
W.-C.~Huang, F.~Sannino and Z.-W.~Wang, \emph{{Gravitational Waves from
  Pati-Salam Dynamics}},
  \href{https://doi.org/10.1103/PhysRevD.102.095025}{\emph{Phys. Rev. D}
  {\bfseries 102} (2020) 095025}
  [\href{https://arxiv.org/abs/2004.02332}{{\ttfamily 2004.02332}}].

\bibitem{Okada:2020vvb}
N.~Okada, O.~Seto and H.~Uchida, \emph{{Gravitational waves from breaking of an
  extra $U(1)$ in $SO(10)$ grand unification}},
  \href{https://doi.org/10.1093/ptep/ptab003}{\emph{PTEP} {\bfseries 2021}
  (2021) 033B01} [\href{https://arxiv.org/abs/2006.01406}{{\ttfamily
  2006.01406}}].

\bibitem{Caprini:2015zlo}
C.~Caprini et~al., \emph{{Science with the space-based interferometer eLISA.
  II: Gravitational waves from cosmological phase transitions}},
  \href{https://doi.org/10.1088/1475-7516/2016/04/001}{\emph{JCAP} {\bfseries
  04} (2016) 001} [\href{https://arxiv.org/abs/1512.06239}{{\ttfamily
  1512.06239}}].

\bibitem{Caprini:2019egz}
C.~Caprini et~al., \emph{{Detecting gravitational waves from cosmological phase
  transitions with LISA: an update}},
  \href{https://doi.org/10.1088/1475-7516/2020/03/024}{\emph{JCAP} {\bfseries
  03} (2020) 024} [\href{https://arxiv.org/abs/1910.13125}{{\ttfamily
  1910.13125}}].

\bibitem{LISA:2017pwj}
{\scshape LISA} collaboration, \emph{{Laser Interferometer Space Antenna}},
  \href{https://arxiv.org/abs/1702.00786}{{\ttfamily 1702.00786}}.

\bibitem{Grojean:2006bp}
C.~Grojean and G.~Servant, \emph{{Gravitational Waves from Phase Transitions at
  the Electroweak Scale and Beyond}},
  \href{https://doi.org/10.1103/PhysRevD.75.043507}{\emph{Phys. Rev. D}
  {\bfseries 75} (2007) 043507}
  [\href{https://arxiv.org/abs/hep-ph/0607107}{{\ttfamily hep-ph/0607107}}].

\bibitem{Huber:2008hg}
S.J.~Huber and T.~Konstandin, \emph{{Gravitational Wave Production by
  Collisions: More Bubbles}},
  \href{https://doi.org/10.1088/1475-7516/2008/09/022}{\emph{JCAP} {\bfseries
  09} (2008) 022} [\href{https://arxiv.org/abs/0806.1828}{{\ttfamily
  0806.1828}}].

\bibitem{Jinno:2016knw}
R.~Jinno and M.~Takimoto, \emph{{Probing a classically conformal B-L model with
  gravitational waves}},
  \href{https://doi.org/10.1103/PhysRevD.95.015020}{\emph{Phys. Rev. D}
  {\bfseries 95} (2017) 015020}
  [\href{https://arxiv.org/abs/1604.05035}{{\ttfamily 1604.05035}}].

\bibitem{Marzola:2017jzl}
L.~Marzola, A.~Racioppi and V.~Vaskonen, \emph{{Phase transition and
  gravitational wave phenomenology of scalar conformal extensions of the
  Standard Model}},
  \href{https://doi.org/10.1140/epjc/s10052-017-4996-1}{\emph{Eur. Phys. J. C}
  {\bfseries 77} (2017) 484}
  [\href{https://arxiv.org/abs/1704.01034}{{\ttfamily 1704.01034}}].

\bibitem{Ellis:2019oqb}
J.~Ellis, M.~Lewicki, J.M.~No and V.~Vaskonen, \emph{{Gravitational wave energy
  budget in strongly supercooled phase transitions}},
  \href{https://doi.org/10.1088/1475-7516/2019/06/024}{\emph{JCAP} {\bfseries
  06} (2019) 024} [\href{https://arxiv.org/abs/1903.09642}{{\ttfamily
  1903.09642}}].

\bibitem{Ellis:2020nnr}
J.~Ellis, M.~Lewicki and V.~Vaskonen, \emph{{Updated predictions for
  gravitational waves produced in a strongly supercooled phase transition}},
  \href{https://doi.org/10.1088/1475-7516/2020/11/020}{\emph{JCAP} {\bfseries
  11} (2020) 020} [\href{https://arxiv.org/abs/2007.15586}{{\ttfamily
  2007.15586}}].

\bibitem{Kierkla:2022odc}
M.~Kierkla, A.~Karam and B.~Swiezewska, \emph{{Conformal model for
  gravitational waves and dark matter: A status update}},
  \href{https://arxiv.org/abs/2210.07075}{{\ttfamily 2210.07075}}.

\bibitem{Levi:2022bzt}
N.~Levi, T.~Opferkuch and D.~Redigolo, \emph{{The Supercooling Window at Weak
  and Strong Coupling}},  \href{https://arxiv.org/abs/2212.08085}{{\ttfamily
  2212.08085}}.

\bibitem{Huber:2015znp}
S.J.~Huber, T.~Konstandin, G.~Nardini and I.~Rues, \emph{{Detectable
  Gravitational Waves from Very Strong Phase Transitions in the General
  NMSSM}}, \href{https://doi.org/10.1088/1475-7516/2016/03/036}{\emph{JCAP}
  {\bfseries 03} (2016) 036}
  [\href{https://arxiv.org/abs/1512.06357}{{\ttfamily 1512.06357}}].

\bibitem{Leitao:2015fmj}
L.~Leitao and A.~M\'egevand, \emph{{Gravitational waves from a very strong
  electroweak phase transition}},
  \href{https://doi.org/10.1088/1475-7516/2016/05/037}{\emph{JCAP} {\bfseries
  05} (2016) 037} [\href{https://arxiv.org/abs/1512.08962}{{\ttfamily
  1512.08962}}].

\bibitem{Megevand:2016lpr}
A.~M\'egevand and S.~Ram\'irez, \emph{{Bubble nucleation and growth in very
  strong cosmological phase transitions}},
  \href{https://doi.org/10.1016/j.nuclphysb.2017.03.009}{\emph{Nucl. Phys. B}
  {\bfseries 919} (2017) 74}
  [\href{https://arxiv.org/abs/1611.05853}{{\ttfamily 1611.05853}}].

\bibitem{Kobakhidze:2017mru}
A.~Kobakhidze, C.~Lagger, A.~Manning and J.~Yue, \emph{{Gravitational waves
  from a supercooled electroweak phase transition and their detection with
  pulsar timing arrays}},
  \href{https://doi.org/10.1140/epjc/s10052-017-5132-y}{\emph{Eur. Phys. J. C}
  {\bfseries 77} (2017) 570}
  [\href{https://arxiv.org/abs/1703.06552}{{\ttfamily 1703.06552}}].

\bibitem{Cai:2017tmh}
R.-G.~Cai, M.~Sasaki and S.-J.~Wang, \emph{{The gravitational waves from the
  first-order phase transition with a dimension-six operator}},
  \href{https://doi.org/10.1088/1475-7516/2017/08/004}{\emph{JCAP} {\bfseries
  08} (2017) 004} [\href{https://arxiv.org/abs/1707.03001}{{\ttfamily
  1707.03001}}].

\bibitem{Ellis:2018mja}
J.~Ellis, M.~Lewicki and J.M.~No, \emph{{On the Maximal Strength of a
  First-Order Electroweak Phase Transition and its Gravitational Wave Signal}},
  \href{https://doi.org/10.1088/1475-7516/2019/04/003}{\emph{JCAP} {\bfseries
  04} (2019) 003} [\href{https://arxiv.org/abs/1809.08242}{{\ttfamily
  1809.08242}}].

\bibitem{Wang:2020jrd}
X.~Wang, F.P.~Huang and X.~Zhang, \emph{{Phase transition dynamics and
  gravitational wave spectra of strong first-order phase transition in
  supercooled universe}},
  \href{https://doi.org/10.1088/1475-7516/2020/05/045}{\emph{JCAP} {\bfseries
  05} (2020) 045} [\href{https://arxiv.org/abs/2003.08892}{{\ttfamily
  2003.08892}}].

\bibitem{Freese:2022qrl}
K.~Freese and M.W.~Winkler, \emph{{Have pulsar timing arrays detected the hot
  big bang: Gravitational waves from strong first order phase transitions in
  the early Universe}},
  \href{https://doi.org/10.1103/PhysRevD.106.103523}{\emph{Phys. Rev. D}
  {\bfseries 106} (2022) 103523}
  [\href{https://arxiv.org/abs/2208.03330}{{\ttfamily 2208.03330}}].

\bibitem{Lewicki:2022pdb}
M.~Lewicki and V.~Vaskonen, \emph{{Gravitational waves from bubble collisions
  and fluid motion in strongly supercooled phase transitions}},
  \href{https://arxiv.org/abs/2208.11697}{{\ttfamily 2208.11697}}.

\bibitem{Badger:2022nwo}
C.~Badger et~al., \emph{{Probing Early Universe Supercooled Phase Transitions
  with Gravitational Wave Data}},
  \href{https://arxiv.org/abs/2209.14707}{{\ttfamily 2209.14707}}.

\bibitem{LIGOScientific:2016aoc}
{\scshape LIGO Scientific, Virgo} collaboration, \emph{{Observation of
  Gravitational Waves from a Binary Black Hole Merger}},
  \href{https://doi.org/10.1103/PhysRevLett.116.061102}{\emph{Phys. Rev. Lett.}
  {\bfseries 116} (2016) 061102}
  [\href{https://arxiv.org/abs/1602.03837}{{\ttfamily 1602.03837}}].

\bibitem{Cutting:2019zws}
D.~Cutting, M.~Hindmarsh and D.J.~Weir, \emph{{Vorticity, kinetic energy, and
  suppressed gravitational wave production in strong first order phase
  transitions}},
  \href{https://doi.org/10.1103/PhysRevLett.125.021302}{\emph{Phys. Rev. Lett.}
  {\bfseries 125} (2020) 021302}
  [\href{https://arxiv.org/abs/1906.00480}{{\ttfamily 1906.00480}}].

\bibitem{Weir:2016tov}
D.J.~Weir, \emph{{Revisiting the envelope approximation: gravitational waves
  from bubble collisions}},
  \href{https://doi.org/10.1103/PhysRevD.93.124037}{\emph{Phys. Rev. D}
  {\bfseries 93} (2016) 124037}
  [\href{https://arxiv.org/abs/1604.08429}{{\ttfamily 1604.08429}}].

\bibitem{Jinno:2016vai}
R.~Jinno and M.~Takimoto, \emph{{Gravitational waves from bubble collisions: An
  analytic derivation}},
  \href{https://doi.org/10.1103/PhysRevD.95.024009}{\emph{Phys. Rev. D}
  {\bfseries 95} (2017) 024009}
  [\href{https://arxiv.org/abs/1605.01403}{{\ttfamily 1605.01403}}].

\bibitem{Hindmarsh:2016lnk}
M.~Hindmarsh, \emph{{Sound shell model for acoustic gravitational wave
  production at a first-order phase transition in the early Universe}},
  \href{https://doi.org/10.1103/PhysRevLett.120.071301}{\emph{Phys. Rev. Lett.}
  {\bfseries 120} (2018) 071301}
  [\href{https://arxiv.org/abs/1608.04735}{{\ttfamily 1608.04735}}].

\bibitem{Bodeker:2017cim}
D.~Bodeker and G.D.~Moore, \emph{{Electroweak Bubble Wall Speed Limit}},
  \href{https://doi.org/10.1088/1475-7516/2017/05/025}{\emph{JCAP} {\bfseries
  05} (2017) 025} [\href{https://arxiv.org/abs/1703.08215}{{\ttfamily
  1703.08215}}].

\bibitem{Hindmarsh:2017gnf}
M.~Hindmarsh, S.J.~Huber, K.~Rummukainen and D.J.~Weir, \emph{{Shape of the
  acoustic gravitational wave power spectrum from a first order phase
  transition}}, \href{https://doi.org/10.1103/PhysRevD.96.103520}{\emph{Phys.
  Rev. D} {\bfseries 96} (2017) 103520}
  [\href{https://arxiv.org/abs/1704.05871}{{\ttfamily 1704.05871}}].

\bibitem{Jinno:2017fby}
R.~Jinno and M.~Takimoto, \emph{{Gravitational waves from bubble dynamics:
  Beyond the Envelope}},
  \href{https://doi.org/10.1088/1475-7516/2019/01/060}{\emph{JCAP} {\bfseries
  01} (2019) 060} [\href{https://arxiv.org/abs/1707.03111}{{\ttfamily
  1707.03111}}].

\bibitem{Megevand:2017vtb}
A.~M\'egevand and S.~Ram\'irez, \emph{{Bubble nucleation and growth in slow
  cosmological phase transitions}},
  \href{https://doi.org/10.1016/j.nuclphysb.2018.01.012}{\emph{Nucl. Phys. B}
  {\bfseries 928} (2018) 38}
  [\href{https://arxiv.org/abs/1710.06279}{{\ttfamily 1710.06279}}].

\bibitem{Konstandin:2017sat}
T.~Konstandin, \emph{{Gravitational radiation from a bulk flow model}},
  \href{https://doi.org/10.1088/1475-7516/2018/03/047}{\emph{JCAP} {\bfseries
  03} (2018) 047} [\href{https://arxiv.org/abs/1712.06869}{{\ttfamily
  1712.06869}}].

\bibitem{Cutting:2018tjt}
D.~Cutting, M.~Hindmarsh and D.J.~Weir, \emph{{Gravitational waves from vacuum
  first-order phase transitions: from the envelope to the lattice}},
  \href{https://doi.org/10.1103/PhysRevD.97.123513}{\emph{Phys. Rev. D}
  {\bfseries 97} (2018) 123513}
  [\href{https://arxiv.org/abs/1802.05712}{{\ttfamily 1802.05712}}].

\bibitem{Niksa:2018ofa}
P.~Niksa, M.~Schlederer and G.~Sigl, \emph{{Gravitational Waves produced by
  Compressible MHD Turbulence from Cosmological Phase Transitions}},
  \href{https://doi.org/10.1088/1361-6382/aac89c}{\emph{Class. Quant. Grav.}
  {\bfseries 35} (2018) 144001}
  [\href{https://arxiv.org/abs/1803.02271}{{\ttfamily 1803.02271}}].

\bibitem{RoperPol:2019wvy}
A.~Roper~Pol, S.~Mandal, A.~Brandenburg, T.~Kahniashvili and A.~Kosowsky,
  \emph{{Numerical simulations of gravitational waves from early-universe
  turbulence}}, \href{https://doi.org/10.1103/PhysRevD.102.083512}{\emph{Phys.
  Rev. D} {\bfseries 102} (2020) 083512}
  [\href{https://arxiv.org/abs/1903.08585}{{\ttfamily 1903.08585}}].

\bibitem{Jinno:2019jhi}
R.~Jinno, H.~Seong, M.~Takimoto and C.M.~Um, \emph{{Gravitational waves from
  first-order phase transitions: Ultra-supercooled transitions and the fate of
  relativistic shocks}},
  \href{https://doi.org/10.1088/1475-7516/2019/10/033}{\emph{JCAP} {\bfseries
  10} (2019) 033} [\href{https://arxiv.org/abs/1905.00899}{{\ttfamily
  1905.00899}}].

\bibitem{Jinno:2019bxw}
R.~Jinno, T.~Konstandin and M.~Takimoto, \emph{{Relativistic bubble
  collisions\textemdash{}a closer look}},
  \href{https://doi.org/10.1088/1475-7516/2019/09/035}{\emph{JCAP} {\bfseries
  09} (2019) 035} [\href{https://arxiv.org/abs/1906.02588}{{\ttfamily
  1906.02588}}].

\bibitem{Hindmarsh:2019phv}
M.~Hindmarsh and M.~Hijazi, \emph{{Gravitational waves from first order
  cosmological phase transitions in the Sound Shell Model}},
  \href{https://doi.org/10.1088/1475-7516/2019/12/062}{\emph{JCAP} {\bfseries
  12} (2019) 062} [\href{https://arxiv.org/abs/1909.10040}{{\ttfamily
  1909.10040}}].

\bibitem{Alanne:2019bsm}
T.~Alanne, T.~Hugle, M.~Platscher and K.~Schmitz, \emph{{A fresh look at the
  gravitational-wave signal from cosmological phase transitions}},
  \href{https://doi.org/10.1007/JHEP03(2020)004}{\emph{JHEP} {\bfseries 03}
  (2020) 004} [\href{https://arxiv.org/abs/1909.11356}{{\ttfamily
  1909.11356}}].

\bibitem{Lewicki:2019gmv}
M.~Lewicki and V.~Vaskonen, \emph{{On bubble collisions in strongly supercooled
  phase transitions}},
  \href{https://doi.org/10.1016/j.dark.2020.100672}{\emph{Phys. Dark Univ.}
  {\bfseries 30} (2020) 100672}
  [\href{https://arxiv.org/abs/1912.00997}{{\ttfamily 1912.00997}}].

\bibitem{Ellis:2020awk}
J.~Ellis, M.~Lewicki and J.M.~No, \emph{{Gravitational waves from first-order
  cosmological phase transitions: lifetime of the sound wave source}},
  \href{https://doi.org/10.1088/1475-7516/2020/07/050}{\emph{JCAP} {\bfseries
  07} (2020) 050} [\href{https://arxiv.org/abs/2003.07360}{{\ttfamily
  2003.07360}}].

\bibitem{Giese:2020rtr}
F.~Giese, T.~Konstandin and J.~van~de Vis, \emph{{Model-independent energy
  budget of cosmological first-order phase transitions\textemdash{}A sound
  argument to go beyond the bag model}},
  \href{https://doi.org/10.1088/1475-7516/2020/07/057}{\emph{JCAP} {\bfseries
  07} (2020) 057} [\href{https://arxiv.org/abs/2004.06995}{{\ttfamily
  2004.06995}}].

\bibitem{BarrosoMancha:2020fay}
M.~Barroso~Mancha, T.~Prokopec and B.~Swiezewska, \emph{{Field-theoretic
  derivation of bubble-wall force}},
  \href{https://doi.org/10.1007/JHEP01(2021)070}{\emph{JHEP} {\bfseries 01}
  (2021) 070} [\href{https://arxiv.org/abs/2005.10875}{{\ttfamily
  2005.10875}}].

\bibitem{Cutting:2020nla}
D.~Cutting, E.G.~Escartin, M.~Hindmarsh and D.J.~Weir, \emph{{Gravitational
  waves from vacuum first order phase transitions II: from thin to thick
  walls}}, \href{https://doi.org/10.1103/PhysRevD.103.023531}{\emph{Phys. Rev.
  D} {\bfseries 103} (2021) 023531}
  [\href{https://arxiv.org/abs/2005.13537}{{\ttfamily 2005.13537}}].

\bibitem{Lewicki:2020jiv}
M.~Lewicki and V.~Vaskonen, \emph{{Gravitational wave spectra from strongly
  supercooled phase transitions}},
  \href{https://doi.org/10.1140/epjc/s10052-020-08589-1}{\emph{Eur. Phys. J. C}
  {\bfseries 80} (2020) 1003}
  [\href{https://arxiv.org/abs/2007.04967}{{\ttfamily 2007.04967}}].

\bibitem{Guo:2020grp}
H.-K.~Guo, K.~Sinha, D.~Vagie and G.~White, \emph{{Phase Transitions in an
  Expanding Universe: Stochastic Gravitational Waves in Standard and
  Non-Standard Histories}},
  \href{https://doi.org/10.1088/1475-7516/2021/01/001}{\emph{JCAP} {\bfseries
  01} (2021) 001} [\href{https://arxiv.org/abs/2007.08537}{{\ttfamily
  2007.08537}}].

\bibitem{Hoche:2020ysm}
S.~H\"oche, J.~Kozaczuk, A.J.~Long, J.~Turner and Y.~Wang, \emph{{Towards an
  all-orders calculation of the electroweak bubble wall velocity}},
  \href{https://doi.org/10.1088/1475-7516/2021/03/009}{\emph{JCAP} {\bfseries
  03} (2021) 009} [\href{https://arxiv.org/abs/2007.10343}{{\ttfamily
  2007.10343}}].

\bibitem{Megevand:2020klf}
A.~M\'egevand and F.A.~Membiela, \emph{{Bubble wall correlations in
  cosmological phase transitions}},
  \href{https://doi.org/10.1103/PhysRevD.102.103514}{\emph{Phys. Rev. D}
  {\bfseries 102} (2020) 103514}
  [\href{https://arxiv.org/abs/2008.01873}{{\ttfamily 2008.01873}}].

\bibitem{Croon:2020cgk}
D.~Croon, O.~Gould, P.~Schicho, T.V.I.~Tenkanen and G.~White,
  \emph{{Theoretical uncertainties for cosmological first-order phase
  transitions}}, \href{https://doi.org/10.1007/JHEP04(2021)055}{\emph{JHEP}
  {\bfseries 04} (2021) 055}
  [\href{https://arxiv.org/abs/2009.10080}{{\ttfamily 2009.10080}}].

\bibitem{Friedlander:2020tnq}
A.~Friedlander, I.~Banta, J.M.~Cline and D.~Tucker-Smith, \emph{{Wall speed and
  shape in singlet-assisted strong electroweak phase transitions}},
  \href{https://doi.org/10.1103/PhysRevD.103.055020}{\emph{Phys. Rev. D}
  {\bfseries 103} (2021) 055020}
  [\href{https://arxiv.org/abs/2009.14295}{{\ttfamily 2009.14295}}].

\bibitem{Jinno:2020eqg}
R.~Jinno, T.~Konstandin and H.~Rubira, \emph{{A hybrid simulation of
  gravitational wave production in first-order phase transitions}},
  \href{https://doi.org/10.1088/1475-7516/2021/04/014}{\emph{JCAP} {\bfseries
  04} (2021) 014} [\href{https://arxiv.org/abs/2010.00971}{{\ttfamily
  2010.00971}}].

\bibitem{Azatov:2020ufh}
A.~Azatov and M.~Vanvlasselaer, \emph{{Bubble wall velocity: heavy physics
  effects}}, \href{https://doi.org/10.1088/1475-7516/2021/01/058}{\emph{JCAP}
  {\bfseries 01} (2021) 058}
  [\href{https://arxiv.org/abs/2010.02590}{{\ttfamily 2010.02590}}].

\bibitem{Balaji:2020yrx}
S.~Balaji, M.~Spannowsky and C.~Tamarit, \emph{{Cosmological bubble friction in
  local equilibrium}},
  \href{https://doi.org/10.1088/1475-7516/2021/03/051}{\emph{JCAP} {\bfseries
  03} (2021) 051} [\href{https://arxiv.org/abs/2010.08013}{{\ttfamily
  2010.08013}}].

\bibitem{Giese:2020znk}
F.~Giese, T.~Konstandin, K.~Schmitz and J.~Van De~Vis, \emph{{Model-independent
  energy budget for LISA}},
  \href{https://doi.org/10.1088/1475-7516/2021/01/072}{\emph{JCAP} {\bfseries
  01} (2021) 072} [\href{https://arxiv.org/abs/2010.09744}{{\ttfamily
  2010.09744}}].

\bibitem{Cai:2020djd}
R.-G.~Cai and S.-J.~Wang, \emph{{Effective picture of bubble expansion}},
  \href{https://doi.org/10.1088/1475-7516/2021/03/096}{\emph{JCAP} {\bfseries
  03} (2021) 096} [\href{https://arxiv.org/abs/2011.11451}{{\ttfamily
  2011.11451}}].

\bibitem{Guo:2021qcq}
H.-K.~Guo, K.~Sinha, D.~Vagie and G.~White, \emph{{The benefits of diligence:
  how precise are predicted gravitational wave spectra in models with phase
  transitions?}}, \href{https://doi.org/10.1007/JHEP06(2021)164}{\emph{JHEP}
  {\bfseries 06} (2021) 164}
  [\href{https://arxiv.org/abs/2103.06933}{{\ttfamily 2103.06933}}].

\bibitem{Ekstedt:2021kyx}
A.~Ekstedt, \emph{{Higher-order corrections to the bubble-nucleation rate at
  finite temperature}},
  \href{https://doi.org/10.1140/epjc/s10052-022-10130-5}{\emph{Eur. Phys. J. C}
  {\bfseries 82} (2022) 173}
  [\href{https://arxiv.org/abs/2104.11804}{{\ttfamily 2104.11804}}].

\bibitem{Gould:2021ccf}
O.~Gould and J.~Hirvonen, \emph{{Effective field theory approach to thermal
  bubble nucleation}},
  \href{https://doi.org/10.1103/PhysRevD.104.096015}{\emph{Phys. Rev. D}
  {\bfseries 104} (2021) 096015}
  [\href{https://arxiv.org/abs/2108.04377}{{\ttfamily 2108.04377}}].

\bibitem{Megevand:2021juo}
A.~Meg\'evand and F.A.~Membiela, \emph{{Gravitational waves from bubble
  walls}}, \href{https://doi.org/10.1088/1475-7516/2021/10/073}{\emph{JCAP}
  {\bfseries 10} (2021) 073}
  [\href{https://arxiv.org/abs/2108.05510}{{\ttfamily 2108.05510}}].

\bibitem{Megevand:2021llq}
A.~M\'egevand and F.A.~Membiela, \emph{{Model-independent features of
  gravitational waves from bubble collisions}},
  \href{https://doi.org/10.1103/PhysRevD.104.123532}{\emph{Phys. Rev. D}
  {\bfseries 104} (2021) 123532}
  [\href{https://arxiv.org/abs/2108.07034}{{\ttfamily 2108.07034}}].

\bibitem{Ai:2021kak}
W.-Y.~Ai, B.~Garbrecht and C.~Tamarit, \emph{{Bubble wall velocities in local
  equilibrium}},  \href{https://arxiv.org/abs/2109.13710}{{\ttfamily
  2109.13710}}.

\bibitem{Gouttenoire:2021kjv}
Y.~Gouttenoire, R.~Jinno and F.~Sala, \emph{{Friction pressure on relativistic
  bubble walls}},  \href{https://arxiv.org/abs/2112.07686}{{\ttfamily
  2112.07686}}.

\bibitem{Hirvonen:2021zej}
J.~Hirvonen, J.~L\"ofgren, M.J.~Ramsey-Musolf, P.~Schicho and T.V.I.~Tenkanen,
  \emph{{Computing the gauge-invariant bubble nucleation rate in finite
  temperature effective field theory}},
  \href{https://doi.org/10.1007/JHEP07(2022)135}{\emph{JHEP} {\bfseries 07}
  (2022) 135} [\href{https://arxiv.org/abs/2112.08912}{{\ttfamily
  2112.08912}}].

\bibitem{Dahl:2021wyk}
J.~Dahl, M.~Hindmarsh, K.~Rummukainen and D.J.~Weir, \emph{{Decay of acoustic
  turbulence in two dimensions and implications for cosmological gravitational
  waves}}, \href{https://doi.org/10.1103/PhysRevD.106.063511}{\emph{Phys. Rev.
  D} {\bfseries 106} (2022) 063511}
  [\href{https://arxiv.org/abs/2112.12013}{{\ttfamily 2112.12013}}].

\bibitem{Dorsch:2021nje}
G.C.~Dorsch, S.J.~Huber and T.~Konstandin, \emph{{A sonic boom in bubble wall
  friction}}, \href{https://doi.org/10.1088/1475-7516/2022/04/010}{\emph{JCAP}
  {\bfseries 04} (2022) 010}
  [\href{https://arxiv.org/abs/2112.12548}{{\ttfamily 2112.12548}}].

\bibitem{Wang:2021dwl}
X.~Wang, F.P.~Huang and Y.~Li, \emph{{Sound velocity effects on the phase
  transition gravitational wave spectrum in the sound shell model}},
  \href{https://doi.org/10.1103/PhysRevD.105.103513}{\emph{Phys. Rev. D}
  {\bfseries 105} (2022) 103513}
  [\href{https://arxiv.org/abs/2112.14650}{{\ttfamily 2112.14650}}].

\bibitem{DeCurtis:2022hlx}
S.~De~Curtis, L.D.~Rose, A.~Guiggiani, A.G.~Muyor and G.~Panico, \emph{{Bubble
  wall dynamics at the electroweak phase transition}},
  \href{https://doi.org/10.1007/JHEP03(2022)163}{\emph{JHEP} {\bfseries 03}
  (2022) 163} [\href{https://arxiv.org/abs/2201.08220}{{\ttfamily
  2201.08220}}].

\bibitem{Cai:2022bcf}
R.-G.~Cai, K.~Hashino, S.-J.~Wang and J.-H.~Yu, \emph{{Gravitational waves from
  patterns of electroweak symmetry breaking: an effective perspective}},
  \href{https://arxiv.org/abs/2202.08295}{{\ttfamily 2202.08295}}.

\bibitem{Cutting:2022zgd}
D.~Cutting, E.~Vilhonen and D.J.~Weir, \emph{{Droplet collapse during strongly
  supercooled transitions}},
  \href{https://doi.org/10.1103/PhysRevD.106.103524}{\emph{Phys. Rev. D}
  {\bfseries 106} (2022) 103524}
  [\href{https://arxiv.org/abs/2204.03396}{{\ttfamily 2204.03396}}].

\bibitem{Auclair:2022jod}
P.~Auclair, C.~Caprini, D.~Cutting, M.~Hindmarsh, K.~Rummukainen, D.A.~Steer
  et~al., \emph{{Generation of gravitational waves from freely decaying
  turbulence}},
  \href{https://doi.org/10.1088/1475-7516/2022/09/029}{\emph{JCAP} {\bfseries
  09} (2022) 029} [\href{https://arxiv.org/abs/2205.02588}{{\ttfamily
  2205.02588}}].

\bibitem{Ajmi:2022nmq}
M.A.~Ajmi and M.~Hindmarsh, \emph{{Thermal suppression of bubble nucleation at
  first-order phase transitions in the early Universe}},
  \href{https://doi.org/10.1103/PhysRevD.106.023505}{\emph{Phys. Rev. D}
  {\bfseries 106} (2022) 023505}
  [\href{https://arxiv.org/abs/2205.04097}{{\ttfamily 2205.04097}}].

\bibitem{Wang:2022lyd}
S.-J.~Wang and Z.-Y.~Yuwen, \emph{{The energy budget of cosmological
  first-order phase transitions beyond the bag equation of state}},
  \href{https://doi.org/10.1088/1475-7516/2022/10/047}{\emph{JCAP} {\bfseries
  10} (2022) 047} [\href{https://arxiv.org/abs/2206.01148}{{\ttfamily
  2206.01148}}].

\bibitem{Tenkanen:2022tly}
T.V.I.~Tenkanen and J.~van~de Vis, \emph{{Speed of sound in cosmological phase
  transitions and effect on gravitational waves}},
  \href{https://doi.org/10.1007/JHEP08(2022)302}{\emph{JHEP} {\bfseries 08}
  (2022) 302} [\href{https://arxiv.org/abs/2206.01130}{{\ttfamily
  2206.01130}}].

\bibitem{Hijazi:2022uzc}
M.~Hijazi, \emph{{Numerical Estimations of the Distribution of the Lifetime of
  Bubbles Emerging from First Order Cosmological Phase Transitions}},
  \href{https://arxiv.org/abs/2208.10636}{{\ttfamily 2208.10636}}.

\bibitem{Lewicki:2022nba}
M.~Lewicki, V.~Vaskonen and H.~Veerm\"ae, \emph{{Bubble dynamics in fluids with
  N-body simulations}},
  \href{https://doi.org/10.1103/PhysRevD.106.103501}{\emph{Phys. Rev. D}
  {\bfseries 106} (2022) 103501}
  [\href{https://arxiv.org/abs/2205.05667}{{\ttfamily 2205.05667}}].

\bibitem{Davis:2003ad}
T.M.~Davis and C.H.~Lineweaver, \emph{{Expanding confusion: common
  misconceptions of cosmological horizons and the superluminal expansion of the
  universe}}, \href{https://doi.org/10.1071/AS03040}{\emph{Publ. Astron. Soc.
  Austral.} {\bfseries 21} (2004) 97}
  [\href{https://arxiv.org/abs/astro-ph/0310808}{{\ttfamily
  astro-ph/0310808}}].

\bibitem{Ellis:2015wdi}
G.F.R.~Ellis and J.-P.~Uzan, \emph{{Causal structures in inflation}},
  \href{https://doi.org/10.1016/j.crhy.2015.07.005}{\emph{Comptes Rendus
  Physique} {\bfseries 16} (2015) 928}
  [\href{https://arxiv.org/abs/1612.01084}{{\ttfamily 1612.01084}}].

\bibitem{Chao:2017ilw}
W.~Chao, W.-F.~Cui, H.-K.~Guo and J.~Shu, \emph{{Gravitational wave imprint of
  new symmetry breaking}},
  \href{https://doi.org/10.1088/1674-1137/abb4cb}{\emph{Chin. Phys. C}
  {\bfseries 44} (2020) 123102}
  [\href{https://arxiv.org/abs/1707.09759}{{\ttfamily 1707.09759}}].

\bibitem{Megevand:2007sv}
A.~M\'egevand and A.D.~S\'anchez, \emph{{Supercooling and phase coexistence in
  cosmological phase transitions}},
  \href{https://doi.org/10.1103/PhysRevD.77.063519}{\emph{Phys. Rev. D}
  {\bfseries 77} (2008) 063519}
  [\href{https://arxiv.org/abs/0712.1031}{{\ttfamily 0712.1031}}].

\bibitem{Darme:2017wvu}
L.~Darm\'e, J.~Jaeckel and M.~Lewicki, \emph{{Towards the fate of the
  oscillating false vacuum}},
  \href{https://doi.org/10.1103/PhysRevD.96.056001}{\emph{Phys. Rev. D}
  {\bfseries 96} (2017) 056001}
  [\href{https://arxiv.org/abs/1704.06445}{{\ttfamily 1704.06445}}].

\bibitem{Hindmarsh:2015qta}
M.~Hindmarsh, S.J.~Huber, K.~Rummukainen and D.J.~Weir, \emph{{Numerical
  simulations of acoustically generated gravitational waves at a first order
  phase transition}},
  \href{https://doi.org/10.1103/PhysRevD.92.123009}{\emph{Phys. Rev. D}
  {\bfseries 92} (2015) 123009}
  [\href{https://arxiv.org/abs/1504.03291}{{\ttfamily 1504.03291}}].

\bibitem{Guth:1979bh}
A.H.~Guth and S.-H.H.~Tye, \emph{{Phase Transitions and Magnetic Monopole
  Production in the Very Early Universe}},
  \href{https://doi.org/10.1103/PhysRevLett.44.631}{\emph{Phys. Rev. Lett.}
  {\bfseries 44} (1980) 631}.

\bibitem{Guth:1981uk}
A.H.~Guth and E.J.~Weinberg, \emph{{Cosmological Consequences of a First Order
  Phase Transition in the SU(5) Grand Unified Model}},
  \href{https://doi.org/10.1103/PhysRevD.23.876}{\emph{Phys. Rev. D} {\bfseries
  23} (1981) 876}.

\bibitem{johnson1939reaction}
W.A.~Johnson, \emph{Reaction kinetics in processes of nucleation and growth},
  {\emph{Am. Inst. Min. Metal. Petro. Eng.} {\bfseries 135} (1939) 416}.

\bibitem{Avrami1}
M.~Avrami, \emph{{Kinetics of Phase Change. I General Theory}},
  \href{https://doi.org/10.1063/1.1750380}{\emph{The Journal of Chemical
  Physics} {\bfseries 7} (1939) 1103}.

\bibitem{Avrami2}
M.~Avrami, \emph{{Kinetics of Phase Change. II Transformation-Time Relations
  for Random Distribution of Nuclei}},
  \href{https://doi.org/10.1063/1.1750631}{\emph{The Journal of Chemical
  Physics} {\bfseries 8} (1940) 212}.

\bibitem{Avrami3}
M.~Avrami, \emph{{Granulation, Phase Change, and Microstructure Kinetics of
  Phase Change. III}}, \href{https://doi.org/10.1063/1.1750872}{\emph{The
  Journal of Chemical Physics} {\bfseries 9} (1941) 177}.

\bibitem{kolmogorov1937statistical}
A.N.~Kolmogorov, \emph{On the statistical theory of metal crystallization},
  {\emph{Izv. Akad. Nauk SSSR, Ser. Math} {\bfseries 1} (1937) 335}.

\bibitem{Athron:2023}
P.~Athron, C.~Bal\'azs, A.~Fowlie, L.~Morris and L.~Wu, \emph{{Cosmological
  Phase Transitions: from perturbative particle physics to gravitational
  waves}},  2023.

\bibitem{fanfoni1998johnson}
M.~Fanfoni and M.~Tomellini, \emph{{The Johnson-Mehl-Avrami-Kolmogorov model: a
  brief review}}, {\emph{Il Nuovo Cimento D} {\bfseries 20} (1998) 1171}.

\bibitem{BURBELKO2005429}
A.A.~Burbelko, E.~Fra\'s and W.~Kapturkiewicz, \emph{{About Kolmogorov's
  statistical theory of phase transformation}},
  \href{https://doi.org/https://doi.org/10.1016/j.msea.2005.08.161}{\emph{Materials
  Science and Engineering: A} {\bfseries 413-414} (2005) 429}.

\bibitem{ALEKSEECHKIN20113159}
N.V.~Alekseechkin, \emph{{Extension of the Kolmogorov–Johnson–Mehl–Avrami
  theory to growth laws of diffusion type}},
  \href{https://doi.org/https://doi.org/10.1016/j.jnoncrysol.2011.05.007}{\emph{Journal
  of Non-Crystalline Solids} {\bfseries 357} (2011) 3159}.

\bibitem{Chodos:1974je}
A.~Chodos, R.L.~Jaffe, K.~Johnson, C.B.~Thorn and V.F.~Weisskopf, \emph{{New
  Extended Model of Hadrons}},
  \href{https://doi.org/10.1103/PhysRevD.9.3471}{\emph{Phys. Rev. D} {\bfseries
  9} (1974) 3471}.

\bibitem{Wu:2019pbm}
J.~Wu, Q.~Li, J.~Liu, C.~Xue, S.~Yang, C.~Shao et~al., \emph{{Progress in
  Precise Measurements of the Gravitational Constant}},
  \href{https://doi.org/10.1002/andp.201900013}{\emph{Annalen Phys.} {\bfseries
  531} (2019) 1900013}.

\bibitem{Enqvist:1991xw}
K.~Enqvist, J.~Ignatius, K.~Kajantie and K.~Rummukainen, \emph{{Nucleation and
  bubble growth in a first order cosmological electroweak phase transition}},
  \href{https://doi.org/10.1103/PhysRevD.45.3415}{\emph{Phys. Rev. D}
  {\bfseries 45} (1992) 3415}.

\bibitem{doi:10.1063/1.1338506}
C.D.~Lorenz and R.M.~Ziff, \emph{{Precise determination of the critical
  percolation threshold for the three-dimensional “Swiss cheese” model
  using a growth algorithm}},
  \href{https://doi.org/10.1063/1.1338506}{\emph{The Journal of Chemical
  Physics} {\bfseries 114} (2001) 3659}.

\bibitem{LIN2018299}
J.~Lin and H.~Chen, \emph{Continuum percolation of porous media via random
  packing of overlapping cube-like particles},
  \href{https://doi.org/https://doi.org/10.1016/j.taml.2018.05.007}{\emph{Theoretical
  and Applied Mechanics Letters} {\bfseries 8} (2018) 299}.

\bibitem{LI2020112815}
M.~Li, H.~Chen and J.~Lin, \emph{Numerical study for the percolation threshold
  and transport properties of porous composites comprising
  non-centrosymmetrical superovoidal pores},
  \href{https://doi.org/https://doi.org/10.1016/j.cma.2019.112815}{\emph{Computer
  Methods in Applied Mechanics and Engineering} {\bfseries 361} (2020) 112815}.

\bibitem{doi:10.1080/00018737100101261}
V.K.S.~Shante and S.~Kirkpatrick, \emph{An introduction to percolation theory},
  \href{https://doi.org/10.1080/00018737100101261}{\emph{Advances in Physics}
  {\bfseries 20} (1971) 325}.

\bibitem{hunt2014percolation}
A.~Hunt, R.~Ewing and B.~Ghanbarian, \emph{Percolation theory for flow in
  porous media}, vol.~880, Springer (2014).

\bibitem{Turner:1992tz}
M.S.~Turner, E.J.~Weinberg and L.M.~Widrow, \emph{{Bubble nucleation in first
  order inflation and other cosmological phase transitions}},
  \href{https://doi.org/10.1103/PhysRevD.46.2384}{\emph{Phys. Rev. D}
  {\bfseries 46} (1992) 2384}.

\bibitem{Heckler:1994uu}
A.F.~Heckler, \emph{{The Effects of electroweak phase transition dynamics on
  baryogenesis and primordial nucleosynthesis}},
  \href{https://doi.org/10.1103/PhysRevD.51.405}{\emph{Phys. Rev. D} {\bfseries
  51} (1995) 405} [\href{https://arxiv.org/abs/astro-ph/9407064}{{\ttfamily
  astro-ph/9407064}}].

\bibitem{Kosowsky:1992vn}
A.~Kosowsky and M.S.~Turner, \emph{{Gravitational radiation from colliding
  vacuum bubbles: envelope approximation to many bubble collisions}},
  \href{https://doi.org/10.1103/PhysRevD.47.4372}{\emph{Phys. Rev. D}
  {\bfseries 47} (1993) 4372}
  [\href{https://arxiv.org/abs/astro-ph/9211004}{{\ttfamily
  astro-ph/9211004}}].

\bibitem{Kamionkowski:1993fg}
M.~Kamionkowski, A.~Kosowsky and M.S.~Turner, \emph{{Gravitational radiation
  from first order phase transitions}},
  \href{https://doi.org/10.1103/PhysRevD.49.2837}{\emph{Phys. Rev. D}
  {\bfseries 49} (1994) 2837}
  [\href{https://arxiv.org/abs/astro-ph/9310044}{{\ttfamily
  astro-ph/9310044}}].

\bibitem{Megevand:2013yua}
A.~M\'egevand and F.A.~Membiela, \emph{{Stability of cosmological deflagration
  fronts}}, \href{https://doi.org/10.1103/PhysRevD.89.103507}{\emph{Phys. Rev.
  D} {\bfseries 89} (2014) 103507}
  [\href{https://arxiv.org/abs/1311.2453}{{\ttfamily 1311.2453}}].

\bibitem{Megevand:2014dua}
A.~M\'egevand, F.A.~Membiela and A.D.~S\'anchez, \emph{{Lower bound on the
  electroweak wall velocity from hydrodynamic instability}},
  \href{https://doi.org/10.1088/1475-7516/2015/03/051}{\emph{JCAP} {\bfseries
  03} (2015) 051} [\href{https://arxiv.org/abs/1412.8064}{{\ttfamily
  1412.8064}}].

\bibitem{Ashoorioon:2009nf}
A.~Ashoorioon and T.~Konstandin, \emph{{Strong electroweak phase transitions
  without collider traces}},
  \href{https://doi.org/10.1088/1126-6708/2009/07/086}{\emph{JHEP} {\bfseries
  07} (2009) 086} [\href{https://arxiv.org/abs/0904.0353}{{\ttfamily
  0904.0353}}].

\bibitem{Huang:2016cjm}
P.~Huang, A.J.~Long and L.-T.~Wang, \emph{{Probing the Electroweak Phase
  Transition with Higgs Factories and Gravitational Waves}},
  \href{https://doi.org/10.1103/PhysRevD.94.075008}{\emph{Phys. Rev. D}
  {\bfseries 94} (2016) 075008}
  [\href{https://arxiv.org/abs/1608.06619}{{\ttfamily 1608.06619}}].

\bibitem{Hashino:2016xoj}
K.~Hashino, M.~Kakizaki, S.~Kanemura, P.~Ko and T.~Matsui, \emph{{Gravitational
  waves and Higgs boson couplings for exploring first order phase transition in
  the model with a singlet scalar field}},
  \href{https://doi.org/10.1016/j.physletb.2016.12.052}{\emph{Phys. Lett. B}
  {\bfseries 766} (2017) 49}
  [\href{https://arxiv.org/abs/1609.00297}{{\ttfamily 1609.00297}}].

\bibitem{Huang:2017jws}
T.~Huang, J.M.~No, L.~Perni\'e, M.J.~Ramsey-Musolf, A.N.~Safonov, M.~Spannowsky
  et~al., \emph{{Resonant di-Higgs boson production in the $b{\bar b}WW$
  channel: Probing the electroweak phase transition at the LHC}},
  \href{https://doi.org/10.1103/PhysRevD.96.035007}{\emph{Phys. Rev. D}
  {\bfseries 96} (2017) 035007}
  [\href{https://arxiv.org/abs/1701.04442}{{\ttfamily 1701.04442}}].

\bibitem{Alves:2018oct}
A.~Alves, T.~Ghosh, H.-K.~Guo and K.~Sinha, \emph{{Resonant Di-Higgs Production
  at Gravitational Wave Benchmarks: A Collider Study using Machine Learning}},
  \href{https://doi.org/10.1007/JHEP12(2018)070}{\emph{JHEP} {\bfseries 12}
  (2018) 070} [\href{https://arxiv.org/abs/1808.08974}{{\ttfamily
  1808.08974}}].

\bibitem{Gould:2019qek}
O.~Gould, J.~Kozaczuk, L.~Niemi, M.J.~Ramsey-Musolf, T.V.I.~Tenkanen and
  D.J.~Weir, \emph{{Nonperturbative analysis of the gravitational waves from a
  first-order electroweak phase transition}},
  \href{https://doi.org/10.1103/PhysRevD.100.115024}{\emph{Phys. Rev. D}
  {\bfseries 100} (2019) 115024}
  [\href{https://arxiv.org/abs/1903.11604}{{\ttfamily 1903.11604}}].

\bibitem{Bian:2019bsn}
L.~Bian, H.-K.~Guo, Y.~Wu and R.~Zhou, \emph{{Gravitational wave and collider
  searches for electroweak symmetry breaking patterns}},
  \href{https://doi.org/10.1103/PhysRevD.101.035011}{\emph{Phys. Rev. D}
  {\bfseries 101} (2020) 035011}
  [\href{https://arxiv.org/abs/1906.11664}{{\ttfamily 1906.11664}}].

\bibitem{Zhou:2019uzq}
R.~Zhou, L.~Bian and H.-K.~Guo, \emph{{Connecting the electroweak sphaleron
  with gravitational waves}},
  \href{https://doi.org/10.1103/PhysRevD.101.091903}{\emph{Phys. Rev. D}
  {\bfseries 101} (2020) 091903}
  [\href{https://arxiv.org/abs/1910.00234}{{\ttfamily 1910.00234}}].

\bibitem{Alves:2020bpi}
A.~Alves, D.~Gon\c{c}alves, T.~Ghosh, H.-K.~Guo and K.~Sinha, \emph{{Di-Higgs
  Blind Spots in Gravitational Wave Signals}},
  \href{https://doi.org/10.1016/j.physletb.2021.136377}{\emph{Phys. Lett. B}
  {\bfseries 818} (2021) 136377}
  [\href{https://arxiv.org/abs/2007.15654}{{\ttfamily 2007.15654}}].

\bibitem{Liu:2021jyc}
W.~Liu and K.-P.~Xie, \emph{{Probing electroweak phase transition with
  multi-TeV muon colliders and gravitational waves}},
  \href{https://doi.org/10.1007/JHEP04(2021)015}{\emph{JHEP} {\bfseries 04}
  (2021) 015} [\href{https://arxiv.org/abs/2101.10469}{{\ttfamily
  2101.10469}}].

\bibitem{Ellis:2022lft}
J.~Ellis, M.~Lewicki, M.~Merchand, J.M.~No and M.~Zych, \emph{{The Scalar
  Singlet Extension of the Standard Model: Gravitational Waves versus
  Baryogenesis}},  \href{https://arxiv.org/abs/2210.16305}{{\ttfamily
  2210.16305}}.

\bibitem{Chiang:2018gsn}
C.-W.~Chiang, Y.-T.~Li and E.~Senaha, \emph{{Revisiting electroweak phase
  transition in the standard model with a real singlet scalar}},
  \href{https://doi.org/10.1016/j.physletb.2018.12.017}{\emph{Phys. Lett. B}
  {\bfseries 789} (2019) 154}
  [\href{https://arxiv.org/abs/1808.01098}{{\ttfamily 1808.01098}}].

\bibitem{Athron:2022jyi}
P.~Athron, C.~Bal\'azs, A.~Fowlie, L.~Morris, G.~White and Y.~Zhang, \emph{{How
  arbitrary are perturbative calculations of the electroweak phase
  transition?}},  \href{https://arxiv.org/abs/2208.01319}{{\ttfamily
  2208.01319}}.

\bibitem{Niemi:2021qvp}
L.~Niemi, P.~Schicho and T.V.I.~Tenkanen, \emph{{Singlet-assisted electroweak
  phase transition at two loops}},
  \href{https://doi.org/10.1103/PhysRevD.103.115035}{\emph{Phys. Rev. D}
  {\bfseries 103} (2021) 115035}
  [\href{https://arxiv.org/abs/2103.07467}{{\ttfamily 2103.07467}}].

\bibitem{Anderson:1991zb}
G.W.~Anderson and L.J.~Hall, \emph{{The Electroweak phase transition and
  baryogenesis}}, \href{https://doi.org/10.1103/PhysRevD.45.2685}{\emph{Phys.
  Rev. D} {\bfseries 45} (1992) 2685}.

\bibitem{Megevand:2003tg}
A.~Meg\'evand, \emph{{First order cosmological phase transitions in the
  radiation dominated era}},
  \href{https://doi.org/10.1103/PhysRevD.69.103521}{\emph{Phys. Rev. D}
  {\bfseries 69} (2004) 103521}
  [\href{https://arxiv.org/abs/hep-ph/0312305}{{\ttfamily hep-ph/0312305}}].

\bibitem{Profumo:2007wc}
S.~Profumo, M.J.~Ramsey-Musolf and G.~Shaughnessy, \emph{{Singlet Higgs
  phenomenology and the electroweak phase transition}},
  \href{https://doi.org/10.1088/1126-6708/2007/08/010}{\emph{JHEP} {\bfseries
  08} (2007) 010} [\href{https://arxiv.org/abs/0705.2425}{{\ttfamily
  0705.2425}}].

\bibitem{Athron:2020sbe}
P.~Athron, C.~Bal\'azs, A.~Fowlie and Y.~Zhang, \emph{{PhaseTracer: tracing
  cosmological phases and calculating transition properties}},
  \href{https://doi.org/10.1140/epjc/s10052-020-8035-2}{\emph{Eur. Phys. J. C}
  {\bfseries 80} (2020) 567}
  [\href{https://arxiv.org/abs/2003.02859}{{\ttfamily 2003.02859}}].

\bibitem{Wainwright:2011kj}
C.L.~Wainwright, \emph{{CosmoTransitions: Computing Cosmological Phase
  Transition Temperatures and Bubble Profiles with Multiple Fields}},
  \href{https://doi.org/10.1016/j.cpc.2012.04.004}{\emph{Comput. Phys. Commun.}
  {\bfseries 183} (2012) 2006}
  [\href{https://arxiv.org/abs/1109.4189}{{\ttfamily 1109.4189}}].

\bibitem{Ham:2004cf}
S.W.~Ham, Y.S.~Jeong and S.K.~Oh, \emph{{Electroweak phase transition in an
  extension of the standard model with a real Higgs singlet}},
  \href{https://doi.org/10.1088/0954-3899/31/8/017}{\emph{J. Phys. G}
  {\bfseries 31} (2005) 857}
  [\href{https://arxiv.org/abs/hep-ph/0411352}{{\ttfamily hep-ph/0411352}}].

\bibitem{OConnell:2006rsp}
D.~O'Connell, M.J.~Ramsey-Musolf and M.B.~Wise, \emph{{Minimal Extension of the
  Standard Model Scalar Sector}},
  \href{https://doi.org/10.1103/PhysRevD.75.037701}{\emph{Phys. Rev. D}
  {\bfseries 75} (2007) 037701}
  [\href{https://arxiv.org/abs/hep-ph/0611014}{{\ttfamily hep-ph/0611014}}].

\bibitem{Ahriche:2007jp}
A.~Ahriche, \emph{{What is the criterion for a strong first order electroweak
  phase transition in singlet models?}},
  \href{https://doi.org/10.1103/PhysRevD.75.083522}{\emph{Phys. Rev. D}
  {\bfseries 75} (2007) 083522}
  [\href{https://arxiv.org/abs/hep-ph/0701192}{{\ttfamily hep-ph/0701192}}].

\bibitem{Barger:2007im}
V.~Barger, P.~Langacker, M.~McCaskey, M.J.~Ramsey-Musolf and G.~Shaughnessy,
  \emph{{LHC Phenomenology of an Extended Standard Model with a Real Scalar
  Singlet}}, \href{https://doi.org/10.1103/PhysRevD.77.035005}{\emph{Phys. Rev.
  D} {\bfseries 77} (2008) 035005}
  [\href{https://arxiv.org/abs/0706.4311}{{\ttfamily 0706.4311}}].

\bibitem{Espinosa:2011ax}
J.R.~Espinosa, T.~Konstandin and F.~Riva, \emph{{Strong Electroweak Phase
  Transitions in the Standard Model with a Singlet}},
  \href{https://doi.org/10.1016/j.nuclphysb.2011.09.010}{\emph{Nucl. Phys. B}
  {\bfseries 854} (2012) 592}
  [\href{https://arxiv.org/abs/1107.5441}{{\ttfamily 1107.5441}}].

\bibitem{No:2013wsa}
J.M.~No and M.J.~Ramsey-Musolf, \emph{{Probing the Higgs Portal at the LHC
  Through Resonant di-Higgs Production}},
  \href{https://doi.org/10.1103/PhysRevD.89.095031}{\emph{Phys. Rev. D}
  {\bfseries 89} (2014) 095031}
  [\href{https://arxiv.org/abs/1310.6035}{{\ttfamily 1310.6035}}].

\bibitem{Fuyuto:2014yia}
K.~Fuyuto and E.~Senaha, \emph{{Improved sphaleron decoupling condition and the
  Higgs coupling constants in the real singlet-extended standard model}},
  \href{https://doi.org/10.1103/PhysRevD.90.015015}{\emph{Phys. Rev. D}
  {\bfseries 90} (2014) 015015}
  [\href{https://arxiv.org/abs/1406.0433}{{\ttfamily 1406.0433}}].

\bibitem{Profumo:2014opa}
S.~Profumo, M.J.~Ramsey-Musolf, C.L.~Wainwright and P.~Winslow,
  \emph{{Singlet-catalyzed electroweak phase transitions and precision Higgs
  boson studies}},
  \href{https://doi.org/10.1103/PhysRevD.91.035018}{\emph{Phys. Rev. D}
  {\bfseries 91} (2015) 035018}
  [\href{https://arxiv.org/abs/1407.5342}{{\ttfamily 1407.5342}}].

\bibitem{Chen:2014ask}
C.-Y.~Chen, S.~Dawson and I.M.~Lewis, \emph{{Exploring resonant di-Higgs boson
  production in the Higgs singlet model}},
  \href{https://doi.org/10.1103/PhysRevD.91.035015}{\emph{Phys. Rev. D}
  {\bfseries 91} (2015) 035015}
  [\href{https://arxiv.org/abs/1410.5488}{{\ttfamily 1410.5488}}].

\bibitem{Sannino:2015wka}
F.~Sannino and J.~Virkaj\"arvi, \emph{{First Order Electroweak Phase Transition
  from (Non)Conformal Extensions of the Standard Model}},
  \href{https://doi.org/10.1103/PhysRevD.92.045015}{\emph{Phys. Rev. D}
  {\bfseries 92} (2015) 045015}
  [\href{https://arxiv.org/abs/1505.05872}{{\ttfamily 1505.05872}}].

\bibitem{Kozaczuk:2015owa}
J.~Kozaczuk, \emph{{Bubble Expansion and the Viability of Singlet-Driven
  Electroweak Baryogenesis}},
  \href{https://doi.org/10.1007/JHEP10(2015)135}{\emph{JHEP} {\bfseries 10}
  (2015) 135} [\href{https://arxiv.org/abs/1506.04741}{{\ttfamily
  1506.04741}}].

\bibitem{Ghosh:2015apa}
S.~Ghosh, A.~Kundu and S.~Ray, \emph{{Potential of a singlet scalar enhanced
  Standard Model}},
  \href{https://doi.org/10.1103/PhysRevD.93.115034}{\emph{Phys. Rev. D}
  {\bfseries 93} (2016) 115034}
  [\href{https://arxiv.org/abs/1512.05786}{{\ttfamily 1512.05786}}].

\bibitem{Kotwal:2016tex}
A.V.~Kotwal, M.J.~Ramsey-Musolf, J.M.~No and P.~Winslow,
  \emph{{Singlet-catalyzed electroweak phase transitions in the 100 TeV
  frontier}}, \href{https://doi.org/10.1103/PhysRevD.94.035022}{\emph{Phys.
  Rev. D} {\bfseries 94} (2016) 035022}
  [\href{https://arxiv.org/abs/1605.06123}{{\ttfamily 1605.06123}}].

\bibitem{Kanemura:2016lkz}
S.~Kanemura, M.~Kikuchi and K.~Yagyu, \emph{{One-loop corrections to the Higgs
  self-couplings in the singlet extension}},
  \href{https://doi.org/10.1016/j.nuclphysb.2017.02.004}{\emph{Nucl. Phys. B}
  {\bfseries 917} (2017) 154}
  [\href{https://arxiv.org/abs/1608.01582}{{\ttfamily 1608.01582}}].

\bibitem{Lewis:2017dme}
I.M.~Lewis and M.~Sullivan, \emph{{Benchmarks for Double Higgs Production in
  the Singlet Extended Standard Model at the LHC}},
  \href{https://doi.org/10.1103/PhysRevD.96.035037}{\emph{Phys. Rev. D}
  {\bfseries 96} (2017) 035037}
  [\href{https://arxiv.org/abs/1701.08774}{{\ttfamily 1701.08774}}].

\bibitem{Chen:2017qcz}
C.-Y.~Chen, J.~Kozaczuk and I.M.~Lewis, \emph{{Non-resonant Collider Signatures
  of a Singlet-Driven Electroweak Phase Transition}},
  \href{https://doi.org/10.1007/JHEP08(2017)096}{\emph{JHEP} {\bfseries 08}
  (2017) 096} [\href{https://arxiv.org/abs/1704.05844}{{\ttfamily
  1704.05844}}].

\bibitem{Ghorbani:2021rgs}
P.~Ghorbani, \emph{{Vacuum stability vs. positivity in real singlet scalar
  extension of the standard model}},
  \href{https://doi.org/10.1016/j.nuclphysb.2021.115533}{\emph{Nucl. Phys. B}
  {\bfseries 971} (2021) 115533}
  [\href{https://arxiv.org/abs/2104.09542}{{\ttfamily 2104.09542}}].

\bibitem{Huang:2022him}
P.~Huang and K.-P.~Xie, \emph{{Primordial black holes from an electroweak phase
  transition}}, \href{https://doi.org/10.1103/PhysRevD.105.115033}{\emph{Phys.
  Rev. D} {\bfseries 105} (2022) 115033}
  [\href{https://arxiv.org/abs/2201.07243}{{\ttfamily 2201.07243}}].

\bibitem{Coleman:1973jx}
S.R.~Coleman and E.J.~Weinberg, \emph{{Radiative Corrections as the Origin of
  Spontaneous Symmetry Breaking}},
  \href{https://doi.org/10.1103/PhysRevD.7.1888}{\emph{Phys. Rev. D} {\bfseries
  7} (1973) 1888}.

\bibitem{Quiros:1999jp}
M.~Quiros, \emph{{Finite temperature field theory and phase transitions}},  in
  \emph{{ICTP Summer School in High-Energy Physics and Cosmology}},
  pp.~187--259, 1, 1999 [\href{https://arxiv.org/abs/hep-ph/9901312}{{\ttfamily
  hep-ph/9901312}}].

\bibitem{Parwani:1991gq}
R.R.~Parwani, \emph{{Resummation in a hot scalar field theory}},
  \href{https://doi.org/10.1103/PhysRevD.45.4695}{\emph{Phys. Rev. D}
  {\bfseries 45} (1992) 4695}
  [\href{https://arxiv.org/abs/hep-ph/9204216}{{\ttfamily hep-ph/9204216}}].

\bibitem{Lerner:2009xg}
R.N.~Lerner and J.~McDonald, \emph{{Gauge singlet scalar as inflaton and
  thermal relic dark matter}},
  \href{https://doi.org/10.1103/PhysRevD.80.123507}{\emph{Phys. Rev. D}
  {\bfseries 80} (2009) 123507}
  [\href{https://arxiv.org/abs/0909.0520}{{\ttfamily 0909.0520}}].

\bibitem{ATLAS:2016neq}
{\scshape ATLAS, CMS} collaboration, \emph{{Measurements of the Higgs boson
  production and decay rates and constraints on its couplings from a combined
  ATLAS and CMS analysis of the LHC pp collision data at $ \sqrt{s}=7 $ and 8
  TeV}}, \href{https://doi.org/10.1007/JHEP08(2016)045}{\emph{JHEP} {\bfseries
  08} (2016) 045} [\href{https://arxiv.org/abs/1606.02266}{{\ttfamily
  1606.02266}}].

\bibitem{Carena:2018vpt}
M.~Carena, Z.~Liu and M.~Riembau, \emph{{Probing the electroweak phase
  transition via enhanced di-Higgs boson production}},
  \href{https://doi.org/10.1103/PhysRevD.97.095032}{\emph{Phys. Rev. D}
  {\bfseries 97} (2018) 095032}
  [\href{https://arxiv.org/abs/1801.00794}{{\ttfamily 1801.00794}}].

\bibitem{CMS:2020gsy-new}
{\scshape CMS} collaboration, \emph{{Combined Higgs boson production and decay
  measurements with up to 137 fb$^{-1}$ of proton-proton collision data at
  $\sqrt s$ = 13 TeV}},  Tech. Rep. CERN, Geneva (2020).

\bibitem{ATLAS:2022vkf}
{\scshape ATLAS} collaboration, \emph{{A detailed map of Higgs boson
  interactions by the ATLAS experiment ten years after the discovery}},
  \href{https://doi.org/10.1038/s41586-022-04893-w}{\emph{Nature} {\bfseries
  607} (2022) 52} [\href{https://arxiv.org/abs/2207.00092}{{\ttfamily
  2207.00092}}].

\bibitem{Croon:2018new}
D.~Croon and G.~White, \emph{{Exotic Gravitational Wave Signatures from
  Simultaneous Phase Transitions}},
  \href{https://doi.org/10.1007/JHEP05(2018)210}{\emph{JHEP} {\bfseries 05}
  (2018) 210} [\href{https://arxiv.org/abs/1803.05438}{{\ttfamily
  1803.05438}}].

\bibitem{Morais:2018uou}
A.P.~Morais, R.~Pasechnik and T.~Vieu, \emph{{Multi-peaked signatures of
  primordial gravitational waves from multi-step electroweak phase
  transition}}, \href{https://doi.org/10.22323/1.364.0054}{\emph{PoS}
  {\bfseries EPS-HEP2019} (2020) 054}
  [\href{https://arxiv.org/abs/1802.10109}{{\ttfamily 1802.10109}}].

\bibitem{Morais:2019fnm}
A.P.~Morais and R.~Pasechnik, \emph{{Probing multi-step electroweak phase
  transition with multi-peaked primordial gravitational waves spectra}},
  \href{https://doi.org/10.1088/1475-7516/2020/04/036}{\emph{JCAP} {\bfseries
  04} (2020) 036} [\href{https://arxiv.org/abs/1910.00717}{{\ttfamily
  1910.00717}}].

\bibitem{Angelescu:2018dkk}
A.~Angelescu and P.~Huang, \emph{{Multistep Strongly First Order Phase
  Transitions from New Fermions at the TeV Scale}},
  \href{https://doi.org/10.1103/PhysRevD.99.055023}{\emph{Phys. Rev. D}
  {\bfseries 99} (2019) 055023}
  [\href{https://arxiv.org/abs/1812.08293}{{\ttfamily 1812.08293}}].

\bibitem{Zhao:2022cnn}
Z.~Zhao, Y.~Di, L.~Bian and R.-G.~Cai, \emph{{Probing the electroweak symmetry
  breaking history with Gravitational waves}},
  \href{https://arxiv.org/abs/2204.04427}{{\ttfamily 2204.04427}}.

\bibitem{Weir:2017wfa}
D.J.~Weir, \emph{{Gravitational waves from a first order electroweak phase
  transition: a brief review}},
  \href{https://doi.org/10.1098/rsta.2017.0126}{\emph{Phil. Trans. Roy. Soc.
  Lond. A} {\bfseries 376} (2018) 20170126}
  [\href{https://arxiv.org/abs/1705.01783}{{\ttfamily 1705.01783}}].

\bibitem{Masoumi:2016wot}
A.~Masoumi, K.D.~Olum and B.~Shlaer, \emph{{Efficient numerical solution to
  vacuum decay with many fields}},
  \href{https://doi.org/10.1088/1475-7516/2017/01/051}{\emph{JCAP} {\bfseries
  01} (2017) 051} [\href{https://arxiv.org/abs/1610.06594}{{\ttfamily
  1610.06594}}].

\bibitem{Athron:2019nbd}
P.~Athron, C.~Bal\'azs, M.~Bardsley, A.~Fowlie, D.~Harries and G.~White,
  \emph{{BubbleProfiler: finding the field profile and action for cosmological
  phase transitions}},
  \href{https://doi.org/10.1016/j.cpc.2019.05.017}{\emph{Comput. Phys. Commun.}
  {\bfseries 244} (2019) 448}
  [\href{https://arxiv.org/abs/1901.03714}{{\ttfamily 1901.03714}}].

\bibitem{Sato:2019wpo}
R.~Sato, \emph{{SimpleBounce : a simple package for the false vacuum decay}},
  \href{https://doi.org/10.1016/j.cpc.2020.107566}{\emph{Comput. Phys. Commun.}
  {\bfseries 258} (2021) 107566}
  [\href{https://arxiv.org/abs/1908.10868}{{\ttfamily 1908.10868}}].

\bibitem{Guada:2020xnz}
V.~Guada, M.~Nemev\v{s}ek and M.~Pintar, \emph{{FindBounce: Package for
  multi-field bounce actions}},
  \href{https://doi.org/10.1016/j.cpc.2020.107480}{\emph{Comput. Phys. Commun.}
  {\bfseries 256} (2020) 107480}
  [\href{https://arxiv.org/abs/2002.00881}{{\ttfamily 2002.00881}}].

\bibitem{No:2009thesis}
J.M.~No, \emph{Aspects of Phenomenology and Cosmology in Hidden sector
  extensions of the Standard Model}, Ph.D. thesis, Universidad Aut\'onoma de
  Madrid, 06, 2009.

\bibitem{Huber:2007vva}
S.J.~Huber and T.~Konstandin, \emph{{Production of gravitational waves in the
  nMSSM}}, \href{https://doi.org/10.1088/1475-7516/2008/05/017}{\emph{JCAP}
  {\bfseries 05} (2008) 017} [\href{https://arxiv.org/abs/0709.2091}{{\ttfamily
  0709.2091}}].

\bibitem{Guth:1982pn}
A.H.~Guth and E.J.~Weinberg, \emph{{Could the Universe Have Recovered from a
  Slow First Order Phase Transition?}},
  \href{https://doi.org/10.1016/0550-3213(83)90307-3}{\emph{Nucl. Phys. B}
  {\bfseries 212} (1983) 321}.

\bibitem{Hawking:1982ga}
S.W.~Hawking, I.G.~Moss and J.M.~Stewart, \emph{{Bubble Collisions in the Very
  Early Universe}}, \href{https://doi.org/10.1103/PhysRevD.26.2681}{\emph{Phys.
  Rev. D} {\bfseries 26} (1982) 2681}.

\bibitem{ParticleDataGroup:2020ssz}
{\scshape Particle Data Group} collaboration, \emph{{Review of Particle
  Physics}}, \href{https://doi.org/10.1093/ptep/ptaa104}{\emph{PTEP} {\bfseries
  2020} (2020) 083C01}.

\bibitem{Elias-Miro:2014pca}
J.~Elias-Mir\'o, J.R.~Espinosa and T.~Konstandin, \emph{{Taming Infrared
  Divergences in the Effective Potential}},
  \href{https://doi.org/10.1007/JHEP08(2014)034}{\emph{JHEP} {\bfseries 08}
  (2014) 034} [\href{https://arxiv.org/abs/1406.2652}{{\ttfamily 1406.2652}}].

\bibitem{Martin:2014bca}
S.P.~Martin, \emph{{Taming the Goldstone contributions to the effective
  potential}}, \href{https://doi.org/10.1103/PhysRevD.90.016013}{\emph{Phys.
  Rev. D} {\bfseries 90} (2014) 016013}
  [\href{https://arxiv.org/abs/1406.2355}{{\ttfamily 1406.2355}}].

\bibitem{Espinosa:2010hh}
J.R.~Espinosa, T.~Konstandin, J.M.~No and G.~Servant, \emph{{Energy Budget of
  Cosmological First-order Phase Transitions}},
  \href{https://doi.org/10.1088/1475-7516/2010/06/028}{\emph{JCAP} {\bfseries
  06} (2010) 028} [\href{https://arxiv.org/abs/1004.4187}{{\ttfamily
  1004.4187}}].

\bibitem{Freitas:2021yng}
F.F.~Freitas, G.~Louren\c{c}o, A.P.~Morais, A.~Nunes, J.a.~Ol\'\i{}via,
  R.~Pasechnik et~al., \emph{{Impact of SM parameters and of the vacua of the
  Higgs potential in gravitational waves detection}},
  \href{https://arxiv.org/abs/2108.12810}{{\ttfamily 2108.12810}}.

\bibitem{Azatov:2022tii}
A.~Azatov, G.~Barni, S.~Chakraborty, M.~Vanvlasselaer and W.~Yin,
  \emph{{Ultra-relativistic bubbles from the simplest Higgs portal and their
  cosmological consequences}},
  \href{https://doi.org/10.1007/JHEP10(2022)017}{\emph{JHEP} {\bfseries 10}
  (2022) 017} [\href{https://arxiv.org/abs/2207.02230}{{\ttfamily
  2207.02230}}].

\end{thebibliography}\endgroup
\bibliographystyle{JHEP}

\end{document}